\documentclass[11pt]{article}
\pdfoutput=1
\usepackage{jheppub}
\usepackage{amssymb,amsfonts,amsmath,amsthm}
\usepackage{mathrsfs}
\usepackage{bbold}
\usepackage{tikz}
\usetikzlibrary{shapes.geometric, arrows}
\usepackage{comment}
\usepackage{tikz}
\usetikzlibrary{shapes.geometric,calc}
\usetikzlibrary{shapes,arrows,chains}
\usetikzlibrary{decorations.markings}
\usetikzlibrary{decorations.pathmorphing}
\usetikzlibrary{shapes.multipart}
\tikzset{snake it/.style={decorate, decoration=snake}}

\makeatletter
\newcommand{\Vast}{\bBigg@{4.75}}
\makeatother

\newcommand{\beq}{\begin{equation}\begin{aligned}}
\newcommand{\eeq}{\end{aligned}\end{equation}}

\newcommand{\ep}{\epsilon}

\newcommand{\x}{\times}

\newcommand{\texr}{\textcolor{red}}
\newcommand{\texb}{\textcolor{blue}}

\newcommand{\ov}{\over}

\newcommand\rH{\mathrm H}

\newcommand\bC{\mathbb{C}}
\newcommand\rU{\mathrm{U}}
\newcommand\cA{\mathcal A}

\newcommand\bZ{\mathbb Z}
\newcommand\cC{\mathcal C}
\newcommand\cM{\mathcal M}

\newcommand\cF{\mathcal F}

\newcommand\cD{\mathcal D}
\newcommand\cZ{\mathcal Z}

\newcommand{\rC}{\mathrm{C}}

\newcommand\bR{\mathbb R}

\newcommand\ol{\overline}

\DeclareMathOperator{\SU}{SU}
\DeclareMathOperator{\Sp}{Sp}

\DeclareMathOperator{\End}{End}
\DeclareMathOperator{\Spin}{Spin}
\DeclareMathOperator{\Ising}{Ising}

\newcommand\longto\longrightarrow
\newcommand\longfrom\longleftarrow

\newtheorem{dummy}{Dummy}[section]

\newtheorem{theorem}[dummy]{Theorem}

\theoremstyle{definition}

\usetikzlibrary{shapes,arrows,chains}
\usetikzlibrary{decorations.markings}
\usetikzlibrary{decorations.pathmorphing}
\usetikzlibrary{shapes.multipart}
\tikzset{snake it/.style={decorate, decoration=snake}}
\usepackage{tikz-cd}
\title{\ \vspace{-1cm} \\
\scalebox{0.8}{Gauging Categorical Symmetries in 3d Topological Orders }
\scalebox{0.8}{ and Bulk Reconstruction}}

\author{
Matthew Yu
}

\emailAdd{myu@perimeterinstitute.ca}
\affiliation{
        Perimeter Institute for Theoretical Physics\\
        31 Caroline St N, Waterloo, ON N2L 2Y5, Canada}
\abstract{We use the language of categorical condensation to give a procedure for gauging nonabelian anyons, which are the manifestations of categorical symmetries in three spacetime dimensions. We also describe how the condensation procedure
can be used in other contexts such as for topological cosets and constructing modular invariants. By studying a generalization of which anyons are condensable, we arrive at representations of congruence subgroups of the modular group. We finally present an analysis for ungauging anyons, which is related to the problem of constructing a Drinfeld center for a fusion category; this procedure we refer to as bulk reconstruction. We introduce a set of consistency relations regarding lines in the parent theory and wall category. Through use of these relations along with the $S$-matrix elements of the child theory, we construct $S$-matrix elements of a parent theory in a number of examples. }
\begin{document}

\maketitle
\vfill\eject

\section{Introduction}\label{sec:Intro}
The study of topological operators in quantum field theories has given many insights into the nature of what a full quantum field theory consists of. The topological operators provide a vast simplification from the space of all possible operators that a theory may possess, and the formalism to understand them is through topological quantum field theories (TQFTs).  A particularly useful feature of TQFTs is their ability to describe, and in some cases classify, the infrared phases of gauge theories and gapped phases of matter.  Among the classification of topological phases are those phases which are nontrivially ordered, also known as ``long range entangled" phases or \textit{topological orders}.  The topological properties of the phase are independent of spacetime or internal symmetries, and only depend on the global structure of the manifold that the phase lives on.   In such long range entangled phases in ($n$+1)-dimensions there exists  extended topological operators, with the structure of an $n$-category,  the classifications for low values of $n$ have been given in 
\cite{wen2016theory,lan2018classification,lan2019classification,Johnson-Freyd:2020usu,Johnson-Freyd:2021tbq}.

In this paper we restrict to topological theories in three spacetime dimensions, with a focus on the line operators that are the \textit{anyons}.  The classification of topological orders in three spacetime dimensions is given by modular tensor categories (MTCs) \footnote{Modular tensor categories technically only classify topological orders up to an invertible phase}; for the purposes of this paper, we will represent these MTCs by 3d Chern-Simons theories, where the details about the framing of our underlying three-manifold is unimportant. Given the spectrum of line operators, one can perform \textit{anyon condensation},
which is an action in three dimensions that also goes by the name of ``gauging a one-form symmetry", or more generally ``gauging a categorical symmetry". 
When an anyon generates a one-form symmetry, it has abelian fusion rules, as higher-form symmetries are always abelian groups \cite{Gaiotto:2014kfa}. The anyon is deemed an \textit{abelian anyon} and the action of condensing abelian anyons is well studied in the literature \cite{Hsin:2018vcg,Bais:2008ni,burnell2018anyon,lan2018thesis,Lou:2020gfq}.

 When the anyon has nonabelian fusion rules, i.e. a \textit{nonabelian anyon}, we must shift to a categorical point of view to understand condensation \cite{Gaiotto:2019xmp,cui2016gauging}.  In the categorical framework we see the anyon, or set of anyons that condense, as being part of an algebra object. More specifically, a \textit{special Frobenius algebra} in the category $\cC$.  From here on out, $\cC$ denotes the uncondensed theory we start with, or in condensed matter parlance the ``parent theory".   Condensing the algebra leads to the ``child" theory $\cD$, where some of the lines in the parent have been projected out, or confined on an interface that arises in the process of going from parent to child.  In the case where the child theory is the vacuum, the interface that separates $\cC$ and the vacuum is deemed to be a \textit{gapped boundary} of $\cC$.  In order to go from $\cC$ to the vacuum one condenses a \textit{Lagrangian} algebra object $\cA_\ell$, where $(\dim \cA_\ell)^2 = \dim \cC = \sum_{\lambda\in \cC } (\text{dim} \lambda)^2$, where the sum ranges over all lines in $\cC$ and we use dimension to mean quantum dimension.  In the literature, the use of the phrase ``anyon condensation" is at times used to apply solely to those integer spin, i.e. bosonic anyons, which give a Lagrangian algebra, and condense $\cC$ to the vacuum \cite{kong2014anyon}. 
 For Lagrangian algebras, it is a theorem that 
\begin{theorem} \cite{davydov2013witt}
For $\mathcal{F}$  a fusion category and $\cC = \cZ(\mathcal F)$. There is a bijection between the sets of Lagrangian algebras in $\cC$ and indecomposable $\cF$-module categories.
\end{theorem}
The role of the fusion category in the above theorem is played by the lines on the interface, that we denote as $\cF$, separating $\cC$ and $\cD$.
While the procedure for determining the lines of the child theory when gauging a one-form symmetry is clear, there are few examples in the literature that perform nonabelian condensation at the level of the spectrum of lines for an MTC.   We set out to outline an algorithm for performing nonabelian condensation, i.e. determining the modules of the condensation algebra in an efficient way, and perform many nontrivial examples of determining not only the spectrum of lines in the child theory but also their quantum dimensions. 

 For our purposes, we will weaken the notion of condensation only being applicable for Lagrangian algebras
 and apply the condensation procedure, which involves finding modules of algebra objects, to a variety of algebras.  The reason for doing this is because the condensation procedure has uses that go beyond just looking for gapped boundaries, and one of our goals is to provide examples that emphasize the other merits.  
It is natural to expect that not all anyons in $\cC$ can be condensed because some do not correspond to an algebra. Using our algorithm we will give examples of how to decide if an algebra is condensable. 
 With the tools for nonabelian condensation developed, we can apply them to verify conformal embeddings given in \cite{davydov2013witt}, and also to other cases where one might ask if two MTCs are Morita equivalent. This gives us a way to construct the interface, i.e. bimodules, between the two theories. Moreover we can use nonabelian condensation to understand the decomposition of characters in 2d topological cosets, which have been useful in describing the IR phases in \cite{Delmastro:2021otj}.
 
 In many instances taking all the bosons anyons and condensing them out may cause lines to split, but in such a way that preserves the quantum dimension.
 As a first step in generalizing beyond bosonic condensation, we look at fermion condensation where by fermion we mean a line with half integer spin. On the other hand, we will use the term \textit{local fermion} to describe the condensed line.
 As we will see, a nonabelian fermion may be split into one that is abelian, and we can furthermore sequentially condense out the abelian part.  We will investigate how this relates to the (super)modular invariants of the parent theory, and see what further insights the condensation algebra can give regarding modular invariants.  
 
 Along the way we will enlarge the notion of which anyons can be condensed, beyond bosons and fermions to a general spin $1/n$ object, if we also couple to an appropriate background $n$-structure \cite{Carqueville:2021cfa}.    We also observe that not all modular invariants correspond to gapped interfaces, like those that arise from Lagrangian algebras, as noted in \cite{Kawahigashi:2015lxa,davydov2016unphysical}.  One way this fails to be true is that there are ``charge conjugation" modular invariants that reflect some symmetry of the parent theory. Furthermore, an algebra that is at least symmetric Frobenius will result in a modular invariant, however, these need not be Lagrangian and therefore the modular invariant is not a truly gapped interface.
  We further supplement the analysis given in the references with more explicit examples of exotic idempotent modular invariants, and relationships between the modular invariants and condensation algebras. In the same manner as for supermodular invariants, we look to the higher modular invariants corresponding to condensing a $1/n$ line to support our claim that these lines can be condensed.  

With a comprehensive understanding of gauging, we next aim to understand how to construct the center of the fusion category on the wall that separates $\cC$ and $\cD$, i.e. reconstructing $\cC$ to some degree by ungauging the algebra used to reach the child theory.  In particular we want to start off with information about the ``wall category", this consists of  the lines that can be confined on the wall. These are the lines that are projected out in going from $\cC \to \cD$, as well as the lines of the child theory $\cD$. We will slightly abuse notation and call this fusion category $\mathcal{F}$ (this is a surface defect, but contains two kinds of lines); note that the lines which are totally confined cannot lift to the child $\cD$, so there is no braided structure on the 2d surface that separates the two phases. The lines in $\cD$ however can be moved to the surface $\cF$ via a functor, and is the reason for our abuse of notation.  We use the  consistency relations mentioned in \cite{Fuchs:2012dt}, and others which we elaborate on, to show in some nontrivial cases that the data of the $S$-matrix elements of $\cC$ can be constructed. The data we start out with involves the $S$-matrix elements of the lines in $\cD$, as well as fusion information of the wall category.  Constructing $\cC$ is not a very methodical process and there is no known procedure that exists in general.  We gain an intuition from the examples in this paper on how much information we can reasonably extract, given our initial data.  

 The layout of the  paper is as follows, in \S\ref{overviewofgauging} we give a mathematical formalism associated to gauging a categorical symmetry in terms of condensation algebras. We follow up by giving explicit examples of how to compute using this formalism by applying it to 3d Chern-Simons theories, and finding the lines of the child theory. More nontrivial examples of gauging are given in appendix \ref{morenonabelianexaples}.  In \S\ref{MIandcondensation} we look at modular invariants and see how in some cases we can identify which algebra objects of the parent theory can lead to a modular invariant. We also introduce supermodular invariants and remark on their feature, as well as discuss generalizations to higher modular invariants that are motivated by the spin of the anyon one can condense. In \S\ref{ungaguginganyons} we give the consistency relations involving the lines on the wall category and see how to determine $S$-matrix elements of the parent theory. We also explain the information that we will provide regarding the fusion category, to be able to determine its center.
 We will put the consistency relations to use in a couple of examples namely in reconstructing the Toric code from the vacuum and $\SU(3)_3$ from $\Spin(8)_1$.  In appendix \ref{reconstructingsu210} we do a nontrivial example with reconstructing the $S$-matrix of $\SU(2)_{10}$ from $\Spin(5)_1$. 
 
 \section{Overview of Gauging}\label{overviewofgauging}

We will perform condensation via a method of introducing \textit{idempotents}. The formalism developed using idempotents and condensation monads in precisely what is needed to do nonabelian condensation, and it furthermore generalizes to higher categories \cite{Gaiotto:2019xmp} \footnote{For a discussion specified to 2-categories see \cite{douglas2018fusion}, where the notion of condensation is referred to as “separable adjunction”.}.  With this rigorous framework in place, the well known notions of anyon condensation in 3d, or simple current extensions in 2d VOAs, can be encapsulated in a common language that generalizes to higher dimensions. To better interpret the mathematical formalism we restrict out attention from general $n$-categories to modular tensor categories, and in particular 3d Chern-Simons. Already here, many of the properties that generalize to $n$-categories are manifest, and computationally tractable.   We will give some examples of performing a familiar task of condensing abelian anyons by this method, while also shedding light on some of the subtleties that traditional methods miss.  Having some familiarity with the steps involved in the procedure will be crucial when we generalize to the nonabelian story.

We first review the properties of idempotents, working just with a linear monoidal 1-category $\cC$.  For an object $X \in \cC$ (we will later use $\cC$ as our parent MTC, and $X$ as our anyons) an idempotent is an endomorphism  $\varphi: X \to X$ such that $\varphi \circ \varphi = \varphi$.
For the purpose of this paper, the categories we will consider are all \textit{idempotent complete}. This means that we can write $\varphi$ using a pair of morphisms $f: X\to Y $ and $g: Y \to X$ as $\varphi = g\circ f$ so that $Y$ is a direct summand of $X$ and is the image of $\varphi$.  We will also work in a finite setting, so that any decomposition into direct sums, is a finite decomposition into simple objects. Such finiteness conditions are a key feature associated with ``topological settings" and generalize to higher categories where the  finiteness properties are captured by the axioms of a multifusion category \footnote{As an example in lower categories, if one is working in representation theory, the finiteness conditions we consider boil down to the axioms when working with a semisimple finite dimensional algebra.}.

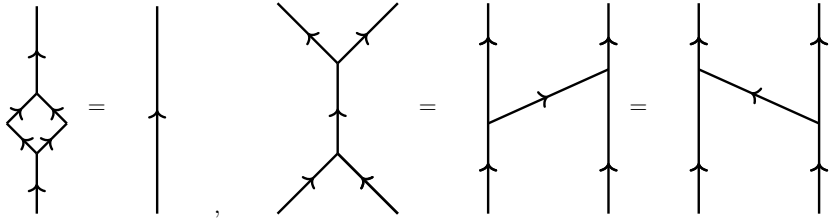
\begin{figure}
    \centering
    \scalebox{.8}{
   \begin{tikzpicture}
   \draw[decoration={markings, mark=at position .5 with
 {\arrow{>}}}, postaction={decorate},line width= .4mm] (-4,0)--(-4,1);
   \draw[decoration={markings, mark=at position .5 with
 {\arrow{>}}}, postaction={decorate},line width= .4mm] (-4,1)--(-3.5,1.5);
 \draw[decoration={markings, mark=at position .5 with
 {\arrow{>}}}, postaction={decorate},line width= .4mm] (-3.5,1.5)--(-4,2);
 \draw[decoration={markings, mark=at position .5 with
 {\arrow{>}}}, postaction={decorate},line width= .4mm] (-4,1)--(-4.5,1.5);
 \draw[decoration={markings, mark=at position .5 with
 {\arrow{>}}}, postaction={decorate},line width= .4mm] (-4.5,1.5)--(-4,2);
 \draw[decoration={markings, mark=at position .5 with
 {\arrow{>}}}, postaction={decorate},line width= .4mm] (-4,2)--(-4,3.45);
 \draw (-3,1.75) node{$=$};
    \draw[decoration={markings, mark=at position .5 with
 {\arrow{>}}}, postaction={decorate},line width= .4mm] (-2,0)--(-2,3.45);
 \draw (-1,0) node{$,$};
       \draw[decoration={markings, mark=at position .5 with
 {\arrow{>}}}, postaction={decorate},line width= .4mm] (1,1)--(1,2.5);
       \draw[decoration={markings, mark=at position .5 with
 {\arrow{<}}}, postaction={decorate},line width = .4mm](1,1)--(2,0);
 \draw[decoration={markings, mark=at position .5 with
 {\arrow{<}}}, postaction={decorate},line width = .4mm](1,1)--(2,0);
 \draw[decoration={markings, mark=at position .5 with
 {\arrow{<}}}, postaction={decorate},line width = .4mm](1,1)--(0,0);
 \draw[decoration={markings, mark=at position .5 with
 {\arrow{>}}}, postaction={decorate},line width = .4mm](1,2.5)--(2,3.5);
 \draw[decoration={markings, mark=at position .5 with
 {\arrow{>}}}, postaction={decorate},line width = .4mm](1,2.5)--(0,3.5);
 \draw (2.5,1.75) node{$=$};
 \draw[decoration={markings, mark=at position .2 with
 {\arrow{<}}}, postaction={decorate},decoration={markings, mark=at position .8 with
 {\arrow{<}}}, postaction={decorate},line width = .4mm](3.5,3.5)--(3.5,0);
 \draw[decoration={markings, mark=at position .5 with
 {\arrow{>}}}, postaction={decorate},line width = .4mm](3.5,1.5)--(5.5,2.4);
 \draw[decoration={markings, mark=at position .2 with
 {\arrow{<}}}, postaction={decorate},decoration={markings, mark=at position .8 with
 {\arrow{<}}}, postaction={decorate},line width = .4mm](5.5,3.5)--(5.5,0);
 \draw (6,1.75) node{$=$};
  \draw[decoration={markings, mark=at position .2 with
 {\arrow{<}}}, postaction={decorate},decoration={markings, mark=at position .8 with
 {\arrow{<}}}, postaction={decorate},line width = .4mm](7,3.5)--(7,0);
 \draw[decoration={markings, mark=at position .5 with
 {\arrow{<}}}, postaction={decorate},line width = .4mm](7,2.4)--(9,1.5);
 \draw[decoration={markings, mark=at position .2 with
 {\arrow{<}}}, postaction={decorate},decoration={markings, mark=at position .8 with
 {\arrow{<}}}, postaction={decorate},line width = .4mm](9,3.5)--(9,0);
   \end{tikzpicture}}
   \caption{The diagram on the left is the axiom that multiplication and comultiplication can be composed into $\varphi$, i.e. all bubbles can be closed. The diagram on the right shows that the composition of comultiplcation and multiplication can be decomposed as a composition of
   ($\text{id}_\varphi\times$multiplcation) and (comultiplcation$\times\text{id}_{\varphi}$) or
  (multiplication$\times \text{id}_\varphi$) and ($\text{id}_\varphi\times$comultiplication)   }\label{firstsetaxioms}
\end{figure}

\begin{figure}
    \centering
    \scalebox{.8}{
    \begin{tikzpicture}[thick, scale = .8]
    \draw[decoration={markings, mark=at position .2 with
 {\arrow{<}}}, postaction={decorate},decoration={markings, mark=at position .8 with
 {\arrow{<}}}, postaction={decorate},line width = .4mm] (3-2,3) -- (1-2,0);
    \draw[decoration={markings, mark=at position .5 with
 {\arrow{<}}}, postaction={decorate},line width = .4mm] (-1-2,3) -- (1-2,0);
    \draw[decoration={markings, mark=at position .6 with
 {\arrow{<}}}, postaction={decorate},line width = .4mm] (1-2,0) -- (1-2,-2);
    \draw[decoration={markings, mark=at position .6 with
 {\arrow{>}}}, postaction={decorate},line width = .4mm] (2-2,1.55) -- (1-2,3);
 
 \draw (1.5,0) node{$=$};
  
     \draw[decoration={markings, mark=at position .5 with
 {\arrow{<}}}, postaction={decorate},line width = .4mm] (3+3,3) -- (1+3,0);
    \draw[decoration={markings, mark=at position .7 with
 {\arrow{<}}}, postaction={decorate},decoration={markings, mark=at position .25 with
 {\arrow{<}}}, postaction={decorate},line width = .4mm] (-1+3,3) -- (1+3,0);
    \draw[decoration={markings, mark=at position .6 with
 {\arrow{<}}}, postaction={decorate},line width = .4mm] (1+3,0) -- (1+3,-2);
    \draw[decoration={markings, mark=at position .5 with
 {\arrow{>}}}, postaction={decorate},line width = .4mm] (0+3,1.55) -- (1+3,3);
 
 \draw (7,-2) node{$,$};

  \draw[decoration={markings, mark=at position .5 with
 {\arrow{<}}}, postaction={decorate},line width = .4mm] (3+9,-2) -- (1+9,1);
    \draw[decoration={markings, mark=at position .7 with
 {\arrow{<}}}, postaction={decorate},decoration={markings, mark=at position .25 with
 {\arrow{<}}}, postaction={decorate},line width = .4mm] (-1+9,-2) -- (1+9,1);
    \draw[decoration={markings, mark=at position .6 with
 {\arrow{<}}}, postaction={decorate},line width = .4mm] (1+9,3) -- (1+9,1);
    \draw[decoration={markings, mark=at position .5 with
 {\arrow{>}}}, postaction={decorate},line width = .4mm] (-0.1+9,-.65) -- (1+9,-2);
 
 \draw (12.5,0) node {$=$};

   \draw[decoration={markings, mark=at position .25 with
 {\arrow{<}}}, postaction={decorate},line width = .4mm] (3+9+5,-2) -- (1+9+5,1);
    \draw[, postaction={decorate},decoration={markings, mark=at position .5 with
 {\arrow{<}}}, postaction={decorate},line width = .4mm] (-1+9+5,-2) -- (1+9+5,1);
    \draw[decoration={markings, mark=at position .6 with
 {\arrow{<}}}, postaction={decorate},line width = .4mm] (1+9+5,3) -- (1+9+5,1);
    \draw[decoration={markings, mark=at position .5 with
 {\arrow{>}}}, postaction={decorate},line width = .4mm] (-0.1+9+5+2.2,-.65) -- (1+9+5,-2);
    \end{tikzpicture}}
    \caption{The left diagram shows that multiplication is associatve. The right diagram shows that comultiplication is coassociative}\label{F-move Postnikov}
\end{figure}
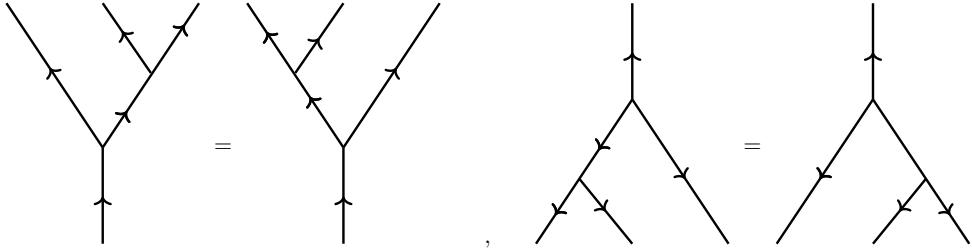

The idempotent $\varphi$ will also be referred to as a condensation algebra in $\cC$, and to perform a condensation, we first must select a finite set of lines to build this semisimple object.
The condensation algebra consists of the data $\varphi \in \cC$ as well as a multiplication map $\varphi \times \varphi \to \varphi$ and a co-multiplication map $\varphi \to \varphi \times \varphi$, and a set of axioms given in figure \ref{firstsetaxioms} and figure \ref{F-move Postnikov} where the line with an arrow denotes $\varphi$. 

It is known that these axioms for $\varphi$ make it into a nonunital special Frobenius algebra.  Condensing this algebra means to flood spacetime with a fine network of lines corresponding to the algebra, and satisfying the axioms of associative (co)multiplication, and composition of comultiplication and multiplication.  The importance of the axioms is to insure that the choice of which network to flood spacetime with is immaterial, i.e. works for any cellulation of spacetime.  With this we can assemble a topological interface, which is two-dimensional interface that is populated by the one dimensional algebra. Since we were able to build this higher dimensional object from the lower dimensional lines in $\cC$, this interface will be called a \textit{condensation descendant} to reflect its fundamental structure, and this notion can be used to classify topological orders as in \cite{Johnson-Freyd:2021tbq,Johnson-Freyd:2020usu}. It is a fact that in 3d Chern-Simons theory, all the surfaces are built out of lines and thus are descendants of the condensation algebra \cite{Carqueville:2017ono}. Two interfaces are isomorphic if the two condensation algebras are Morita equivalent. This fact about interfaces will play a role later in our discussion of modular invariants for $\cC$. The term anyon condensation is sometimes used in the literature to refer to the case when we condense an $\varphi$ with the lines that comprise $\varphi$ actually forming a Lagrangian algebra in $\cC$ \cite{2015,kong2014anyon}.  

Physically speaking, condensing out a Lagrangian algebra creates a gapped boundary for $\cC$ \cite{Kaidi:2021gbs,Lan:2014uaa}. We will refer to anyon condensation in a looser manner that can be done for any ``reasonable" condensation object, and not necessarily a Lagrangian algebra.  In addition, we will consider condensing out anyons that are not only bosons, but have spin $1/n$ for $n\geq 2$.
By enlarging the definition we will be able to employ our algorithm for anyon condensation to $\cC$ that do not have Lagrangian algebras and gain insight into modular invariants that do not correspond to gapped boundaries, as well as constructing the lines of the child theory.  It will also highlight how our computational methods naturally generalize. 

While the physical interpretation of condensation corresponding to filling a submanifold with a network of lines is inspiring, we still have yet to fill in the technical details of computing the new spectrum using the condensation algebra and the data of the lines in $\cC$. 
Let us explain by considering the spectrum of a $G_k$ Chern-Simons theory which is given by all the integral representations at level $k$. Such representations are labeled by their highest weight $\lambda$. which can be expanded in a basis of fundamental weights as 
\begin{equation}
    \lambda = \sum_{i=0}^r\lambda_i \, \omega_i\,,
\end{equation}
where $[\lambda_0,\lambda_1,\ldots,\lambda_r]$ are the Dynkin labels of $\lambda$, and $r \equiv \text{rank}\, \mathfrak{g}$.  The spectrum of $G_k$ consists of all non-negative integer solutions to the equation 
\begin{equation}
    \lambda_0 + (\lambda, \theta) \equiv k\,,
\end{equation}
where $( \cdot, \cdot)$ is the scalar product of $\mathfrak{g}$, and $\theta$ is the highest root vector.
A line given by a representation $\lambda$ has topological spin and quantum dimension given by 
\begin{align}
    h_\lambda = \frac{(\lambda,\lambda+2\rho)}{2(k+h^{\vee})}\,, \quad  \text{q-dim}_\lambda = \prod_{\alpha \in \Delta_+} \frac{\sin\left(\frac{\pi(\lambda+\rho,\alpha)}{k+h^{\vee}}\right)}{\sin\left(\frac{\pi(\rho,\alpha)}{k+h^{\vee}}\right)}\,,
\end{align}
where $\Delta_{+}$ denotes the positive roots of $\mathfrak{g}$, $\rho$ is the Weyl vector, and $h^\vee$ is the dual Coxeter number. 
We will first focus on the case where the lines in the algebra have a grouplike fusion structure.  This is known as gauging a one-form symmetry group. A well known method of gauging a one-form symmetry is to select the anyon generator $a$ for the cyclic group, and compute the monodromy charge (induced by $a$) defined by 
\begin{equation}
    Q(\lambda) = h_\lambda +h_a - h_{\lambda\times a}
\end{equation}
for all the other anyons $\lambda$.
When the charge between the generator and a line is nontrivial mod 1, then that line is projected out of the spectrum and does not survive the gauging. Of the lines that remain, we break them up into orbits of the symmetry. While this procedure works for $\cC$ with one-form symmetries, it does not generalize well to nonabelian anyons. Our understanding of why lines split is also obscured by computing monodromy charges, and in certain cases that we will see later on, projecting out lines based on their monodromy charge hides some of the subtleties of finding orbits when gauging a one-form symmetry, especially if the generator of the one-form symmetry is not bosonic. 
Furthermore, if we wanted to condense a general set of abelian anyons, this method becomes inefficient.

In order to formalize gauging one-form symmetries we consider a group homomorphism $\mu: G \to \cC^\times$ from a finite group $G$ to the set of invertible topological lines, denoted $\cC^\times$, one can produce the norm element 
\begin{equation}\label{normforgroups}
    N = \bigoplus_{g\in G} \mu(g)\,\in \mathcal{C}^\times,
\end{equation}
which has the structure of a categorified idempotent. To see this structure, we first introduce the notion of a \textit{fiber functor} $F: \mathrm{\textbf{Vec}}[G] \to \mathrm{\textbf{Vec}}$. The objects in the domain of $F$ are $G$-graded vector spaces, and written as formal sums $\underset{g\in G}{\bigoplus} \mathrm{V}_g \cdot  g$
where $\mathrm{V}_g \in \mathrm{\textbf{Vec}}$.
The homomorphism $F$ from the group algebra to the one-dimensional vector space  is a choice of one dimensional representation for the group. There is also an adjoint of the fiber functor 
\begin{equation}\label{adjointfiber}
    F^*: \mathrm{\textbf{Vec}} \to \mathrm{\textbf{Vec}}[G], \,\, \text{with}\,\, F^*(\mathbb{1}) = \underset{g\in G}{\bigoplus}\, g\,, 
\end{equation}
 which is sensible since the one-dimensional vector in $\mathrm{\textbf{Vec}}$ is an algebra, the 
  map $F^*$ takes it to another algebra.  The element $\underset{g\in G}{\bigoplus} g \in  \mathrm{\textbf{Vec}}[G] $ is an idempotent whose image is $\mathrm{\textbf{Vec}}$, and the homomorphism $\mu$ is equivalent to giving a monoidal functor from $\mathrm{\textbf{Vec}}[G]$ to $\cC$ and preserves idempotents, therefore $\mu \left(\underset{g\in G}{\bigoplus} g \right)$ is also idempotent.

What this fiber functor does at the level of lines is that it takes the algebra built out of lines and sends it to the vacuum. This makes manifesting the idempotent nature of the condensation algebra as products of the vacuum with itself again gives the vacuum.  Furthermore, a physical way to view $F^*$ in the realm of topological phases described by 3d Chern-Simons is that it builds a phase by starting from the vacuum $\mathbb 1$ and inserting the algebra of lines, similar in spirit to the construction of phases via the methods in \cite{Levin:2004mi}.

As an example, take the object $\varphi = 0+1$ where $1$ is a $\bZ_2$ object, and we know an isomorphism $1\times 1 \overset{F}{\simeq} 0$. In an attempt to make $\varphi$ into an algebra, we need a map from
\begin{equation}
    (0+1) \times (0+1) \overset{m}{\longrightarrow} (0+1)
\end{equation}
the only interesting data is the map from $1\times 1\to (0+1)$, as the other values from distribution take a canonical value. One can use some multiple of the isomorphism for $1\times 1 \simeq 0$ to write 
\begin{equation}
    1 \times 1 \overset{(\lambda F,\,0)}{\longrightarrow} (0+1)\,, \quad \lambda \in \mathbb{C}\,,
\end{equation}
for each of the components of $\varphi$. It appears that there are infinitely many unital multiplicative  maps $m$ one can use, but up to isomorphism there is only a single map.

  
  The result of condensing the norm in \eqref{normforgroups}, as per the prescription of flooding spacetime by the algebra, is the familiar notion of summing over $G$-bundles on spacetime, or the ways to insert $G$-flux to each wall of the cellulation of spacetime. After the condensation, we get a new phase which we denote as the child theory $\cD$.
In a 3d theory, the surface operators serve as interfaces between the vacuum $\mathbb 1$ and itself, and thus given by $\End_\cC (\mathbb 1)$.  Applying this same intuition to the line operators that are the actual objects of the 1-category $\cC$ tells us that they also exist as endomorphisms.  To take into account also the $G$-group action, we note that by the map $\mu$, the lines of $\cC$ are a $G$-module by right multiplication.  We can form $\cC \underset{\mathrm{\textbf{Vec}[G]}}{\otimes} \mathbb{1}$ i.e. by tensoring with the one-dimensional module, which identifies operators that have the same image under the fiber functor; the result is still a $\cC$ module by left multiplication.
We therefore see that the objects of $\cD$ are given by 
\begin{equation}\label{operatorcontent}
    \End_{\cC}\left(\cC \underset{\mathrm{\textbf{Vec}[G]}}{\otimes} \mathbb{1}  \right)\,,
\end{equation}
where we are taking $\cC$-linear endomorphisms.
Formula \eqref{operatorcontent} is equivalent to   
\begin{equation}\label{operatorcontent2}
    \left(\cC \underset{\mathrm{\textbf{Vec}[G]}}{\otimes} \mathbb{1} \right)^G\,,
\end{equation}
which are the $G$-invariant operators in $\cC \underset{\mathrm{\textbf{Vec}[G]}}{\otimes} \mathbb{1}$.  The $G$-invariant operators are reasonable to consider because $\cC \underset{\mathrm{\textbf{Vec}[G]}}{\otimes} \mathbb{1}$ itself still had a residual $G$-action. 

We now tell an analogous story for condensing nonabelian anyons, which is sometimes known as gauging a categorical symmetry, as the fusion rules of nonabelian anyons do not exhibit a grouplike structure.  We therefore replace $G$ by a  fusion category $\mathcal{G}$, which has an action by the topological lines of $\cC$, and a monoidal fiber functor $\mathcal{F}: \mathcal{G}\to \mathrm{\textbf{Vec}}$. An idempotent in $\mathcal{G}$ takes the form of a sum of nonabelian anyons, and the fiber functor again identifies it with the vacuum. The operators after the condensation is formally given by %
\begin{equation}\label{nonabeliancondensateformula}
    \left(\cC \underset{\mathrm{\mathcal{G}}}{\otimes} \mathrm{\textbf{Vec}} \right)^\mathcal{G}\,.
\end{equation}
Suppose we had another fusion category $\mathcal{G}'$, with a map $\mathcal{G}' \to \mathrm{\textbf{Vec}}$, that is Morita equivalent to $\mathcal{G}$ given by the $\mathcal G$-linear endomorphisms of $\mathrm{\textbf{Vec}}$, i.e. $\mathcal{G}' = \End_{\mathcal{G}}(\mathrm{\textbf{Vec}})$. We can then consider $\cC /\!/ \mathcal{G} /\!/\mathcal{G}' = \cC$, this gives the notion of ``ungauging" the categorical symmetry and reconstructing $\cC$. Ungauging is in practice difficult to do at the level of MTCs, and amounts to being as difficult as constructing the Drinfeld center of another fusion category \footnote{Note that in the case of gauging a regular symmetry, ungauging amounts to gauging the ``dual symmetry". Since the notion of a dual symmetry does not exist for categorical symmetries, then the analogue of ungauging becomes a hard problem of reconstructing the parent theory in the bulk.}. 
We will study  ungauging in more depth in a later section when we attempt to reconstruct the $S$-matrix of a parent theory, starting with a collection of data from the child theory.

\subsection{Condensing Abelian Anyons}\label{S:condensingabelian}
We now put the formalism into practice by consider some examples of condensating an abelian anyon, or equivalently gauging a one-form symmetry. We start with two elementary examples $\SU(3)_3$ and $\SU(4)_4$, where the generator of the one-form symmetry in the former is a boson, and the latter is a fermion \cite{Wan:2016php,Aasen:2017ubm}. In the latter case, the child theory will contain a local fermion and we must couple to spin structure. 
The data of the spectrum for  $\SU(3)_3$ is
given by the integer solutions to $\lambda_0+ \lambda_1+\lambda_2\equiv 3$. Thus we have the lines
\begin{align}
    \begin{array}{c|ccc}
    \SU(3)_3 & \lambda &h & \text{q-dim}\\\hline
         0& [0,0,3] & 0 & 1 \notag \\
         1& [0,3,0] & 1 & 1  \notag\\
         2& [3,0,0] & 1 & 1 \notag\\
         3& [0,1,2] & 2/9 & 2 \notag\\
         4& [1,2,0] & 8/9 & 2 \notag\\
         5& [2,0,1] & 5/9 & 2 \notag\\
         6& [0,2,1] & 5/9 & 2 \notag\\
         7& [2,1,0] & 8/9 & 2 \notag\\
         8& [1,0,2] & 2/9 & 2 \notag\\
         9& [1,1,1] & 1/2 & 3 \,,
    \end{array}
\end{align}
where the first column assigns a number to label each of the Dynkin labels, the third column gives the spins, and the final column gives the quantum dimension.  The notation we adopt for naming the lines is the same as that used in the KAC program \cite{KAC}. As directed by \eqref{normforgroups} we form the idempotent $\varphi = 0+1+2$; by applying the fiber functor we identify this as the new vacuum. Now we use \eqref{operatorcontent2} to compute the operator content of the gauged theory. There is a monoidal functor that moves a line $\ell \in \cC$ to the surface formed out of a network $\varphi$ by multiplying $\ell$ with the newly condensed vacuum, i.e. $\varphi \times \ell$.
 Physically, what this functor does is to take a line in the bulk and zoom out so that the line is very close to the surface. Everything in this setting is topological except for the distance from the line to the surface.
This is the same as finding the modules of $\varphi$, given by: 
\begin{align}\label{findmodules}
    \varphi \times 0 &= \varphi\,, \notag \\
    \varphi \times 1 &= 1+2+0\,, \notag \\
    \varphi \times 2 &= 2+0+1\,, \notag \\
    \varphi \times 3 &= 3+4+5\,, \notag \\
    \varphi \times 4 &= 4+5+3\,, \notag \\
    \varphi \times 5 &= 5+3+4\,, \notag \\
    \varphi \times 6 &= 6+7+8\,, \notag \\
    \varphi \times 7 &= 7+8+6\,, \notag \\
    \varphi \times 8 &= 8+6+7\,, \notag \\
    \varphi \times 9 &= 9_1+9_2+9_3\,, 
\end{align}
where we have used the fusion rules for the lines in $\SU(3)_3$. Since we do not write down all the elements $m\in \mathcal{C}$ such that there is a map $\varphi \times m \to m$,
 what we mean here and for the rest of the paper by the ``modules of $\varphi$" is actually the free modules  $m=\varphi \times \ell$ for some $\ell \in \mathcal{C}$.  By ``free", we mean that the map $\varphi \times m \to m$ is multiplication in $\varphi$. The free modules generate the category of all modules,  in particular if $\varphi$ is separable, then the category of $\varphi$-modules is semisimple, and every module is a direct sum of simple summands of free modules. Therefore, writing down \eqref{findmodules} is sufficient information to be able to tell what are all the simple summands of $\varphi \times \ell$.

Not all of the lines define a
different representation of $\SU(3)_3/\bZ_3$. When we mod out by the group $\bZ_3$, two lines in $\SU(3)$ may be indistinguishable in the child theory because any set of lines which differ by a gauge transformation, should be identified.  For this example where the lines that condense are bosons, lines which differ by a gauge transformation, but have different spins (mod 1) should not be identified because one could still tell them apart via the individual spins. Said more precisely, the lines are grouped into orbits, and all the lines in a given orbit have the same spin and quantum dimension, as is expected from lines that are indistinguishable.
Note that the fusion of $\varphi \times 9$ involves three copies of 9. In this case 9 is said to fit into a \textit{short orbit} because it is fixed by some elements of $\varphi$.
We therefore ``split" the line $9$ giving a degeneracy index, up to the order of the stabilizer of 9 in $\varphi$, with the constraint that the sum of the quantum dimensions or the split lines is conserved.

In terms of the free modules, the set of module maps 
\begin{equation}
   \hom_\varphi(\varphi \times \ell,\, \varphi \times k) =\hom(\ell , \,\varphi \times k)\,, \quad k\in \cC.
\end{equation}
This allows us to answer the question of which simple summands of $\ell$ in $\varphi \times \ell$ match which simple summands of  $\varphi \times k$. In the case of $\ell=9$ and $k=9$, then we have 
\begin{equation}
    \hom_\varphi(\varphi \times 9, \varphi \times 9) = \hom(9, \varphi \times 9)\,, 
\end{equation}
where the copies of 9 in $\varphi 
\times 9$  index the simple summands of $\varphi \times 9$.

We have that the semisimple objects (or orbits) are 
\begin{equation}
    \{\varphi,\, (3+4+5),\, (6+7+8),\, 9_1,\,9_2,\,9_3\}\,,
\end{equation}
 but there is still the task to take the $G$-invariant operators. Thus, $(3+4+5)$ and $(6+7+8)$ are projected out, and lines that are degenerate are never grouped into the same semisimple object. 
Thus lines of the gauged theory are 
\begin{equation}\label{childforsu3}
    \{\varphi,\, 9_1,\,9_2,\,9_3\}\,,
\end{equation}
and they correspond to the lines of $\Spin(8)_1$.

We end this example by noting that there exists a conformal embedding $\SU(3)_3 \subset \Spin(8)_1$ at the level of affine Lie algebras. At the level of 3d Chern Simons, the subalgebra plays the role of the parent theory, and the lines of the child $\Spin(8)_1$ are direct sums of parent theory lines. The natural way to see this is to treat the 3d MTC as $\textbf{Rep}(V)$ and $\textbf{Rep}(W)$, for $W \subset V$ as 2d VOAs and $V$ a $W$-module. 
One might also want to make an analogy to the 2d GKO coset picture for $\frac{\Spin(8)_1}{\SU(3)_3}$, where the characters of $\Spin(8)_1$ decompose as sums of characters of $\SU(3)_3$ by the formula 
\begin{equation}
\chi_\lambda(q)=\sum_{\Lambda}b^\Lambda_\lambda(q)\chi_\Lambda(q)\,, \quad \lambda={s,v,c} \in \Spin(8)_1,\quad  \Lambda \in \SU(3)_3\,.
\end{equation}
Since the coset is topological, the $q$-expansion of the branching  function $b^\Lambda_\lambda(q)$ is finite, and in particular  
\begin{align}
 \chi_s = \chi_v=\chi_c=\chi_{[1,1,1]}\,,   
\end{align}
which gives us a check that the three characters $\chi_s, \chi_v, \chi_{c}$ corresponding to the three spinors of $\Spin(8)$ correspond to the line 9 which split into three copies.  The triality symmetry also shows up in the fact that anyon condensation cannot tell apart which of the $9_i$ should be the two spinors or the vector.  

We now move onto $\SU(4)_4$ with the main goal to point out some of the subtleties when the generator is a fermion. We also use this opportunity to introduce the notion of \textit{sequential condensation}, which will be important when we move onto nonabelian condensation. 
The data of the spectrum for  $\SU(4)_4$ is
given by the integer solutions to $\lambda_0+ \lambda_1+\lambda_2+\lambda_3\equiv 4$. Thus we have 35 lines \footnote{See KAC for the full spectrum. The program also has the ability to produce the spectrum after condensing an abelian boson.}:
\begin{align}
    \begin{array}{c|ccc}
    \SU(4)_4 & \lambda &h & \text{q-dim}\\\hline
         0& [0,0,0,4] & 0 & 1 \notag \\
         1& [0,0,4,0] & 3/2 & 1  \notag\\
         2& [0,4,0,0] & 2 & 1 \notag\\
         3& [4,0,0,0] & 3/2 & 1 \notag\\
         4& [0,0,1,3] & 15/64 & 2.613125929753 \notag\\
         5& [0,1,3,0] & 95/64 & 2.613125929753 \notag\\
         6& [1,3,0,0] & 111/64 & 2.613125929753 \notag\\
         \vdots& \\
         34 & [1,1,1,1] & 15/16 & 9.656854249492\,.
    \end{array}
\end{align}
To gauge the one-form $\bZ_4$ symmetry generated by line 3, we proceed with a two step process.  We first condense out the abelian boson which is line 2 by forming $\varphi = 0+2$ and performing the procedure in \eqref{findmodules}.  The unconfined lines in the following table are listed in the first column, with their constituent $\SU(4)_4$ lines in the second column:
\begin{align}
    \begin{array}{c|ccc}
   \SU(4)_4 \overset{\varphi}{\rightarrow}  & \SU(4)_4 &h & \text{q-dim}\\\hline
         0& 0 & 0 & 1 \notag \\
         1& 1+3 & 3/2 & 1  \notag\\
         2& 8+10 & 9/16 & 3.414213562373 \notag\\
         3& 9+11 & 9/16 & 3.414213562373 \notag\\
         4& 16+18 & 5/16 & 3.414213562373 \notag\\
         5& 17+19 & 21/16 & 3.414213562373 \notag\\
         6& 24+26 & 1 & 5.828427124746 \notag\\
         7& 25+27 & 1/2 & 5.828427124746 \notag\\
         8& 28_1 & 3/4 & 2.414213562373 \notag\\
         9& 28_2 & 3/4 & 2.414213562373 \notag\\
         10& 29_1 & 5/4 & 2.414213562373 \notag\\
         11& 29_2 & 5/4 & 2.414213562373 \notag\\
         12& 34_1 & 15/16 & 4.828427124746 \notag\\
         13& 34_2 & 15/16 & 4.828427124746\,.
    \end{array}
\end{align}
We are left with an abelian spin $1/2$ line, which is also condensible. The caveat to the use of the fiber functor~$F$, is that now $F$ passes onto a super fiber functor $F: \mathrm{\textbf{Vec}}[G]\to \mathrm{\textbf{SVec}}$ \cite{deligne2002}. Physically, this makes the line into a local fermion and also requires the child theory to couple to spin structure.
 By forming the condensation algebra~$\tilde \varphi = 0+1$ in the table for $\SU(4)_4 \overset{\varphi}{\rightarrow}$ we find that 
\begin{align}\label{findmodulessu4}
    \tilde \varphi\times 0 &=  \tilde \varphi\,,    & \tilde  \varphi\times 7 &= 7+6\,,   \notag \\  
    \tilde \varphi\times 1 &= \tilde \varphi\,,  & \tilde \varphi\times 8 &= 8+10\,, \notag \\  
    \tilde \varphi\times 2 &= 2+3\,,  &  \tilde \varphi\times 9 &= 9+11\,, \notag \\ 
    \tilde \varphi\times 3 &= 3+2\,,   & \tilde \varphi\times 10 &= 10+8\,, \notag \\ 
    \tilde \varphi\times 4 &= 4+5\,,  & \tilde \varphi\times 11 &= 11+9\,, \notag \\ 
    \tilde \varphi\times 5 &= 5+4\,,  & \tilde \varphi\times 12 &= 12_1+12_2\,, \notag \\  
    \tilde \varphi\times 6 &= 6+7\,,  & \tilde \varphi\times 13 &= 13_1+13_2\,,
\end{align}
with the lines that are unconfined
\begin{align}\label{sequentialcondense1}
    \begin{array}{c|c}
        \ell& \text{q-dim}     \\ \hline
        \varphi & 1 \\
         (6+7)& 5.828427124746 \\
         (8+10) & 2.414213562373\\
         (9+11) & 2.414213562373\,.
    \end{array}
\end{align}
In terms of the Dynkin indices of $\SU(4)_4$ the lines above read 
\begin{align}\label{linessplit}
    \begin{array}{c|c}
        \ell& h     \\ \hline
        \varphi & 1 \\
         6 = ([1,0,1,2]+[1,2,1,0])& 1 \\
         7 = ([0,1,2,1]+ [2,1,0,1]) & 1/2\\
         8 = 9 = [0,2,0,2] & 3/4\\
         10=11=[2,0,2,0] & 1/4\,.
    \end{array}
\end{align}
  After condensing the fermion, the algebra gives a natural grouping where lines with spins that differ by $1/2$ are identified. The semisimple objects now have simple components which differ by $1/2$, i.e. equivalence up to a fermion. 
  
\begin{figure}[ht]
\centering
\scalebox{.7}{
    \begin{tikzpicture}[thick,scale=.8]
        \def\Depth{5}
        \def\Height{6}
        \def\Width{4}
        \coordinate (O) at (3,0-1,0-1);
        \coordinate (A) at (3,\Width+1,0-1);
        \coordinate (B) at (3,\Width+1,\Height+1);
        \coordinate (C) at (3,0-1,\Height+1);
        \coordinate (D) at (\Depth,0-1,0-1);
        \coordinate (E) at (\Depth,\Width+1,0-1);
        \coordinate (F) at (\Depth,\Width+1,\Height+1);
        \coordinate (G) at (\Depth,0-1,\Height+1);
        
        \draw[above] (\Depth/2,\Width+1/2+1.5,\Height/2);
        \draw[left] (-1, \Width/2, \Height/2);
        \draw[right] (\Depth+.5, \Width/2, \Height/2);

        \draw[black,line width =.3mm,fill=green!10] (O) -- (A) -- (B) -- (C) -- cycle;
       \draw[black, line width = 1mm] (O)--(A);
       \draw[black] (O)--(3,0,0);
       \draw[black] (3,0,0)--(3,1,0)--(3,2,-1);
       \draw[black] (3,0,0)--(3,-1,1);
       \draw[black] (3,1,0)--(3,2,1);
       \draw[black](3,-1,1)--(3,0,2);
       \draw[black] (3,0,2)--(3,1,2)--(3,2,1);
       \draw[black] (3,0,2)--(3,-1,3);
       \draw[black] (3,1,2)--(3,2,3);
       \draw[black] (3,2,1)--(3,3,0)--(3,4,0)--(3,5,-1);
       \draw[black]  (3,2,-1)--(3,3,0);
       \draw[black] (3,4,0)--(3,5,1);
       \draw[black] (3,2,3)--(3,3,2)--(3,4,2)--(3,5,1);
       \draw[black] (3,2,1)--(3,3,2); 
 \draw[black] (3,4,2)--(3,5,3);    
  \draw (3,3.8,4) node {\scalebox{0.5}{$\bullet$}};
  \draw (3,3.8,4.5) node {\scalebox{0.5}{$\bullet$}};
  \draw (3,3.8,5) node {\scalebox{0.5}{$\bullet$}};
  \draw (3,3.8-3,4) node {\scalebox{0.5}{$\bullet$}};
  \draw (3,3.8-3,4.5) node {\scalebox{0.5}{$\bullet$}};
  \draw (3,3.8-3,5) node {\scalebox{0.5}{$\bullet$}};
    \end{tikzpicture}}
    \caption{The physical picture of condensation looks like inserting a fine mesh of the algebra that takes the form of a surface when zoomed out. The dark line at the boundary represents a module for the algebra. }\label{boundarymodule}
\end{figure}
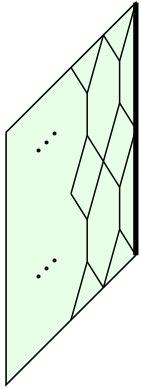

  When we were only focused on bosonic condensation, then the lines of any child theory must have constituent objects that are all of equivalent spin mod 1 in the parent, in order to be in the unconfined sector.  A subtlety to mention here is that in doing identifications up to spin 1/2 lines, the lines now do not have a definite spin. One way to understand this is that the algebra which includes a fermion is only associative and not commutative, and thus loses the braided structure that condensation algebras with bosons would have. This forces the algebra to only be able to fill in two-dimensions as shown in figure \ref{boundarymodule}.

More precisely, an associative multiplication that takes place in one space dimension, when given to a one dimensional particle worldline in the time direction, grants a way for the line to fill in two-dimensions.  Taking $\varphi$ with its associative multiplication is a two-dimensional surface and the modules for the algebra look like a boundary condition whereas a bimodule is an interface on the surface. It is therefore also natural to view gauging an associative algebra as gauging a 2d surface operator that implements a zero-form global symmetry. We elaborate more explicitly on this point in \S\ref{S:tensoredtheoryMV}.
If in addition the algebra also had a braiding, then there are two directions for multiplication, and the algebra can fill in three-dimensions.  In the new phase given by flooding with the commutative algebra, one can reasonably ask about the spins of the lines. But without the knowledge of how to flood 3d space, then it is not sensible to talk about spins of modules or bimodules. 

One could also perform the two step condensation in one step, by choosing the algebra $\varphi = 0+1+2+3$ in $\SU(4)_4$, which generates the full $\bZ_4$ symmetry.  This algebra consists of two lines that are spin 0 and two that are spin $\frac{1}{2}$ mod 1, so this is regarded as a fermion condensation.
One can check that the unconfined lines for this algebra are
\begin{align}
    \begin{array}{c|c}
        \ell& \text{q-dim}     \\ \hline
        \varphi=0+1+2+3 & 1 \\
         (24+25+26+27)& 5.828427124746 \\
         (28_1+29_1) & 2.414213562373\\
         (28_2+29_2) & 2.414213562373
    \end{array}
\end{align}
which matches the data in equation \eqref{sequentialcondense1}, upon matching the labels for lines. It is important to note that while fusing $\varphi$ with line 28 (and 29) technically gives four lines $(28_1+29_1+28_2+29_2)$, the largest grouping we could have is $(28_1+29_1)$ and $(28_2+29_2)$ because the same line can not be grouped with itself.   
We end this example by noting that there are nonabelian bosons in the spectrum. By condensing those boson out, using the details in the next section, we find the embedding~$\SU(4)_4 \subset \Spin(15)_1$. The lines of $\Spin(15)_1$ in terms of the dynkin labels of $\SU(4)_4$ are given by 
\begin{align}
    0 &= [0,0,0,4]+[0,4,0,0]+[0,1,2,1]+[2,1,0,1]\,,\notag \\
    1 & = [0,0,4,0]+[4,0,0,0]+[1,2,1,0]+[1,0,1,2]\,,\notag \\
    2 & = 2 [1,1,1,1]\,.
\end{align}
Since the spectrum is large, another way to arrive at the same result is from the coset perspective. This is by considering $\frac{\Spin(15)_1}{\SU(4)_4}$, which is topological in the sense that the central charge of the numerator matches that of the denominator. The three characters of $\Spin(15)_1$ exactly decompose into the characters of $\SU(4)_4$ with the labels on the right hand side of the equality in the above equations.

The one-form generators need not be bosonic nor fermionic, as was the case in the last two examples.  The one-form generator could have a more general rational value for its spin. Just like how we moved from integer spin lines to half integer spin lines we introduced a $\bZ_2$ grading by enlarging the fiber functor to map to supervector spaces, a general $\frac{1}{n}$ anyon when condensed would lead to a $\bZ_n$ graded vector space.  This might be at odds physically with what is natural, due to the fact that one demands a \textit{Hilbert pairing} in a physical Hilbert space. This is a pairing with no null vectors i.e. $\langle x| x  \rangle > 0$ for $x \neq 0$ in the Hilbert space.  Applying the Hilbert pairing to a vector purely in the $i$-th graded piece of  the Hilbert space pairs it with another vector in the $i$-th graded piece and returns a real number. However, tensoring two purely $i$-th graded vectors should give a  vector in the $2i$-th graded piece.  Therefore, introducing a Hilbert pairing would be an unnatural morphism in our category of $\bZ_n$-graded vector spaces.  Nevertheless, one can still make use of \eqref{operatorcontent2} for a condensation algebra that includes the one-form generator, and perform condensation as purely an algebraic manipulation.  Sequential condensation can also be generalized this way, to include a boson and a spin $1/n$ anyon with the resulting object having simple components with spins that differ by $1/n$.
We will show an example with $\SU(2)_4$ here; the spectrum for this theory consists of 5 lines given by \begin{align}
    \begin{array}{c|ccc}
    \SU(2)_{4} & \lambda &h & \text{q-dim}\\\hline
         0& [0,4] & 0 & 1 \notag \\
         1& [4,0] & 1 & 1 \notag\\
         2& [1,3] & 1/8 & 1.732050807569  \notag\\
         3& [3,1] & 5/8 & 1.732050807569 \notag\\
         4& [2,2] & 1/3 & 2\,. \notag\\
    \end{array}
\end{align}
Condensing the abelian boson splits the spin $1/3$ line into two copies. Similar to how we can pass to a super fiber functor, we now let $F: \mathrm{\textbf{Vec}}[G]\to \mathrm{\textbf{r-Vec}}$ which sends $\varphi = 0+4_1+4_2$ to the new vacuum, while coupling to a $r$-spin structure. 
Two other examples where a similar effect takes place is $\Sp(8)_1$ and $\Spin(7)_2$.

\subsection{Condensing Nonabelian Anyons}\label{condensingNaA}
The formalism for finding the operators after gauging a categorical symmetry ``generated" by a nonabelian anyon bears resemblance to the case of a regular symmetry, however due to the potentially complicated fusion structure of the MTCs, the nonabelian condensation can have complicated modules to work out.  We will present an algorithm that is useful in practice to find the lines of the child theory. While this algorithm in principle works for any number of lines, the process quickly becomes complicated when the number of lines is large, the condensation algebra involves multiple lines, or when the fusion of nonabelian lines decomposes into many simple objects.  The difficulty in performing the computation comes from assigning the proper quantum dimensions to each of the child lines, and grouping the lines from the parent that are equivalent under the fiber functor as in \eqref{nonabeliancondensateformula}. We believe the best way to proceed is through examples.
We begin with a well known and considerably elementary example of condensing the nonabelian boson in $\SU(2)_{10}$.
In Appendix \ref{morenonabelianexaples} we give more nontrivial examples of performing nonabelian condensation by using this algorithm. 

We align with the notation commonly used in the anyon condensation literature for this example instead of using KAC's notation. The data of the spectrum of $\SU(2)_{10}$ consists of 11 lines given by  
\begin{align}
    \begin{array}{c|ccc}
    \SU(2)_{10} & \lambda &h & \text{q-dim}\\\hline
         0& [0,10] & 0 & 1 \notag \\
         1& [1,9] & 1/16 & 1.931851652578  \notag\\
         2& [2,8] & 1/6 & 2.732050807569  \notag\\
         3& [3,7] & 5/16 & 3.346065214951 \notag\\
         4& [4,6] & 1/2 & 3.732050807569 \notag\\
         5& [5,5] & 35/48 & 3.863703305156 \\ 
         6& [6,4] & 1 & 3.732050807569 \notag\\
         7& [7,3] & 21/16 & 3.346065214951 \notag\\
         8& [8,2] & 5/3 & 2.732050807569 \notag\\
         9& [9,1] & 33/16 & 1.931851652578 \notag\\
        10& [10,0] & 5/2 & 1\,.
    \end{array}
\end{align}
the condensation algebra we take is $\varphi=0+6$. Interestingly, the lowest-energy eigenspace of this anyon is the \textbf{7}-dimensional representation of $\SU(2)$.  There is a well known ``cross product" map $\textbf{7} \otimes \textbf{7} \to \textbf{7}$, and correspondingly we get a multiplication map $6 \times 6 \to 6$. The condensation algebra above is therefore a version of the octonions.
The modules are
  \begin{align}\label{FuseWithVac}
      \varphi \times 0 &= \varphi\, &  \varphi \times 6 &= 6 + (0 + 2 + 4 + 6 + 8) \notag\\
      \varphi \times 1 &= 1 + (5 + 7)\,, &  \varphi \times 7 &= 7 + (1 + 3 + 5 + 7) \notag\\
      \varphi \times 2 &= 2 + (4 + 6 + 8 )\,, & \varphi \times 8 &= 8 + (2+ 4 + 6) \notag\\
      \varphi \times 3 &= 3 + (3 + 5 + 7+9 )\,, &  \varphi \times 9 &= 9 + (3 + 5) \notag\\
      \varphi \times 4 &= 4 + (2 + 4 + 6 + 8 + 10)\,, & \varphi \times 10 &= 10 +  (4) \,. \notag\\
      \varphi \times 5 &= 5 + (1+3+5 + 7 +9)\,,
  \end{align}
We use parenthesis  to denote the lines which came from fusing with 6 in $\varphi$.
The lines that split in $\SU(2)_{10}$ are the lines that appear multiple times when fused with the vacuum $\varphi$.  The multiplicity dictates the number of copies the line splits up into, just as in the abelian case.  Therefore we have 
  \begin{align}
      3 &\to 3_1 + 3_2\,, & 6 &\to 6_1 + 6_2 \notag \\
      4 &\to 4_1 + 4_2\,, & 7 &\to 7_1 + 7_2 \,. \notag \\
      5 &\to 5_1 + 5_2\,.
  \end{align}
By using our knowledge that the quantum dimension should be conserved in the condensed phase, we work our way down the list of lines assigning a subscript label to the lines which split.
 Without loss of generality, we are free to assign the subscript so that the larger subscript values appear first in the list of lines, when reading right to left, in \eqref{FuseWithVac}. As an example, we write the subscripts in  \eqref{FuseWithVac} as 
 \begin{align}\label{FuseWithVacsu210}
      \varphi \times 0 &= \varphi\, &  \varphi \times 6 &= 6_1 + (0 + 2 + 4_2 + 6_2 + 8_2) \notag\\
      \varphi \times 1 &= 1 + (5_2 + 7_2)\,, &  \varphi \times 7 &= 7_1 + (1 + 3_2 + 5_2 + 7_2) \notag\\
      \varphi \times 2 &= 2 + (4_2 + 6_2 + 8 )\,, & \varphi \times 8 &= 8 + (2+ 4_2 + 6_2) \notag\\
      \varphi \times 3 &= 3_1 + (3_2 + 5_2 + 7_2+9 )\,, &  \varphi \times 9 &= 9 + (3_2 + 5_2) \notag\\
      \varphi \times 4 &= 4_1 + (2 + 4_2 + 6_2 + 8 + 10)\,, & \varphi \times 10 &= 10 + ( 4_2 )\,. \notag\\
      \varphi \times 5 &= 5_1 + (1+3_2+5_2 + 7_2 +9)\,,
  \end{align}
Notice that while lines 5 and 7 both split, in our convention we only take $5_2$ and $7_2$ to be group, which is indicated by the parenthesis. A similar story goes for 4 and 6. Now we need to assign quantum dimensions to the lines the split and  group together the lines that have the same quantum dimension. Since the line 1 does not split and itself has quantum dimension 1.93$\ldots$, let us greedily assign this value to $5_2$ and $7_2$ because 1 appears with $5_2$ and $7_2$ frequently when we find the modules of $\varphi$. 
Then we form a grouping of lines $(1+5_2+7_2)$.
Next, suppose we greedily assign the quantum dimension 2.73$\ldots$\,, which is that of line 2 and 8, to both $4_2$ and $6_2$. Then we form the group $(2+4_2+6_2+8)$ of lines. We now consider $\varphi \times 3$, where we have the group $(5_2+ 7_2)$ from earlier, and we can form the group $(3_2+9)$ by assigning quantum dimension 1.93$\ldots$ to $3_2$, which is the quantum dimension of 9. This leaves $1.41\ldots $ for the quantum dimension of $3_1$, by conservation.  From $\varphi \times 1$ we learned that $(1+5_2+7_2)$ are condensed to the same line in the child theory, and we just learned that line 9 and $3_2$ should also be condensed to the same group.  We will keep these two lines separate, even though they share the same quantum dimension. We will subsequently see why we do not join them when we look at $\varphi \times 5$.  For now, consider $\varphi \times 4$ which again contains $(2+4_2+6_2+8)$, something we already determined from $\varphi \times 2$ should be grouped, due to quantum dimension.  This leaves $4_1$ with q-dim 1, which is exactly the same quantum dimension as 10, so we condense them into the same line and have $(4_1+10)$.  From $\varphi\times 5$ we see that since $5_2$ was assigned q-dim 1.93$\ldots$ then $5_1$ also has q-dim 1.93$\ldots$ by the conservation of quantum dimension.  However, since lines that spit should not be condensed into the same line, $5_1$ gets condensed into $(3_2+5_1+9)$ while $5_2$ gets condensed into $(1+5_2+7_2)$. Because $5_1$ and $5_2$ have the same quantum dimension, we can exchange the two lines, so it is irrelevant whether we take $5_1$ or $5_2$ to be grouped with the former or the latter. We proceed to $\varphi \times 6$ and $\varphi\times 7$ and from here we learn that $6_1$ should have q-dim 1, and $7_1$ should have q-dim $1.41\ldots\,$. We will slightly abuse notation and denote the actual vacuum of the child theory as $\varphi = 0+6_1$, which makes sense as an abelian object coming from grouping 0 and $6_1$, and can be given the properties of an idempotent. After the condensation we have the lines
\begin{align}
    \begin{array}{c|c}
        \ell& \text{q-dim}     \\ \hline
        \varphi=0+6_1 & 1 \\
         (4_1+10)& 1 \\
         (3_1+7_1) & 1.41421356237\\
         (1+5_2+7_2) & 1.931851652578 \\
         (3_2+5_1+9) & 1.931851652578 \\
         (2+4_2+6_2+8) & 2.732050807569\,.
    \end{array}
\end{align}
 The final step is to project out the lines in which the spins from the parent theory do not agree.  Therefore we only have
  $$\{(0+6_1),\,(4_1+10), \,(3_1+7_1)\}$$ 
at the end of bosonic condensation, which correspond to the three lines in $\Spin(5)_1$.  The nonabelian spin 1/2 line labeled 4 in $\SU(2)_{10}$ is now abelian after condensing the nonabelian boson, so we can further sequentially condense out $(4_1+10)$ and only be left with the vacuum line.  It can be checked that the full algebra $\mathcal{A}_\ell=(0+6+4+10)$ in $\cC=\SU(2)_{10}$ is a Lagrangian algebra object, and therefore condensing the algebra leads to a gapped interface \cite{Hung:2015hfa}.  Furthermore, since a fermion was condensed out the last step, the resulting theory couples in spin structure.

We will run through another example of using the algorithm with $(G_2)_3$. The spectrum consists of 6 lines given by 
\begin{align}
    \begin{array}{c|ccc}
    (G_2)_3 & \lambda &h & \text{q-dim}\\\hline
         0& [0,0,3] & 0 & 1 \notag \\
         1& [0,1,2] & 2/7 & 3.791287847478  \notag\\
         2& [0,2,1] & 2/3 & 5.791287847478 \notag\\
         3& [0,3,0] & 8/7 & 3.791287847478 \notag\\
         4& [1,0,1] & 4/7 & 3.791287847478 \notag\\
         5& [1,1,0] & 1 & 4.791287847478 
    \end{array}
\end{align}
We condense the algebra $\varphi = 0+5$ and see that in the modules the lines that repeat are 2 and 5, and splits into 
\begin{equation*}
    2 \to 2_1+2_2+2_3\,, \quad 5\to 5_1 +5_2\,.
\end{equation*}
The lines with subscripts written using our previous prescription is listed on the right:
\begin{align}
\varphi\times 0 &= \varphi\,, & \varphi\times 0 &= \varphi\,, \notag \\ 
\varphi\times 1 &= 1 +(2 +3  +4  +5)\,, & \varphi\times 1 &= 1 +(2_3 +3  +4  +5_2)\,,\notag \\
\varphi \times 2 &= 2 +(1+2+2  +3 +4 +5)\,, & \varphi \times 2 &= 2_1 +(1+2_2+2_3  +3 +4 +5_2)\,,\notag \\
\varphi \times 3 & = 3  +(1 +2 +4+5)\,,& \varphi \times 3 & = 3  +(1 +2_3 +4+5_2)\,, \notag \\
\varphi \times 4 & = 4  +(1 +2 +3+5)\,, & \varphi \times 4 & = 4  +(1 +2_3 +3+5_2)\,, \notag \\
\varphi \times 5 & = 5  +(0 +1 +2+3 +4+5)\,, & \varphi \times 5 & = 5_1  +(0 +1 +2_3+3 +4+5_2)\,.
 \end{align}
We start with $\varphi\times 1 $ and greedily assigning the quantum dimension of lines 1, 3, and 4 to $2_3$ and $5_2$; this gives us the group
$(1+2_3+3+4+5_2)$.
When we look at $\varphi\times 2$ we notice that some of the lines in parenthesis already appeared in $\varphi \times 1$, where we decided to group them together.  We leave $2_1$ and $2_2$ separated and not grouped, due to the fact stated earlier that we do not group lines together which split from the same parent line.  When we consider $\varphi\times 5$ there is $5_1$ which we group with 0, since the q-dim is 1, and again we have $(1+2_3+3+4+5_2)$ reappearing. At the end of the condensation we have the lines
\begin{align}
    \begin{array}{c|c}
        \ell& \text{q-dim}     \\ \hline
        \varphi=0+5_1 & 1 \\
         (4_1+10)& 1 \\
         2_1 & 1\\
         2_2 & 1\\
         (1+2_3+3+4+5_2) & 3.791287847478\,,
    \end{array}
\end{align}
 but we project out $(1+2_3+3+4+5_2)$ because the lines do not all have the same spin.  We see that condensing the line 5 in the parent theory results in $5_1$ being identified with the vacuum. Furthermore, the lines $2_1,2_2$ have the right q-dim to both be abelian lines, which they must be or else one of them will have a quantum dimension that is less than 1.

One may wonder how to determine if our choice of condensation algebra is valid, in the sense that it will lead to a consistent child phase? In order for the child phase to be consistent, it must be true that
the lines within the modules can be consistently assigned quantum dimension, while obeying the conservation requirement.
In the process of  constructing the modules of an algebra, if the quantum dimension for a line that has been split is reduced to a value that is smaller than the smallest number on the list of q-dim from the original spectrum, yet still not abelian, then our algorithm can rule out the condensation algebra. We stress that to generalize the notion of ``condensability", a canonical way of being able to assign quantum dimensions is key.

 As a tractable example consider $(G_2)_2$ which has a simple spectrum given by 
\begin{align}
    \begin{array}{c|ccc}
    (G_2)_2 & \lambda &h & \text{q-dim}\\\hline
         0& [0,0,2] & 0 & 1 \notag \\
         1& [0,1,1] & 1/3 & 2.879385241572  \notag\\
         2& [0,2,0] & 7/9 & 2.532088886238 \notag\\
         3& [1,0,0] & 2/3 & 1.879385241572
    \end{array}
\end{align}
We can consider three algebras $\varphi_1 = 0+1$, $\varphi_2=0+2$, and $\varphi_3=0+3$.  The three modules are given by 
\begin{align}
    \varphi_1 \times 0 & = \varphi_1 & \varphi_2 \times 0 & = \varphi_2 & \varphi_3 \times 0 &= \varphi_3 \notag \\
    \varphi_1 \times 1 & = 1_1+(0+1_2+2_2+3) & \varphi_2 \times 1& =1_1+(1_2+2_2+3_2) & \varphi_3 \times 1 &= 1_1+(1_2+2) \notag \\
    \varphi_1 \times 2 &= 2_1+(1_2+2_2+3) & \varphi_2 \times 2 &= 2_1+(0+1_2+2_2) &
    \varphi_3 \times 2 &= 2+(1_2+3) \notag \\
    \varphi_1 \times 3 &= 3+(1_1+2_2) & \varphi_2 \times 3 &= 3_1+(1_2+3_2) &
    \varphi_3 \times 3 &= 3+(0+2)\,, \notag 
\end{align}
each one having issues that we now point out.
In the module for $\varphi_1$, the grouping $(1_2+2_2+3)$ that we give the q-dim $1.87\ldots$ means that the quantum dimension of $2_1$ is less than 1. In the module for $\varphi_2$ the grouping $(1_2+2_2+3_2)$ that we assign q-dim $1.53\ldots$ means that the quantum dimension of $3_1$ is less than 1.  
The module for $\varphi_3$ does not make 3 into an abelian line to join with the vacuum 0.

Another useful application of this notion of condensibility based on quantum dimenions is that we can see that the proper way to condense out nonabelian spin $\frac{1}{n}$ lines is to do so sequentially. In some cases, trying to pick an algebra that only includes a fermion, alike how we did for a nonabelian boson, will lead to quantum dimensions not being able to split properly. However if we condense the boson first resulting in an abelian fermion, then the quantum dimensions will be able to split properly\footnote{There are examples where condensing out a nonabelian fractional spin anyon is possible, namely in $\Sp(16)_1$.}  As an example consider $(F_4)_3$, the data of which is presented in appendix \ref{morenonabelianexaples}. If we wanted to just naively condense the nonabelian fermion, the condensation algebra one can choose is $\varphi = 0+ 1$, which leads to the modules 
\begin{align}
    \varphi \times 0 &= \varphi\,,  & \varphi \times 5 &= 5_1+(2_2+3_2+4_2+5_2 \notag \\
    \varphi \times 1 &= 1_1+ (0+1_2+2_2+4_2+7)\,, & &\hspace{15mm} +5_3+6_2+8_2)\,,\notag \\ 
    \varphi \times 2 & = 2_1+(1_2+2_2+3_2+4_2+5_3+8_2)\,, & \varphi \times 6 &= 6_1+(4_2+5_3+6_2)\,, \notag \\
    \varphi \times 3 & = 3_1+(2_2+3_2+5_3)\,, & \varphi \times 7 &= 7+(1_2+4_2+8_2)\,, \notag \\
    \varphi \times 4 & = 4_1+(1_2+2_2+4_2+5_3+6_2+7+8_2)\,, & \varphi \times 8 &= 8_1+(2_2+4_2+5_3+6_2\notag \\
     & & &\hspace{15mm}+7+8_2)\,.
\end{align}
Greedily assigning the q-dim $4.49\ldots$ of 7 to the group $(1_2+2_2+3_2+4_2+5_3+6_2+7+8_2)$ leaves $3_1$ with zero quantum dimension which contradicts the fact that the line $3$ splits. To distribute $4.49\ldots$ among $3_1$ and $3_2$ would result in both of the lines being simple objects in the gauged theory, yet at least one would be nonabelian carrying q-dim less than $4.49\ldots$. In appendix \ref{morenonabelianexaples} we will show that by condensing the nonabelian boson first, that the spin 1/2 line becomes abelian, and we can seqentially condense it.

\section{Modular Invariants and Condensation}\label{MIandcondensation}

Having done a couple of examples where we find the lines of the child theory in the previous section, we now present some of the modular invariants of those theories, and others. 
It is well known that the modular invariants should correspond to the Frobenius algebra objects up to Morita equivalence. So in particular, there are modular invariants that correspond to nonabelian bosonic condensation; we will refer to them as ``extension" modular invariants.  This is not the end of the story as there also exists ``permutation" modular invariants that pair up the lines with the same spin and in certain cases displays some symmetry of the theory. This is also referred to in the literature as the ``charge conjugation" modular invariant. 
One might expect that these modular invariants arise from an algebra that includes a boson, but we can also find these permutation invariants in theories with no bosons at all! In this case, finding the condensation algebra for these invariants can be complicated. When the fusion rules are grouplike, it is more likely that we are able to determine what is the algebra that gives the permutation invariant.  
For abelian Chern-Simons theories, their unitary symmetries, documented in \cite{Delmastro:2019vnj}, is reflected by the modular invariants. Furthermore for $\SU(2)_k$ theories where there is an ADE classification of modular invariants \cite{Cappelli:1987xt,kirillov2002q}, it can be checked that the modular data as well as the $F$- and $R$-symbols reflect the symmetries given by the permutation modular invariants.
Motivated by this, one could study the modular invariants that are not of the extension type, to reveal a subset of the symmetries of the nonabelian Chern-Simons, even though we are unable to check these symmetries entirely since we do not have knowledge of the $F$- and $R$-symbols for a general theory.

As an example of an algebra associated to a permutation, consider the toric code ($=\Spin(16)_1$). There are two bosons and a fermion and there is a global $\bZ_2$ symmetry which is usually called ``electromagnetic duality" but which might as well be called charge conjugation. It is implemented by (the Morita equivalence class of) an algebra whose underlying object is $1+\text{fermion}$. 
As another example one can consider is $\Spin(4)_1 = \SU(2)_1^2 = \text{semion}^2$. Its particles are the vacuum, a fermion, and two semions, and again 1+fermion is an algebra who implements a $\bZ_2$ global symmetry. In this case that global symmetry switches the two semions.  We will give more nontrivial examples  such as $\SU(N)_1, (E_6)_1 $, where we can explicitly see the association of a permutation modular invariant to an algebra.

{While the modular invariants for the Lagrangian algebra correspond to gapped boundaries, the permutation types do not give gapped boundaries. 
This fact is manifest when we consider the embedding $\SU(3)_1\times (E_6)_1 \subset (E_8)_1$. The product theory is abelian and contains 9 lines given by the following table, where spins of the $\SU(3)_1$ lines are on the horizontal axis, and the spins of the $(E_6)_1$ lines are on the vertical axis:
\begin{equation}
    \begin{array}{c|ccc}
        \SU(3)_1\times (E_6)_1  & 0& 1/3& 1/3  \\ \hline
         0&0 & 1/3 & 1/3 \\
         2/3& 2/3&1&1\\
         2/3& 2/3&1&1\,.
    \end{array}
\end{equation}
The two Lagrangian algebras are given by the three lines on the diagonal, and the line 0 with the two off diagonal bosons. The nondiagonal modular invariants however are 
\begin{align}
 \left( \begin{tabular}{ccccccccc}
     1&0&0&0&0&0&0&0&0\\
     0&0&1&0&0&0&0&0&0\\
     0&1&0&0&0&0&0&0&0\\
     0&0&0&0&0&0&1&0&0\\
     0&0&0&0&0&0&0&0&1\\
     0&0&0&0&0&0&0&1&0\\
     0&0&0&1&0&0&0&0&0\\
     0&0&0&0&0&1&0&0&0\\
     0&0&0&0&1&0&0&0&0\\
    \end{tabular}\right)\,,
   \quad 
     \left( \begin{tabular}{ccccccccc}
     1&0&0&0&0&0&0&0&0\\
     0&0&1&0&0&0&0&0&0\\
     0&1&0&0&0&0&0&0&0\\
     0&0&0&1&0&0&0&0&0\\
     0&0&0&0&0&1&0&0&0\\
     0&0&0&0&1&0&0&0&0\\
     0&0&0&0&0&0&1&0&0\\
     0&0&0&0&0&0&0&0&1\\
     0&0&0&0&0&0&0&1&0\\
     \end{tabular}\right)\,,
\end{align}
\begin{equation}
         \left( \begin{tabular}{ccccccccc}
     1&0&0&0&0&0&0&0&0\\
     0&1&0&0&0&0&0&0&0\\
     0&0&1&0&0&0&0&0&0\\
     0&0&0&0&0&0&1&0&0\\
     0&0&0&0&0&0&0&1&0\\
     0&0&0&0&0&0&0&0&1\\
     0&0&0&1&0&0&0&0&0\\
     0&0&0&0&1&0&0&0&0\\
     0&0&0&0&0&1&0&0&0\\
    \end{tabular}\right)\,,
\end{equation}
with the rows labeled by \{\{0,0\}, \{0,1\},\{0,2\},\{1,0\},\{1,1\},\{1,2\},\{2,0\},\{2,1\},\{2,2\}\} with the first entry a line in $\SU(3)_1$ and the second entry a line in $(E_6)_1$. Each matrix squares to the identity, and none corresponds to either of the Lagrangian algebras. In particular, the last two modular invariants correspond to the algebra $(0+2/3+2/3)$ and $(0+1/3+1/3)$ from the two separate theories. Therefore, they do not give gapped boundaries.  
In  cases when the Lagrangian algebra contains a fermion we have to couple to spin structure in order to get the gapped boundary; this is because Lagrangian algebras require not only associativity but also commutativity.  Therefore the gapped boundary will have to be seen through the super modular invariant.
It is a natural generalization that coupling to higher spin structures can also make an algebra composed of $1/n$-spin anyons commutative. 
 }
 \subsection{Modular invariants for spin $1/n$ anyons}
The first of these new modular invariants arising when $n=2$ is recognized as supermodular invariants. These are matrices $M$ such that
\begin{equation}
\begin{aligned}
     \begin{cases}
     [M,S] = [M,T^2]=0\,, \\
         T\,MT\,^{-1} \quad \text{is integral,} \\
         (ST) M (ST)^{-1} \quad \text{has positive integral values.}
     \end{cases}
     \end{aligned}
\end{equation}
These exist when there are extension modular invariants coming from condensing a fermion. There are also supermodular invariants which are permutation matrices, but permute the lines with spins differing by $1/2$. 

 {Given the fact that some super modular invariants correspond to condensing out a fermion, let us consider $(E_7)_1$, which has an abelian line but is spin 3/4.  When we tensor this theory with itself we get a fermion which generates a center $\bZ_2$ one-form symmetry in the overall $\bZ_2 \times \bZ_2$ symmetry, and also an extension type super modular invariant
 \begin{equation}
    \left(\begin{tabular}{cccc}
     1&0&0&1\\
     0&0&0&0\\
     0&0&0&0\\
     1&0&0&1\,
    \end{tabular}\right).
\end{equation}
 This indicates that the fermion composed of the two 3/4 lines should be condensable. But then to allow the constituent lines of the fermion to also be condensable, we should allow the original abelian 3/4 lines to be ``condensable", at least when we couple to proper background $r$-structure.  Thus, the super modular invariant motivates us to naturally enlarge the notion of the fiber functor beyond $\textbf{\text{SVec}}$, as was discussed at the end of \S\ref{S:condensingabelian}.}

 We can generalize the conditions for a supermodular invariant further to  matrices $\cM$, which pair up lines that differ by spin $1/n$, such that 
\begin{equation}\label{beyondsuper}
    [\cM,\mathcal{T}^n\,] = [\cM,T^n]=0.
\end{equation}
Such nontrivial $\cM$ of extension type would fit in conjointly with the discussion in \S\ref{S:condensingabelian} about the possibility to condense a spin $1/n$ anyon. We denote $\mathcal{T}=T.S.T$ as the operation what replaces $S$ in the search for (super)modular invariants.  This is motivated by the fact that we can take our three dimensional theory and compactify the two spatial dimensions on a torus. The Hilbert space for the 3d theory restricted to the torus, has a basis given by conformal blocks i.e. the spectrum of lines, and comes with an action of a mapping class group of the torus.

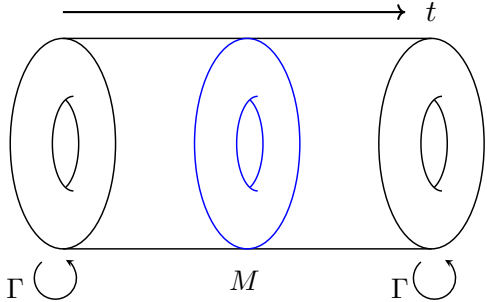
\begin{figure}[t!]
    \centering
    \scalebox{.7}{
    \begin{tikzpicture}
    \draw[thick] (0,0) ellipse (1cm and 2cm);
    \draw[thick,domain=180:360,smooth,variable=\x] plot ({.2+.4*sin(\x)},{.9*cos(\x)});
\draw[thick,domain=23:157,smooth,variable=\x] plot ({-.1+.4*sin(\x)},{.9*cos(\x)});
\draw[thick] (0,2) -- (3.5,2);
\draw[thick] (0,-2) -- (3.5,-2);
\draw[thick] (3.5,2) -- (7,2);
\draw[thick] (3.5,-2) -- (7,-2);
 \draw[thick,blue] (3.5,0) ellipse (1cm and 2cm);
    \draw[thick,blue,domain=180:360,smooth,variable=\x] plot ({3.5+.2+.4*sin(\x)},{.9*cos(\x)});
\draw[thick,blue,domain=23:157,smooth,variable=\x] plot ({3.5+-.1+.4*sin(\x)},{.9*cos(\x)});
 \draw[thick] (7,0) ellipse (1cm and 2cm);
    \draw[thick,domain=180:360,smooth,variable=\x] plot ({7+.2+.4*sin(\x)},{.9*cos(\x)});
\draw[thick,domain=23:157,smooth,variable=\x] plot ({7+-.1+.4*sin(\x)},{.9*cos(\x)});
\draw[decoration={markings, mark=at position .999 with
 {\arrow{>}}}, postaction={decorate},line width = .4mm] (0,2.5)--(6.5,2.5);
\node at (7,2.5) {\scalebox{1.5}{$\color{black}{t}$}};
{\draw[->,thick,>=stealth] (-.4,-2.25) arc (130:420:.4cm);}
{\draw[->,thick,>=stealth] (-.5+7.3,-2.25) arc (130:420:.4cm);}
\node at (-.9,-2.75) {\scalebox{1.4}{$\Gamma$}};
\node at (-.3+3.75,-2.6) {\scalebox{1.4}{$M$}};
\node at (-.9+7.3,-2.75) {\scalebox{1.4}{$\Gamma$}};
    \end{tikzpicture}}
    \caption{Each of the black tori represents the spatial dimensions of the 3d theory, with time running horizontally. The blue torus indicates a defect that can be placed in this quantum mechanics model at an instant in time. The black tori are both acted on by the modular group, so the defect $M$ intertwines the two actions. The 2d theory on the black tori can in particular be the chiral or anti-chiral half of a WZW model. }
    \label{mappingCylinder}
\end{figure}

We insert a defect along the time direction, as in figure \ref{mappingCylinder}, which intertwines the representation of the modular  group $\Gamma =\text{SL}_2(\bZ)$ acting on the torus on each side of the defect.
In particular, the matrices 
\begin{equation}
    \mathcal{T} : \begin{pmatrix}
     a\\
     b
    \end{pmatrix} \rightarrow
    \begin{pmatrix}
     a\\
     a+b
    \end{pmatrix}\,,\quad
    T: \begin{pmatrix}
     a\\
     b
    \end{pmatrix} \rightarrow
    \begin{pmatrix}
     a+b\\
     b
    \end{pmatrix}
\end{equation}
give the Dehn twists on the torus. 
The matrices $\mathcal{T}^n$ and $T^n$ also belong to the group 
\begin{equation}
    \Gamma(n) = \left\{ \begin{pmatrix}
     a&b\\
     c&d
    \end{pmatrix} \in \text{SL}_2(\bZ)\Big| \,a\equiv d \equiv 1 \mod n, \quad b\equiv c \equiv 0 \mod n\right\}\,,
\end{equation}
which is a congruence subgroup of $\Gamma$. {In the set of matrices $\cM$ that satisfy \eqref{beyondsuper}, some might not correspond to interfaces that are built via true commutative algebra objects and thus do not contain the same physical interpretation as a modular invariant that came from condensing a Lagrangian algebra. These $\cM$ only take the interpretation of intertwiners for $\Gamma(n)$ representations, in the same spirit as how there can exist modular invariants $M$ that are intertwiners for $\Gamma$, i.e. matrices that commute with the modular actions, but do not come from Lagrangian algebras.

Nevertheless, to put these $\cM$ into context, let us change perspectives from asking the categorical questions one can pose regarding the data of MTCs. If we look solely from a representation theory point of view, it is surprising that matrices in the representation of $\Gamma(n)$ can appear when we study MTCs.
Given a representation of $\Gamma$, there can be endomorphisms of this representation as well as endomorphisms when we restrict to a subgroup $\Gamma(n)$. A reasonable question to ask is how one can construct the endomorphisms of $\Gamma(n)$, and where did they come from. It appears the condensation procedure we use can be useful to answering this question.
Moreover, even the motivation for restricting to $\Gamma(n)$ representations is also clear as it came from observing the spins in the spectrum of anyons.}

To make the discussion of using anyon condensation to find $\Gamma(n)$ representations more concrete, we give the explicit form of $\cM$ in the examples $\SU(4)_2/\bZ_2$, $(E_6)_1 \times (E_7)_1$ with the boson condensed out, and  $\Spin(5)_1$. 
 The spectrum of $\SU(4)_2$ is given by the following table on the left, and we can condense the boson:
\begin{align}
    \begin{array}{c|ccc}
    \SU(4)_2 & \lambda &h & \text{q-dim}\\\hline
         0& [0,0,0,2] & 0 & 1 \notag \\
         1& [0,0,2,0] & 3/4 & 1  \notag\\
         2& [0,2,0,0] & 1 & 1 \notag\\
         3& [2,0,0,0] & 3/4 & 1 \notag \\
          4& [0,0,1,1] & 5/16 & 1.732050807569 \notag \\
         5& [0,1,1,0] & 13/16 & 1.732050807569 \notag\\
         6& [1,1,0,0] & 13/16 & 1.732050807569 \notag\\
         7& [1,0,0,1] & 5/16 & 1.732050807569 \notag \\
          8& [1,0,1,0] & 2/3 & 2\notag\\
         9& [0,1,0,1] & 5/12 & 2 \notag \\
    \end{array}
    \quad \overset{\varphi=(0+2)}{\longrightarrow}
    \begin{array}{c|c|c}
        \SU(4)_2/\bZ_2&\ell& \text{q-dim}     \\ \hline
        0&\varphi=(0+2) & 1 \\
         1&(1+3)& 1 \\
         2&8_1 & 1\\
         3&8_2 & 1 \\
         4&9_1 & 1 \\
         5&9_2 &1\,.
    \end{array}
\end{align}
We notice that $8_{1,2}$ and $9_{1,2}$ differ by $3/4\equiv -1/4 \mod 1$, so we consider the following matrices for $\cM$ that pair up lines with spins that differ by $-1/4$
\begin{equation}\label{SU42M}
   \left( \begin{tabular}{cccccc}
     1&1&0&0&0&0\\
     1&1&0&0&0&0\\
     0&0&1&0&0&1\\
     0&0&0&1&1&0\\
     0&0&0&1&1&0\\
     0&0&1&0&0&1
    \end{tabular}\right)\,,
    \quad 
   \left( \begin{tabular}{cccccc}
     1&1&0&0&0&0\\
     1&1&0&0&0&0\\
     0&0&1&0&1&0\\
     0&0&0&1&0&1\\
     0&0&1&0&1&0\\
     0&0&0&1&0&1
    \end{tabular}\right)\,,
\end{equation}
and one can check that both commute with $\mathcal T^4=(T^{-1}.S.T^{-1})^4$ and $(T^{-1})^4$. Here, $T$ and $S$ are those of the theory after condensing the boson i.e. $\SU(4)_2/\bZ_2$. If we proceed in our usual manner of finding modules for an algebra object, we can consider the modules of $\varphi = 0+1$ in the table for $\SU(4)_2/\bZ_2$ and we get 
\begin{align}
    \varphi \times 0 &= \varphi  &  \varphi \times 3 &= 3+4 \notag   \\
    \varphi \times 1 &= \varphi & \varphi \times 4 &= 4+3 \notag \\
    \varphi \times 2 &= 2+5 & \varphi \times 5 &= 2+5\,.
\end{align}
Therefore, the first of the two matrices in \eqref{SU42M} corresponds to this $\varphi$.

The spectrum of $(E_6)_1\times (E_7)_1$ contains 6 lines given by 
\begin{equation}
 \begin{array}{c|ccc}
    (E_6)_1\times (E_7)_1 & \ell &h & \text{q-dim}\\\hline
         0& \{0,0\} & 0 & 1 \notag \\
         1& \{1,1\} & 17/12 & 1  \notag\\
         2& \{2,0\} & 2/3 & 1 \notag\\
         3& \{0,1\} & 3/4 & 1 \notag \\
         4& \{1,0\} & 2/3 & 1 \notag \\
         5& \{2,1\} & 17/12 & 1 \notag\\
    \end{array}
\end{equation}
and we see that by condensing $\varphi = 0+3$ the other lines are grouped as $(1+4)$ and $(2+5)$. The explicit matrix that corresponds to this condensation is  
\begin{equation}
     \cM=\left(\begin{tabular}{cccccc}
     1&0&0&1&0&0\\
     0&1&0&0&1&0\\
     0&0&1&0&0&1\\
     1&0&0&1&0&0\\
     0&1&0&0&1&0\\
     0&0&1&0&0&1
    \end{tabular}\right)\,,
\end{equation}
which can be checked commutes with $\mathcal{T}^4$ and $(T^{-1})^4$. Just as with $\SU(4)_2/\bZ_2$, we can construct another $\cM$ by grouping the lines by $(1+2)$ and $(4+5)$, but this is not what $\varphi$ produces, so is unphysical.
 
 The spectrum of $\Spin(5)_2$ is given by the  following table on the left, where the boson can be condensed 
  \begin{align}
    \begin{array}{c|ccc}
    \Spin(5)_2 & \lambda &h & \text{q-dim}\\\hline
         0& [0,0,2] & 0 & 1 \notag \\
         1& [2,0,0] & 1 & 1  \notag\\
         2& [0,1,1] & 1/4 & 2.236067977500 \notag\\
         3& [1,1,0] & 3/4 & 2.236067977500 \notag \\
          4& [0,2,0] & 3/5 & 2 \notag \\
         5& [1,0,1] & 2/5 & 2 \notag\\
    \end{array}
    \quad \overset{\varphi=(0+1)}{\longrightarrow}
    \begin{array}{c|c}
        \ell& \text{q-dim}     \\ \hline
        \varphi=(0+1) & 1 \\
         4_1 & 1\\
         4_2 & 1 \\
         5_1 & 1 \\
         5_2 &1\,.
    \end{array}
\end{align}
In this case, the spins of the child theory are all fifth roots of unity, and thus $T^5= \text{id}$. We also find that $\mathcal{T}^5$ is  proportional to the identity, and thus all matrices satisfy \eqref{beyondsuper}, indicating there is a plethora of possible condensable algebras if we couple to background $r$-structure \footnote{In addition to the matrices that correspond to algebras, we also get matrices that do not correspond to algebras since any general $6\times6$ matrix satisfies \eqref{beyondsuper}.}.  

We end the discussion on generalizing modular invariants with the
case of $(G_{2})_2$, which does not have such an $\cM$ as in \eqref{beyondsuper}. Even though the spectrum contains two lines that differ by $1/3$, the spin $1/3$ line here is nonabelian. It was shown earlier that this spin $\frac{1}{3}$ was also not condensable, by the criterion we gave for a condensation in \S\ref{condensingNaA}. If one were to consider the matrices that paired up the lines differing by spin 1/3 such as 
\begin{equation}
   \cM= \left(\begin{tabular}{cccc}
     1&0&0&0\\
     0&0&0&1\\
     0&0&1&0\\
     0&1&0&0\\
    \end{tabular}\right)\,,
\end{equation}
one would find that none commute with $\mathcal{T}^3$. This further supports our claim that there are no condensations possible, and that if one were to condense a spin $1/n$ line, then it must be abelian.

\subsection{Modular invariants of tensored theories}\label{S:tensoredtheoryMV}
We now consider in more depth what modular invariants one finds when we tensor theories. In this case, some of the lines may become bosons when combined with other lines, but are still not condensable algebras. This reinforces the fact that it is not the anyon necessarily that is crucial, but the algebra object. Just because some anyons might be nonabelian bosons, does not mean they  belong to a condensation algebra, e.g. the Fibonacci category has no gapped boundary for any tensor product of the theory with itself \cite{Davydov:2011pp}. When one considers a tensored theory such as $(G_k)^n$, there is an inherent symmetry group with order $n!$ that permutes the theories among themselves and is also reflected in the modular invariants of the tensored theory.
 From a physical point of view, recall that automorphisms of the theory are zero-form symmetries and therefore enacted by surface operators for our purposes. We will illustrate this explicitly in the example $(E_7)_1^3$. In a Reshetikhin- Turaev type theory, all of the surfaces arise as condensation descendants of lines by means described in \S\ref{overviewofgauging}. In this way we can think of the permutation modular invariants as being built from algebras.

To make contact with the previous section, we first look at the nondiagonal modular invariants of $\SU(3)_3$ given by 
\begin{equation}
    \left(\begin{tabular}{cccccccccc}
     1&0&0&0&0&0&0&0&0&0\\
     0&0&1&0&0&0&0&0&0&0\\
     0&1&0&0&0&0&0&0&0&0\\
     0&0&0&0&0&0&0&0&1&0\\
     0&0&0&0&0&0&0&1&0&0\\
     0&0&0&0&0&0&1&0&0&0\\
     0&0&0&0&0&1&0&0&0&0\\
     0&0&0&0&1&0&0&0&0&0\\
     0&0&0&1&0&0&0&0&0&0\\
     0&0&0&0&0&0&0&0&0&1\\
    \end{tabular}\right)\,,
   \quad 
    \left(\begin{tabular}{cccccccccc}
     1&1&1&0&0&0&0&0&0&0\\
     1&1&1&0&0&0&0&0&0&0\\
     1&1&1&0&0&0&0&0&0&0\\
     0&0&0&0&0&0&0&0&0&0\\
     0&0&0&0&0&0&0&0&0&0\\
     0&0&0&0&0&0&0&0&0&0\\
     0&0&0&0&0&0&0&0&0&0\\
     0&0&0&0&0&0&0&0&0&0\\
     0&0&0&0&0&0&0&0&0&0\\
     0&0&0&0&0&0&0&0&0&3\\
     \end{tabular}\right)\,,
\end{equation}
which is a permutation invariant and the extension invariant, from gauging the one-form symmetry. There is a new nondiagonal super modular invariant given by %
\begin{equation}
    \left(\begin{tabular}{cccccccccc}
    1&1&1&0&0&0&0&0&0&1\\
     1&1&1&0&0&0&0&0&0&1\\
     1&1&1&0&0&0&0&0&0&1\\
     0&0&0&0&0&0&0&0&0&0\\
     0&0&0&0&0&0&0&0&0&0\\
     0&0&0&0&0&0&0&0&0&0\\
     0&0&0&0&0&0&0&0&0&0\\
     0&0&0&0&0&0&0&0&0&0\\
     0&0&0&0&0&0&0&0&0&0\\
     1&1&1&0&0&0&0&0&0&1\\
     \end{tabular}\right)\,,
\end{equation}
which is the result of sequentially condensing out either of the three 
fermions in \eqref{childforsu3}. 

Moving onto $\SU(2)_{10}$, the nondiagonal modular invariants are 
\begin{equation}
\left(
  \begin{tabular}{ccccccccccc}
     1&0&0&0&0&0&0&0&0&0&0 \\
     0&0&0&0&0&0&0&0&0&1&0\\
     0&0&1&0&0&0&0&0&0&0&0\\
     0&0&0&0&0&0&0&1&0&0&0\\
     0&0&0&0&1&0&0&0&0&0&0\\
     0&0&0&0&0&1&0&0&0&0&0\\
     0&0&0&0&0&0&1&0&0&0&0\\
     0&0&0&1&0&0&0&0&0&0&0\\
     0&0&0&0&0&0&0&0&1&0&0\\
     0&1&0&0&0&0&0&0&0&0&0\\
     0&0&0&0&0&0&0&0&0&0&1\\
    \end{tabular}\right)\,,
    \quad
    \left(
    \begin{tabular}{ccccccccccc}
     1&0&0&0&0&0&0&0&0&0&0\\
     0&0&1&0&0&0&0&0&0&0&0\\
     0&1&0&0&0&0&0&0&0&0&0\\
     0&0&0&0&0&0&0&0&1&0&0\\
     0&0&0&0&0&0&0&1&0&0&0\\
     0&0&0&0&0&0&1&0&0&0&0\\
     0&0&0&0&0&1&0&0&0&0&0\\
     0&0&0&0&1&0&0&0&0&0&0\\
     0&0&0&1&0&0&0&0&0&0&0\\
     0&0&0&0&0&0&0&0&0&1&0\\
     0&0&0&0&0&0&0&0&0&1&0\\
    \end{tabular}\right)\,.
\end{equation}
There also exist super modular invariants for this theory, given by 
\begin{equation}
\left(
  \begin{tabular}{ccccccccccc}
     1&0&0&0&0&0&0&0&0&0&1 \\
     0&0&0&0&0&0&0&0&0&0&0\\
     0&0&1&0&0&0&0&0&1&0&0\\
     0&0&0&0&0&0&0&0&0&0&0\\
     0&0&0&0&1&0&1&0&0&0&0\\
     0&0&0&0&0&0&0&0&0&0&0\\
     0&0&0&0&1&0&1&0&0&0&0\\
     0&0&0&0&0&0&0&0&0&0&0\\
     0&0&1&0&0&0&0&0&1&0&0\\
     0&0&0&0&0&0&0&0&0&0&0\\
     1&0&0&0&0&0&0&0&0&0&1\\
    \end{tabular}\right)\,,
    \quad
    \left(
    \begin{tabular}{ccccccccccc}
     1&0&0&0&1&0&1&0&0&0&1\\
     0&0&0&0&0&0&0&0&0&0&0\\
     0&0&0&0&0&0&0&0&0&0&0\\
     0&0&0&0&0&0&0&0&0&0&0\\
     1&0&0&0&1&0&1&0&0&0&1\\
     0&0&0&0&0&0&0&0&0&0&0\\
     1&0&0&0&1&0&1&0&0&0&1\\
     0&0&0&0&0&0&0&0&0&0&0\\
     0&0&0&0&0&0&0&0&0&0&0\\
     0&0&0&0&0&0&0&0&0&0&0\\
     1&0&0&0&1&0&1&0&0&0&1\\
    \end{tabular}\right)\,,
\end{equation}
the first is the permutation type that corresponds to condensing out the algebra $\varphi=0+10$. By computing the modules of $\varphi$ one can see that indeed the lines are paired as given by the left matrix:
  \begin{align}\label{FuseWithfVac}
      \varphi \times 0 &= \varphi\, &  \varphi \times 6 &= 6 + 4 \notag\\
      \varphi \times 1 &= 1 + 9\,, &  \varphi \times 7 &= 7 + 3\notag\\
      \varphi \times 2 &= 2 +8 \,, & \varphi \times 8 &= 8 + 2 \notag\\
      \varphi \times 3 &= 3 + 7\,, &  \varphi \times 9 &= 9 + 1 \notag\\
      \varphi \times 4 &= 4 +  6 \,, & \varphi \times 10 &= \varphi \,. \notag\\
      \varphi \times 5 &= 5_1 + 5_2\,,
  \end{align}
The latter modular invariant
corresponds to condensing out the Lagrangian algebra, which included a fermion.

 We give another example of finding the algebra that gives a permutation invariant by considering $\SU(N)_1$ with $N=2n+1$. This is an abelian theory with $\bZ_{2n+1}$ fusion rules, and associator $\kappa \in \rH^3(\bZ_{2n+1}; \rU(1))$ that is trivial.  The algebras up to Morita equivalence i.e.  the modules of the fusion category with fusion rules $G$ and associator $\kappa$, are in bijection with subgroups $H \subset G$ 
 and $\beta \in \rC^2(H; \rU(1))$ with $d\beta = \kappa|_H$. 
 There is always the trivial subgroup, and the whole group itself. These give the  diagonal modular invariant, and the permutation modular invariant -- with the condensation algebra built by all of the lines $\varphi = 0+1+\ldots+2n$. For $\SU(N)_1$ with $N=2n$, the associator is nontrivial and given by $n$ mod $2n$. This is an obstruction to creating an algebra out of all the anyons, but we can form an associative algebra from the even anyons $\varphi = 0+2+\ldots+2n-2$ which corresponds to the charge conjugation modular invariant. 
 
We now present a theory that is formed as a tensor product of three copies of $(E_7)_1$. 
The spectrum is given by 
\begin{align}
    \begin{array}{c|ccc}
    (E_7)^3_1& \{\ell_1,\ell_2,\ell_3\} &h & \text{q-dim}\\\hline
         0& \{0,0,0\} & 0 & 1 \notag \\
         1& \{0,0,1\} & 3/4 & 1  \notag\\
         2& \{0,1,0\} & 3/4 & 1 \notag\\
         3& \{0,1,1\} & 3/2 & 1 \notag\\
         4& \{1,0,0\} & 3/4 & 1 \notag\\
         5& \{1,0,1\} & 3/2 & 1 \notag\\
         6& \{1,1,0\} & 3/2 & 1 \notag\\
         7& \{1,1,1\} & 9/4 & 1\,. \notag
    \end{array}
\end{align}
There are indeed five nondiagonal modular invariants of $(E_7)_1^3$, three of which give $\bZ_2$ symmetries
\begin{equation}
    \left(\begin{tabular}{cccccccc}
     1&0&0&0&0&0&0&0\\
     0&0&0&0&1&0&0&0\\
     0&0&1&0&0&0&0&0\\
     0&0&0&0&0&0&1&0\\
     0&1&0&0&0&0&0&0\\
     0&0&0&0&0&1&0&0\\
     0&0&0&1&0&0&0&0\\
     0&0&0&0&0&0&0&1\\
    \end{tabular}\right)\,,
   \quad 
    \left(\begin{tabular}{cccccccc}
     1&0&0&0&0&0&0&0\\
     0&0&1&0&0&0&0&0\\
     0&1&0&0&0&0&0&0\\
     0&0&0&1&0&0&0&0\\
     0&1&0&0&1&0&0&0\\
     0&0&0&0&0&0&1&0\\
     0&0&0&0&0&1&0&0\\
     0&0&0&0&0&0&0&1\\
    \end{tabular}\right)\,,
\end{equation}
\begin{equation}
     \left(\begin{tabular}{cccccccc}
     1&0&0&0&0&0&0&0\\
     0&1&0&0&0&0&0&0\\
     0&0&0&0&1&0&0&0\\
     0&0&0&0&0&1&0&0\\
     0&0&1&0&0&0&0&0\\
     0&0&0&1&0&0&0&0\\
     0&0&0&0&0&0&1&0\\
     0&0&0&0&0&0&0&1\\
     \end{tabular}\right)\,,
\end{equation}
and two which give a $\bZ_3$ symmetry 
\begin{equation}
    \left(\begin{tabular}{cccccccc}
     1&0&0&0&0&0&0&0\\
     0&0&0&0&1&0&0&0\\
     0&1&0&0&0&0&0&0\\
     0&0&0&0&0&1&0&0\\
     0&0&1&0&0&0&0&0\\
     0&0&0&0&0&0&1&0\\
     0&0&0&1&0&0&0&0\\
     0&0&0&0&0&0&0&1\\
    \end{tabular}\right)\,,
   \quad 
    \left(\begin{tabular}{cccccccc}
     1&0&0&0&0&0&0&0\\
     0&0&1&0&0&0&0&0\\
     0&0&0&0&1&0&0&0\\
     0&0&0&0&0&0&1&0\\
     0&1&0&0&0&0&0&0\\
     0&0&0&1&0&0&0&0\\
     0&0&0&0&0&1&0&0\\
     0&0&0&0&0&0&0&1\\
    \end{tabular}\right)\,.
\end{equation}
The three $\bZ_2$'s are interfaces between any two of the three $(E_7)_1$ theories, and the $\bZ_3$ symmetry allows us to cyclically go between the three $(E_7)_1$'s. For more discussion on these surface defects see \cite{Carqueville:2017ono,Koppen:2021kry}. As for the super modular invariants of this product theory, we find 15 in total: 6 that were already mentioned and 9 new ones. Of the new matrices are idempotents:
\begin{equation}\label{idempotentSMM}
    \left(\begin{tabular}{cccccccc}
     1&0&0&0&0&0&1&0\\
     0&0&0&0&0&0&0&0\\
     0&0&0&0&0&0&0&0\\
     1&0&0&0&0&0&1&0\\
     0&1&0&0&0&0&0&1\\
     0&0&0&0&0&0&0&0\\
     0&0&0&0&0&0&0&0\\
     0&1&0&0&0&0&0&1\\
   \end{tabular}\right)\,,
   \quad 
     \left(\begin{tabular}{cccccccc}
     1&0&0&1&0&0&0&0\\
     0&0&0&0&1&0&0&1\\
     0&0&0&0&0&0&0&0\\
     0&0&0&0&0&0&0&0\\
     0&0&0&0&0&0&0&0\\
     0&0&0&0&0&0&0&0\\
     1&0&0&1&0&0&0&0\\
     0&0&0&0&1&0&0&1\\
    \end{tabular}\right)\,,
\end{equation}
formed from $\overline{(0+3)}(0+6)+\overline{(4+7)}(1+7)$ and its transpose,
\begin{equation}\label{idempotentSM2}
    \left(\begin{tabular}{cccccccc}
     1&0&0&0&0&1&0&0\\
     0&0&0&0&0&0&0&0\\
     0&0&0&0&0&0&0&0\\
     1&0&0&0&0&1&0&0\\
     0&0&1&0&0&0&0&1\\
     0&0&0&0&0&0&0&0\\
     0&0&0&0&0&0&0&0\\
     0&0&1&0&0&0&0&1\\
    \end{tabular}\right)\,,
    \quad 
    \left(\begin{tabular}{cccccccc}
     1&0&0&1&0&0&0&0\\
     0&0&0&0&0&0&0&0\\
     0&0&0&0&1&0&0&1\\
     0&0&0&0&0&0&0&0\\
     0&0&0&0&0&0&0&0\\
     1&0&0&1&0&0&0&0\\
     0&0&0&0&0&0&0&0\\
     0&0&0&0&1&0&0&1\\
      \end{tabular}\right),
\end{equation}
formed from $\overline{(0+3)}(0+5)+\overline{(4+7)}(2+7)$ and its conjugate, and
\begin{equation}\label{idempotentSM3}
   \left(\begin{tabular}{cccccccc}
     1&0&0&0&0&1&0&0\\
     0&0&1&0&0&0&0&1\\
     0&0&0&0&0&0&0&0\\
     0&0&0&0&0&0&0&0\\
     0&0&0&0&0&0&0&0\\
     0&0&0&0&0&0&0&0\\
     1&0&0&0&0&1&0&0\\
     0&0&1&0&0&0&0&1\\
     \end{tabular}\right)\,,
     \quad 
   \left(\begin{tabular}{cccccccc}
     1&0&0&0&0&0&1&0\\
     0&0&0&0&0&0&0&0\\
     0&1&0&0&0&0&0&1\\
     0&0&0&0&0&0&0&0\\
     0&0&0&0&0&0&0&0\\
     1&0&0&0&0&0&1&0\\
     0&0&0&0&0&0&0&0\\
     0&1&0&0&0&0&0&1\\
    \end{tabular}\right)\,,
\end{equation}
formed from $\overline{(0+5)}(0+6)+\overline{(2+7)}(1+7)$ and its conjugate.
There are furthermore matrices that are not idempotent, but whose elements grow as $2^{n-1}$ where $n$ is the power in which the matrix is raised
\begin{equation}\label{nonidempotentSM}
   \left(\begin{tabular}{cccccccc}
     1&0&0&0&0&1&0&0\\
     0&0&0&0&0&0&0&0\\
     0&0&1&0&0&0&0&1\\
     0&0&0&0&0&0&0&0\\
     0&0&0&0&0&0&0&0\\
     1&0&0&0&0&1&0&0\\
     0&0&0&0&0&0&0&0\\
     0&0&1&0&0&0&0&1\\
     \end{tabular}\right)\,,
   \quad 
    \left(\begin{tabular}{cccccccc}
     1&0&0&0&0&0&1&0\\
     0&1&0&0&0&0&0&1\\
     0&0&0&0&0&0&0&0\\
     0&0&0&0&0&0&0&0\\
     0&0&0&0&0&0&0&0\\
     0&0&0&0&0&0&0&0\\
     1&0&0&0&0&0&1&0\\
     0&1&0&0&0&0&0&1\\
    \end{tabular}\right)\,,
\end{equation}
\begin{equation}\label{nonidempotentSM2}
    \left(\begin{tabular}{cccccccc}
     1&0&0&1&0&0&0&0\\
     0&0&0&0&0&0&0&0\\
     0&0&0&0&0&0&0&0\\
     1&0&0&1&0&0&0&0\\
     0&0&0&0&1&0&0&1\\
     0&0&0&0&0&0&0&0\\
     0&0&0&0&0&0&0&0\\
     0&0&0&0&1&0&0&1\\
     \end{tabular}\right)\,.
\end{equation}
The three total matrices in \eqref{nonidempotentSM} and \eqref{nonidempotentSM2} are formed from 
\begin{subequations}\label{nonidempotentexpression}
\begin{align}
    &\vert 0+6\vert^2+\vert 1+7\vert^2\,, \\
    &\vert 0+5\vert^2+\vert 2+7\vert^2\,, \\
    &\vert 0+3\vert^2+\vert 4+7\vert^2\,.
\end{align}
\end{subequations}
It is natural to expect that the lines in the super modular invariant are grouped such that they differ by $\frac{1}{2}$ in spin. One can check that by condensing out $(0+3)$, $(0+5)$, $(0+6)$, that the lines which remain are $(4+7)$, $(2+7)$, and $(4+7)$ respectively.
Thus the equations in \eqref{idempotentSMM} \eqref{idempotentSM2} \eqref{idempotentSM3} are the ones that ``mix" two choices of condensation, and the expressions in \eqref{nonidempotentexpression} take each condensate individually.
\section{Ungauging Anyons}\label{ungaguginganyons}
We now consider starting off with some child theory $\mathcal{D}$ which is obtained from condensing some algebra in a parent $\cC$, and present a method for studying the $S$-matrix elements of $\cC$. The two MTCs $\cC$ and $\cD$ separated by an interface $\cF$, and both acting on $\cF$ by a braided monoidal map $\cC \boxtimes \overline \cD \rightarrow \cZ(\cF)$. Here, $\overline{\cD}$ means the category with opposite braiding, and $\cZ$ means Drinfeld center. 
This has the structure of a braided monoidal category \cite{kassel2012quantum,etingof2005fusion} with braiding given by
\begin{align}
    \cZ(\cF) : =\{ (w,\beta_x)\,|&\, w,x \in \cF \,\,\text{and}\,\, \beta_x: w\,\otimes\, x \to x \,\otimes\, w \,\,\notag  \\
   &\quad  \text{is natural in $x$, such that}\,\,  \beta_{x \otimes y} = \beta_y \,\otimes \,\beta_x  \}.
\end{align}
Thus there are two actions ${\cC} \to \cZ(\cF)$ and $ \overline \cD \to \cZ(\cF)$, which commute \footnote{Actually, the map ${\cC} \boxtimes \overline \cD \to \cZ(\cF)$ is an equivalence.}.    This implies that $\cC$ and $\overline \cD$ are each other's commutants in $\cZ(\cF)$ i.e. if we know $\cD$ and $\cF$ and $\overline{\cD} \to \cZ(\cF)$, then we can compute $\cC$. It is precisely the subcategory of $\cZ(\cF)$ of all objects that braid trivially with everything in $\cD$, and similarly in the other order. In this way it is possible to reconstruct $\cC$ from its ``boundary" $\cF$ \footnote{The term boundary is used a bit loosely because I don't mean a true boundary condition, but an interface to some other TFT.  A boundary condition is a special case where it is an interface to the vacuum.}.  The composition $\cC \rightarrow \cZ(\cF) \rightarrow \cD $ is dominant, in that every object is a direct summand of objects in the image.  On the other hand $\cD \rightarrow \cZ(\cF) \rightarrow \cC $ is not dominant, as we have already learned from gauging condensation algebras.

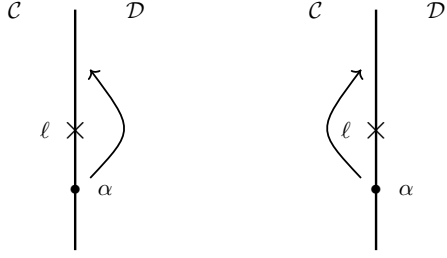
\begin{figure}
    \centering
    \scalebox{.8}{
    \begin{tikzpicture}
    \draw[line width = .4mm] (0,0)--(0,4);
    \draw (0,2) node {\scalebox{1.5}{$\times$}};
    \draw (0,1) node {$\bullet$};
    \draw[decoration={markings, mark=at position .999 with
 {\arrow{>}}}, postaction={decorate}, line width =.3mm] (.25,1.2).. controls (1.,2) .. (.25,3);
 \draw(-1,4) node {$\cC$};
 \draw (1,4) node{$\cD$};
 \draw(-1+5,4) node {$\cC$};
 \draw (1+5,4) node{$\cD$};
  \draw[decoration={markings, mark=at position .999 with
 {\arrow{>}}}, postaction={decorate}, line width =.3mm] (-.25+5,1.2).. controls (-1.+5,2) .. (-.25+5,3);
 \draw[line width = .4mm] (0+5,0)--(0+5,4);
    \draw (0+5,2) node {\scalebox{1.5}{$\times$}};
    \draw (0+5,1) node {$\bullet$};
    \draw (.5,1) node {$\alpha$};
    \draw (-.5+5,2) node {$\ell$};
        \draw (.5+5,1) node {$\alpha$};
    \draw (-.5,2) node {$\ell$};
    \end{tikzpicture}}
    \caption{We give a top down view of the interface, which is represented by the solid line, that separates theories $\cC$ and $\cD$. Suppose that $\alpha$ is a line that exists in the parent theory, but lifts off to the child theory. Then it can pass by the totally confined object in two equivalent ways.}
    \label{confinednotconfined}
\end{figure}

We now review the details of the consistency relations that we will be using to reconstruct $S$ of the parent.
Consider a boundary line $\ell$ that is confined to the interface $\mathcal{F}$, and another line $\alpha$ on the wall brought in by moving it from the bulk $\cD$.  There is strictly speaking more information that $\alpha$ carries in the bulk, which might have been forgotten by moving to the boundary, but we can still uplift $\alpha$ from the wall back in to the bulk $\cD$.  Since $\alpha$ exists as a child line, it can be restricted back to the parent, where it can pass around $\ell$.
In particular, if on the wall we have the configuration $\alpha$ then $\ell$, we can commute the two lines by lifting $\alpha$ into either of the bulks, which gives it a dimension to move around $\ell$.  So we have a configuration of $\ell$ then $\alpha$, as summarized in figure \ref{confinednotconfined}.  On the $\mathcal{D}$ side, $\alpha$ is passing an invisible line since $\ell$ does not lift off the wall.  On the parent side, both $\alpha$ and $\ell$ can be restricted to their respective lines belonging to the theory $\cC$. In general, both $\alpha$ and $\ell$ are semisimple with respect to the lines of $\cC$, thus there can be multiple choices for restrictions. 
Since the two ways of $\alpha$ passing $\ell$ are equivalent, then $S_{\alpha,\ell}=0$ in the parent, where the 0 denotes the fact that the braiding in the child theory is trivial among these two lines. 

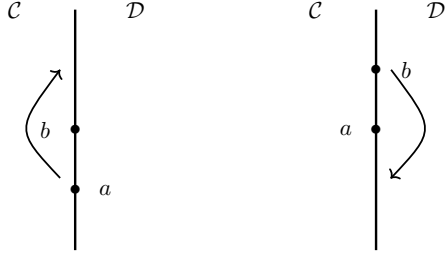
\begin{figure}
    \centering
    \scalebox{.8}{
    \begin{tikzpicture}
    \draw[line width = .4mm] (0,0)--(0,4);
    \draw (0,2) node {$\bullet$};
    \draw (0,1) node {$\bullet$};
    \draw[decoration={markings, mark=at position .05 with
 {\arrow{<}}}, postaction={decorate}, line width =.3mm] (.25+5,1.2).. controls (1.+5,2) .. (.25+5,3);
 \draw(-1,4) node {$\cC$};
 \draw (1,4) node{$\cD$};
 \draw(-1+5,4) node {$\cC$};
 \draw (1+5,4) node{$\cD$};
 \draw[line width = .4mm] (0+5,0)--(0+5,4);
    \draw (0+5,2) node {{$\bullet$}};
    \draw (0+5,3) node {$\bullet$};
    \draw (.5,1) node {$a$};
    \draw (-.5+5,2) node {$a$};
        \draw (.5+5,3) node {$b$};
    \draw (-.5,2) node {$b$};
     \draw[decoration={markings, mark=at position .999 with
 {\arrow{>}}}, postaction={decorate}, line width =.3mm] (-.25,1.2).. controls (-1.,2) .. (-.25,3);
    \end{tikzpicture}}
    \caption{Since either $a$ or $b$ may lift off the wall, the configuration obtained from
    passing either one around the other by going into the respective bulk is equivalent.}
    \label{abcommute}
\end{figure}

This is just stressing that the functor from $\cD \to \mathcal{F}$ is also central.  The compatibility for the lines in $\cC$ with the lifting procedure is if
\begin{equation}
    \cC = \{\text{relative center of}  \, \cZ(\cF ; \cD) \}\,.
\end{equation}
%
      By definition, an object $X \in \cZ(\mathcal{F}; \cD)$ is an underlying object $\underline{X} \in \mathcal{F}$ together with half-braidings $\underline X \otimes Y \sim Y \otimes \underline X$ for all $Y \in \mathcal{F}$, monoidality, and commutativity with $\mathcal{D} \subset \mathcal F$.
   Furthermore, as can be seen in figure \ref{abcommute} given $a,b$ lines on the wall where $a \in \cC$ and $b \in \cD$ originally, if we move $a$ around $b$, then we move $b$ around $a$, the two actions commute. In this case the $S$-matrix of the child can directly give the $S$-matrix elements of the parent, and we just need to ``pull-back" the data.

Already in the case where $\cD$ is the child theory as a result of condensing an abelian line from $\cC$, it is nontrivial to use the consistency relations explained above to construct the $S$-matrix of $\cC$. One could ask the obvious question which is ``what is the minimum data of $\cF$ and $\cD$ that needs to be given to determind $\cC$ uniquely?"  This question goes beyond the scope and this paper, and perhaps does not even have a general answer for any MTC $\cC$.  For our purposes we will provide the content of the line spectrum and fusion rules on the interface $\cF$, as well as the $S$-matrix of the child theory which can be calculated as in \cite{DiFrancesco:1997nk}, all in terms of the simple objects of $\cC$.

We consider an example where the fusion information of the category $\mathcal{F}$ is not enough to construct the exact parent theory (even though we might be able to attain the $S$-matrix), and we also need to give extra data in form of the associator. 
Let us suppose that $\mathcal{D}$ is trivial, and let $\mathcal{F} = \textbf{Vec}^\omega[\bZ_p]$ for $p$ an odd prime.  The fusion rules are independent of the cocycle $\omega \in \rH^3(\bZ_p; \rU(1))$ known as the associator.  By the Bockstein homomorphism for the short exact sequence 
\begin{equation}
    0 \rightarrow \bZ \rightarrow \bR \rightarrow \rU(1) \rightarrow 0\,,
\end{equation}
$\omega$ is mapped to $\rH^4(\bZ_p; \bZ )$, so $\beta(\omega) \in \rH^4(\bZ_p; \bZ ) $. There also exists a ``squaring" map that goes from 
\begin{equation}
    \rH^2(\bZ_p; \bZ) \to \rH^4(\bZ_p; \bZ)\,.
\end{equation}
Furthermore, the automorphisms of $\bZ_p$ permute the entries in $  \rH^2(\bZ_p; \bZ)$ and so permute the $\omega$ such that $\beta(\omega) = \text{Square} $.  The three possibilities that $\omega$ can take are, 
\begin{equation}
    \omega = 0, \quad \beta(\omega) = \text{Square}, \quad \beta(\omega) = \text{non-Square}.
\end{equation}
The parent is just the Drinfeld center of $\mathcal{F}$, so when $\omega = 0$, we denote $\mathcal{C}_0 = \bZ_p \times \bZ_p$, and for both of the other values of $\omega$, the parent is $\mathcal{C}_1 = \bZ_{p^2}$. At the level of groups, the map $\mathcal{C}_0 \to \mathcal{F}$ takes  $(a,b) \to [b]$, where $a$ and $b$ are valued mod $p$.  In other words, the line labeled $[j] \in \mathcal{F}$ is 
\begin{equation}
    [j] = \{(0,j), (1,j), \ldots , (p-1, j)\}\,,
\end{equation}
i.e. comes from $p$ many lines in the parent. In the case of $\mathcal{C}_1$ the map takes $(ap+b) \to [b]$.  Given this, one could not tell the case of $\mathcal{C}_0$ and $\mathcal{C}_1$ apart because in either of the ways that we label lines in the two parents, the label shows up as $[b]$ when you move to the wall \footnote{If we are also given some fusion information about the parent, then we  could at least distinguish $\mathcal{C}_0$ from $\mathcal{C}_1$.}. 
 Thus without giving the associator for the wall category~$\mathcal{F}$, the fusion of the lines on $\cF$ is not sufficient to give a unique parent in this example.

 \subsection{Analysis of Ising $\boxtimes$ $\overline{\text{Ising}}$}
 Before we explicitly reconstruct $S$-matrix elements, it is useful to use the consistency relations and apply them to evaluate $B$ elements where by $B(a,b)$ we mean the result $\frac{S_{ab}}{S_{1b}}$, where $S_{ab}$ is the trace of the full braiding of lines $a,b$.
Let $\cC$ be $\Ising \boxtimes\, \ol \Ising$, and by condensing $\varphi =\mathbb{1} \overline{\mathbb{1}}+\epsilon\overline{\epsilon}$, the child theory is the Toric code with
\begin{align}
    (\mathbb{1} \overline{\mathbb{1}}+\epsilon\overline{\epsilon}) &= {1}\,, \quad (\mathbb{1}\overline{\epsilon}+\epsilon\overline{\mathbb{1}}) = f, \notag \\
    \sigma \overline{\sigma}_1 + \sigma\overline{\sigma}_2 &= e + m\,.
\end{align}
The lines that are totally confined are given by 
\begin{equation}
    c_1 = \mathbb 1 \ol \sigma+\ep \ol \sigma\,, \quad  c_2=\sigma \ol{\mathbb{1}} + \sigma \ol \ep\,.
\end{equation}
The picture one should have in mind is given by figure \ref{isingising}.
\begin{figure}
    \centering
    \scalebox{.8}{
    \begin{tikzpicture}[thick,scale=.8]
        \def\Depth{5}
        \def\Height{6}
        \def\Width{4}
        \coordinate (O) at (3,0-1,0-1);
        \coordinate (A) at (3,\Width+1,0-1);
        \coordinate (B) at (3,\Width+1,\Height+1);
        \coordinate (C) at (3,0-1,\Height+1);
        \coordinate (D) at (\Depth,0-1,0-1);
        \coordinate (E) at (\Depth,\Width+1,0-1);
        \coordinate (F) at (\Depth,\Width+1,\Height+1);
        \coordinate (G) at (\Depth,0-1,\Height+1);
        
        \draw[above] (\Depth/2,\Width+1/2+1.5,\Height/2);
        \draw[left] (-1, \Width/2, \Height/2);
        \draw[right] (\Depth+.5, \Width/2, \Height/2);
        \draw[black,line width =.3mm,fill=green!10] (O) -- (A) -- (B) -- (C) -- cycle;
       \draw[black] (O)--(A);
       \draw (6,4) node {\underline{Toric Code}}; 
       \draw (6,3) node {$(\mathbb{1}\overline{\mathbb{1}}+\varepsilon \overline{\varepsilon})=1$};
       \draw (6,2.5) node {$(\mathbb{1}\overline{\varepsilon}+\varepsilon \overline{\mathbb{1}})=f$};
       \draw (6,2) node {$(\sigma \overline{\sigma})_1=e$};
       \draw (6,1.5) node {$(\sigma \overline{\sigma})_2=m$};
       \draw (1.75,2-1) node {$c_1=\mathbb{1}\overline{\sigma}+\epsilon \overline{\sigma}$};
       \draw (1.75,1.5-1) node {$c_2=\sigma \mathbb{1}+\sigma \overline{\epsilon}$};
        \draw (-3,4) node {$\underline{\text{Ising} \boxtimes \overline{\text{Ising}}}$}; 
    \end{tikzpicture}}
    \caption{All lines of the fusion category on the wall are written in terms of the data of the parent theory.
    The lines that can not lift off the wall are $c_1$ and $c_2$. The data of the Toric Code is drawn in the bulk but can be brought to the wall. }\label{isingising}
\end{figure}
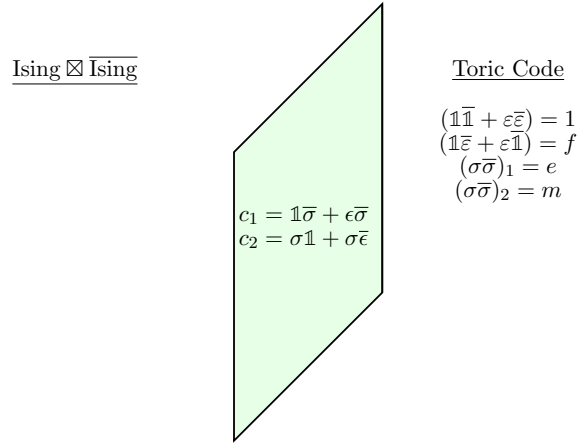
\begin{itemize}
    \item { \textbf{Braiding of lines that exist in the child theory}}
\end{itemize}
 
 Suppose we wanted to determine $S_{\sigma \ol \sigma ,\sigma \ol \sigma}$ in the parent.  There is a relationship between the $S$-matrix of Toric code and $\Ising \boxtimes \, \ol \Ising$.  This is like a ``restriction" map onto the parent theory from the child theory, and is a less expensive way of recovering the some of the $S$-matrix elements of the parent, without needing the full machinery of the fusion category on the wall.   This says that if we can build a line in the parent theory, as some data that comes from the child theory, then we can restrict the $S$-matrix from the child MTC to get the $S$-matrix of the parent.
 We know that the line $e+m$ in the child restricts to $\sigma \ol \sigma$ in the parent. 
Since we know $S_{e+m,e+m} = 0$, the restriction of this across the boundary is zero.   Indeed with knowledge of $\cC$ we find \footnote{The $S$-matrix elements $S_{a,b}$ are given by $R^{a,b}_i R^{b,a}_i$.} 
\begin{equation}
    S_{\sigma \ol \sigma, \sigma \ol \sigma} = (R^{\sigma \ol \sigma, \sigma \ol \sigma}_{\mathbb{1} \ol{ \mathbb{1}}})^2 \, d_{\mathbb{1} \ol{\mathbb{1}}} +  (R^{\sigma \ol \sigma, \sigma \ol \sigma}_{\ep \ol \ep})^2 \,d_{\ep \ol \ep}
+ (R^{\sigma \ol \sigma, \sigma \ol \sigma}_{\mathbb{1} \ol \ep})^2\, d_{\mathbb{1} \ol \ep} + (R^{\sigma \ol \sigma, \sigma \ol \sigma}_{\ep \ol{\mathbb{1}}})^2\, d_{\ep \ol{\mathbb{1}}} = 0\,.
\end{equation}
{While some elements can be restricted, in general we will need to have more knowledge of the fusion rules of the totally confined lines to understand the braiding in the parent theory. Thus we need to know $c_1 \times c_2 = e+m$. } 
From the values of $B(1,f)$, $B(e+m,1)$ and $B(e+m,f)$ in the Toric code, by restriction we get $B(1,f)$ restricts to 
\begin{align}
    B(\mathbb{1}\overline{\mathbb{1}},\mathbb{1} \overline{\epsilon})=1\,,\quad  B(\epsilon\overline{\epsilon},\mathbb{1}\overline{\epsilon})=1\notag \\
    B(\mathbb{1}\overline{\mathbb{1}},\epsilon\overline{\mathbb{1}})=1\,,\quad B(\epsilon\overline{\epsilon},\epsilon\overline{\mathbb{1}})=1\,.
\end{align}
Furthermore $B(e+m,1)$ and $B(e+m,f)$ restrict to 
\begin{align}
    B(\sigma \overline{\sigma}, \mathbb{1}\overline{\epsilon} )=-2\,, \quad B(\sigma \overline{\sigma},\epsilon\overline{\mathbb{1}})-2\,, \notag 
\end{align}
and 
\begin{align}
    B(\sigma \overline{\sigma}, \mathbb{1}\overline{\mathbb{1}} )=2\,, \quad B(\sigma \overline{\sigma},\epsilon\overline{\epsilon})=2\,. \notag 
\end{align}

\begin{itemize}
    \item { \textbf{Braiding of totally confined lines in the parent}}
\end{itemize}

The next task to understand is how the confined lines on the wall, $\mathbb{1} \ol \sigma+\ep \ol \sigma $ and $\sigma \ol{\mathbb{1}} + \sigma \ol \ep $, braid in the parent theory.  These two lines do not lift to the Toric code side, so we can not simply restrict the $S$-matrix from the Toric code to get the braiding. To answer this, suppose the line $\mathbb{1} \ol \ep + \ep \ol{\mathbb{1}}$ is brought in from the child theory to the wall.  On the wall, $(\mathbb{1} \ol \ep + \ep \ol{\mathbb{1}}) \times (\mathbb{1} \ol \sigma+\ep \ol \sigma ) \overset{\cong}{\rightarrow}     (\mathbb{1} \ol \sigma+\ep \ol \sigma ) \times(\mathbb{1} \ol \ep + \ep \ol{\mathbb{1}})$ because $(\mathbb{1} \ol \ep + \ep \ol{ \mathbb{1}})$ lifts off to the Toric code side as $f$, and so we can bring it around $\mathbb{1} \ol \sigma+\ep \ol \sigma$.   Furthermore $\mathbb{1} \ol \sigma+\ep \ol \sigma$ restricted to the parent becomes $\mathbb{1} \ol \sigma$ or $\epsilon \ol \sigma$ and similarly $\mathbb{1} \ol \ep + \ep \ol{\mathbb{1}}$ becomes $\mathbb{1} \ol \ep$ or $\ep \ol{\mathbb{1}}$.  Thus, we consider the braidings $B(\mathbb{1} \ol \sigma, \mathbb{1} \ol \ep),\, B(\ep \ol \sigma, \mathbb{1} \ol \ep)$.  An important fact to notice is that the lines $\{1,f,c_1\}$, as a subcategory of the wall fusion category, have the same fusion rules as the Ising category.  Here, $c_1$ has the fusion rules as the $\sigma$ line.  Therefore, $B(\mathbb{1}\ol \sigma, 1\ol \ep)=-\sqrt{2}$ in the parent theory 
to reflect the fact that $B(\sigma,f) =-\sqrt{2} $ in Ising.  We notice that the spin of $\ep \ol{\mathbb{1}}$ is the negative of the spin of $\mathbb{1} \ol \ep$ in the parent, so the braiding should have a relative negative i.e. $B(\mathbb{1} \ol \sigma ,\ep \ol{\mathbb{1}} )=\sqrt{2}$. Due to the restriction of $\mathbb{1}\ol \sigma +\sigma \ol \epsilon$ from the wall to the parent, then 
\begin{equation}
    B(\ep \ol \sigma ,\mathbb{1}\ol \ep)=-\sqrt{2},\quad \quad  B(\ep \ol \sigma, \ep \ol{\mathbb{1}}) = \sqrt{2}\,.
\end{equation}
 The next object to consider is  $B(\sigma \ol{\mathbb{1}} , \mathbb{1} \ol \ep )$, which is natural to consider after lifting $c_2$ to the parent.  Similar to before, we notice that $\{1,f,c_2\}$ also can be used to create a Ising subcategory. Therefore
 \begin{align*}
      B(\sigma \ol{\mathbb{1}} , \ep \ol{\mathbb{1}}   ) &= B(\sigma,f)=-\sqrt{2}\,, \quad  B(\sigma \ol{\mathbb{1}},  \mathbb{1} \ol \ep )=\sqrt{2}\,,\\
      B(\sigma \ol \ep, \ep \ol{\mathbb{1}})&= -\sqrt{2}\,, \qquad \qquad \quad  B(\sigma \ol \ep, \mathbb{1} \ol \ep) = \sqrt{2}\,.
 \end{align*}
 We now consider the braiding of $\mathbb{1} \ol \sigma$ and $\sigma \ol{\mathbb{1}}$, or in general the braiding of two lines both comprising of $\sigma$ in the parent theory.  The braiding of $B(\sigma \ol{\mathbb{1}}, \sigma \ol \ep)$ in the parent is the restriction of $c_1$ and $c_2$ from the wall.  This is analogous to asking about the braiding of two particles that behave like $\sigma$ in the Ising category, but we know $B(\sigma,\sigma) =0,$ so  $B(\sigma \ol{\mathbb{1}}, \sigma \ol \ep)=0$.   The next braidings to consider is $B(\sigma \ol \sigma, \sigma \ol{\mathbb{1}})$ and $B(\sigma \ol \sigma, \sigma \ol \ep)$.   First examine the fusion of $\sigma \ol \sigma $ with $c_1$ and $c_2$ on the wall fusion category, and notice that $c_1 \times c_2 = e+m$ and so can be moved off the wall to the Toric code side.
 If we consider on the wall $B(\sigma \ol \sigma, c_1\times c_2)$, which after moving to the Toric code is $B(e+m,e+m) = 0$, this implies that one of $ B(\sigma \ol \sigma,\, c_1),\, B(\sigma \ol \sigma, c_2)$ is equal to zero.  But $c_1$ and $c_2$ should be symmetric as particles because they play the same role in the subIsing category, and so both braidings in the parent theory should be zero. Thus we have 
 \begin{align}
     B(\sigma \ol \sigma , \mathbb{1} \ol \sigma)&= B(\sigma \ol \sigma , \epsilon \ol \sigma)=0\,, \\
      B(\sigma \ol \sigma , \sigma \ol{\mathbb{1}})&= B(\sigma \ol \sigma , \sigma \ol \epsilon)=0\,. 
 \end{align}

 \subsection{Reconstructing the Toric Code}
 We will now apply the consistency relations to a simple example of the Toric code to solve for actual $S$-matrix elements. This MTC consists of four simple objects $\{1,e,m,f\}$. It has following fusion and braiding rules
\begin{align}
    ~&e\times e=1,\quad m\times m=1,\quad e\times m=f\,,
\notag \\&B(e,e)=B(m,m)=1,\quad B(e,m)=-1\,. \notag 
\end{align}
 To help with computing the matrix elements, we give some $S$-matrix identities involving products and linearity; for $a,b,c,d$ simple lines we have 
 \begin{subequations}
 \begin{align}
    S_{a,b\times c}  &= \sum_\ell {S_{a,\ell}N^{\ell}_{b,c}}= \frac{S_{a,b} S_{a,c}}{S_{a,0}}\,, \label{verlinde} \\
       S_{a,b+c} &=S_{a,b}+S_{b,c}\,,\quad    S_{a+b,c}=S_{a,c}+S_{b,c}\,.
\end{align}
 \end{subequations}
The Toric code has two kinds of bosonic anyon condensation given by $\varphi=1+ e$ or $\varphi=1+ m$.
If we condense with $\varphi=1+ m$, the remaining aynons $\{e,f\}$ will be confined on the wall, unable to lift to the child theory. Hence the child phase $\mathcal D$ is just the vacuum $\varphi$. On the other hand, the wall category which is just a fusion category consists of wall vacuum $1+ m$ (which in this case is identical to the condensed vacuum) and the remaining confining anyons are grouped into a single module, $e+ f$. Now let us try to reconstruct the Toric code from the above condensed phase $\mathcal D$ and the wall category; the confined lines on the wall have a natural embedding in the Toric code.  We assume the fusion rules  of the confined line with $\varphi$ are known:
\begin{equation}
m\times f=e ,\quad m\times e=f\,.
\end{equation}
 
From the lifting property of $1 + m$ to be able to go to the $\cD$ side of the wall, we start off with the fact that 
\begin{equation}\label{wallandbulk}
    \frac{S_{1+m,e+f}}{S_{1,e+f}}= 1+\frac{S_{m,e}+S_{m,f}}{S_{1,e}+S_{1,f}} = 0\,.
\end{equation}
using \eqref{verlinde} we see that 
\begin{subequations}
\begin{align}
        S_{m,e\times f} &= \frac{S_{m,e}S_{m,f}}{S_{m,1}} = {S_{m,m}}\,, \\
         S_{m,e} &= {S_{m,m\times f}} = \frac{S_{m,m}S_{m,f}}{S_{1,m}}\,,\quad  S_{m,f} = {S_{m,m\times e}} = \frac{S_{m,m}S_{m,e}}{S_{1,m}}\,. \label{Sme:toriccode} 
\end{align}
\end{subequations}
From \eqref{Sme:toriccode} we have the two equations
\begin{align}
    S_{m,f}S_{1,m} &= S_{m,m}S_{m,e}\,, \\
    S_{m,e}S_{1,m}&= S_{m,m}S_{m,f}\,,
\end{align}
and combining the two equations we have
 \begin{align*}
      S_{m,m}-S_{1,m}=0, \quad  \text{or} \quad S_{m,e,}+S_{m,f}=0\,.
 \end{align*}
But by \eqref{wallandbulk}, the latter can not be zero, thus we have $ S_{m,m}=S_{1,m}$.
Another important relationship is 
\begin{equation*}
    S_{1,e\times m}=\frac{S_{1,e}S_{1,m}}{S_{1,f}} \to S^2_{1,f} = S_{1,e}S_{1,m}.
\end{equation*}
but $S_{1,e}$ and $S_{1,m}$ are equivalent, and $S_{1,e}\neq -S_{1,f}$ by \eqref{wallandbulk}, so the only consistent choice is 
\begin{equation}\label{S1ftoriccode}
    S_{1,f} =  S_{1,e}= S_{1,m}.
\end{equation}
  We now use a fact from the $S$-matrix of the child theory, which is the value of 
\begin{align}\label{child:toriccode}
    S_{1+m,1+m} = S_{1,1}+2S_{1,m}+S_{m,m} = 1\,.
\end{align}
To get the value of $S_{1,1}$ we use 
\begin{align}\label{S11toriccode}
    S_{1,m\times m} = \frac{S^2_{1,m}}{S_{1,1}} \rightarrow S^2_{1,1} = S^2_{1,m}\,.
\end{align}
But there are two choices to be made for the value in \eqref{S11toriccode}. Suppose we take 
\begin{equation}\label{choices:toriccode}
    S_{1,1}=S_{1,m}.
\end{equation}
We see immediately from \eqref{child:toriccode} that $S_{11}=\frac{1}{2}$. Then by using \eqref{wallandbulk} and \eqref{Sme:toriccode}
we see 
\begin{align*}
    S_{m,e}+S_{m,f}&=-1\,,\\
    S_{m,e}&=S_{m,f}\,,
\end{align*}
thus $S_{m,e} = S_{m,f}=-\frac{1}{2}$. Finally, to get $S_{f,f}$ notice that 
\begin{equation*}
    S_{f,f} = S_{f,e\times m}= \frac{S_{f,e}S_{f,m}}{S_{1,f}}\,,
\end{equation*}
so $S_{f,f}=\frac{1}{2}$.
With this and the other equations relating different $S$-matrix elements, as well as the symmetry between $e$ and $m$, we can fully determine $S$ of the Toric code parent theory.  One could wonder what happens if we had made the other choice in \eqref{choices:toriccode} by taking $S_{1,1}=-S_{1,m}$. If we consider  
\begin{equation}\label{Sqef:toriccode}
   S_{1,m}= S_{1,e\times f} =\frac{S_{1,e}S_{1,f}}{S_{1,1}}
\end{equation}
we get that $S_{1,1}=S_{1,f}$, coupled with the  earlier fact that $S_{1,f}=S_{1,m}$, leads to a contradiction.  

It is important to remark that in our reconstruction of the parent $S$-matrix we assumed the associator with respect to the fusion ring of $(1+m)$ and $(e+m)$ was trivial. However, because the lines are the group ring for the group $\bZ_2$ and $\rH^3(\bZ_2;\rU(1))=\bZ_2$, there also exists a nontrivial associator.  Had we chosen the nontrivial associator, the parent theory would be $\SU(2)_1\boxtimes \overline{\SU(2)_1}$ aka the semion anti-semion theory.  Let $x$ denote the nontrivial element confined on the wall such that $(xx)=1$.  Giving $x$ a central structure amounts to defining $\beta_{x,-}: x \times - \to  - \times x $, in which the only data is $\beta_{x,x} \in \bC$.  We require that braiding with the trivial element is trivial 
   \[\begin{tikzcd}
	{\tilde{x}(xx)} && {(xx)\tilde x}
	\arrow["{\beta_{x,1}=1}", from=1-1, to=1-3]\,,
\end{tikzcd}\]
and also the hexagon identity applies
\[\begin{tikzcd}
	& {\tilde{x}  (xx)} && {(\tilde{x}x)x} \\
	{(xx)\tilde{x}} &&&& {(x\tilde{x})x} \\
	& {x(x\tilde{x})} && {x(\tilde{x}x)}
	\arrow["{\alpha=-1}", from=1-2, to=1-4]
	\arrow["{\beta_{x,\tilde{x}}}", from=1-4, to=2-5]
	\arrow["{\beta_{\tilde{x},1}}"', from=1-2, to=2-1]
	\arrow["{\alpha=-1}"', from=2-1, to=3-2]
	\arrow["{\beta_{x,\tilde{x}}}"', from=3-2, to=3-4]
	\arrow["{\alpha=-1}", from=2-5, to=3-4]\, \quad .
\end{tikzcd}\]
This implies that $\beta^2_{x,x} = -1$ so $\beta_{x,x} = \pm \sqrt{-1}$.  If the associator was trivial, then $\beta_{x,x} = \pm \sqrt{1}$ and that's why $x$ would have lifted to either a boson or a fermion in the toric code. This implies that when we choose different associators that the $S$-matrix in the parent theory will be different. 

If the fusion rules on the wall are not a group, then there is a set of associators, which are solutions to some polynomial equation. In general none of the solutions have to be trivial.  In contrast, for grouplike fusion rules, one of the solutions is just a constant and deserves to be called trivial. In the examples that we will consider the fusion category of the wall as well as the child theory will be bosonic, and $H_k$ the parent theory conformally embeds into $G_{1}$ of the child. Therefore, the natural algebra object of the parent is a sum of bosonic anyons. We will use this fact to
reconstruct the $S$-matrix elements of the parent, without the need to solve for the possible associators of the wall fusion category; it is surprising that it suffices to only utilize facts about relative centers and the fusion rules on the wall. In general, given a theory with finitely many anyons, there can be infinitely many fusion rings, but there are only a finite number of categorifications. The fact that in our examples we are reconstructing a parent that comes from a conformal embedding may contribute to the fact that we did not have to give the associator, yet still landed on equations that consistently produced an $S$-matrix.


 \subsection{Reconstructing $\SU(3)_3$}
 For the case of reconstructing SU(3)$_3$ from Spin(8)$_1$ we will use the consistency relations to show the relationships among $S$-matrix elements, we will then comment on how to obtain the explicit values. As we did for the Toric code, we will split up finding the $S$-matrix into different cases.
\begin{itemize}
    \item { \textbf{$S$-matrix element with only the vacuum line}}
\end{itemize}

From \S\ref{S:condensingabelian} the lines $\varphi=0+1+2$ was the condensation algebra, so it can lift off the wall to the parent or child theory. 
  On the wall, there are three ways for the line ${(0+1+2)}$ to lift into the parent side, and go around the $(0+1+2)$ on the wall. This is like saying we have three equations from restricting $S_{\varphi,\varphi}$ to the parent (restricting ${(0+1+2)}$ back to parent), namely
\begin{equation}\label{vacuumrestrict}
    S_{{0},(0+1+2)} = \tfrac{1}{2},\, S_{{1},(0+1+2)} = \tfrac{1}{2},\, S_{{2},(0+1+2)} = \tfrac{1}{2}.
\end{equation}
 In more colloquial terms, for each one of the lift to the parent side $\{0,1, 2 \}$, we could have taken that ``lift element", moved it to the child where it becomes $\varphi$, and then gone around $\varphi$ in the child theory where $S_{\varphi,\varphi}=\frac{1}{2}$.   Since each element of $\{0,1, 2 \}$ is treated on ``equal footing" in terms of being in $\varphi$, then each element $S_{\texr i j}$ in \eqref{vacuumrestrict} should be equal to $\tfrac{1}{6}$, by distribution.  
 
 \begin{itemize}
    \item { \textbf{S-matrix elements containing the line 9}}
\end{itemize}

   From the wall to the child side, $ 9 $ has three lifts as  $(9_1+9_2+9_3)$, resulting in the other three nontrivial lines of $\Spin(8)_1$.  Each of the lifts has an $S$-matrix element $S_{\varphi, {9_j}} = \tfrac{1}{2}$ in the child, thus $S_{\varphi, {9_1}}+ S_{\varphi, {9_2}}+S_{\varphi, {9_3}}= \tfrac{3}{2}$.  When we restrict back to the parent side ${(9_1+9_2+9_3)}$ restricts to 9, and ${(0+1+2)}$ has three ways to restrict to the parent; 
   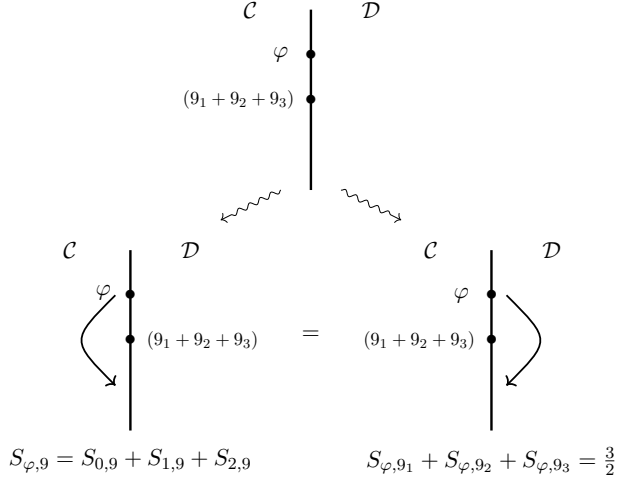
\begin{figure}
    \centering
    \scalebox{.8}{
    \begin{tikzpicture}
    \draw[line width = .4mm] (0,0)--(0,3);
    \draw (0,2.25) node {$\bullet$};
    \draw (0,1.5) node {$\bullet$};
 \draw(-1,3) node {$\cC$};
 \draw (1,3) node{$\cD$};
    \draw (-1.2,1.5) node {\scalebox{.8}{$(9_1+9_2+9_3)$}};
    \draw (-.5,2.25) node {$\varphi$};
 \draw[line width = .4mm] (3,-4)--(3,-1);
  \draw (3,2.25-4) node {$\bullet$};
    \draw (3,1.5-4) node {$\bullet$};
     \draw(-1+3,3-4) node {$\cC$};
 \draw (1+3,3-4) node{$\cD$};
    \draw (-1.2+3,1.5-4) node {\scalebox{.8}{$(9_1+9_2+9_3)$}};
    \draw (-.5+3,2.25-4) node {$\varphi$};
     \draw[decoration={markings, mark=at position .999 with
 {\arrow{>}}}, postaction={decorate}, line width =.3mm] (.25+3,2.25-4).. controls (1.+3,1.5-4) .. (.25+3,.75-4);
  \draw[line width = .4mm] (-3,-4)--(-3,-1);
  \draw (-3,2.25-4) node {$\bullet$};
    \draw (-3,1.5-4) node {$\bullet$};
     \draw(-1-3,3-4) node {$\cC$};
 \draw (1-3,3-4) node{$\cD$};
    \draw (1.2-3,1.5-4) node {\scalebox{.8}{$(9_1+9_2+9_3)$}};
    \draw (-.45-3,2.3-4) node {$\varphi$};
     \draw[decoration={markings, mark=at position .999 with
 {\arrow{>}}}, postaction={decorate}, line width =.3mm] (-.25-3,2.25-4).. controls (-1.-3,1.5-4) .. (-.25-3,.75-4);
 \draw (-3,-4.5) node {$S_{\varphi,9}=S_{0,9}+S_{1,9}+S_{2,9}$}
 ;
 \draw(3,-4.5) node {$S_{\varphi,9_1}+S_{\varphi,9_2}+S_{\varphi,9_3}=\frac{3}{2}$};
 \draw (0,-2.4) node {$=$};
 \draw[->,decorate,decoration={snake,amplitude=.4mm,segment length=2mm,post length=1mm}] (-.5,0) -- (-1.5, -.5);
  \draw[->,decorate,decoration={snake,amplitude=.4mm,segment length=2mm,post length=1mm}] (.5,0) -- (1.5, -.5);
    \end{tikzpicture}}
    \caption{The two ways of passing $\varphi$ around the totally confined line on the wall are equivalent, and this relates the $S$-matrix elements.}
    \label{twoequivmovements}
\end{figure}
   figure \ref{twoequivmovements} therefore gives the equations 
  \begin{equation}
     S_{\varphi,\texb {9_1}}+ S_{\varphi,\texb {9_2}}+S_{\varphi,\texb {9_3}}=S_{\varphi,(9_1+9_2+9_3)} \overset{\text{parent}}{\longrightarrow}S_{\varphi,9} = S_{0, 9}+S_{1, 9}+S_{2, 9} = \tfrac{3}{2}\,,
  \end{equation}
 and so  $$S_{0, 9}=S_{1, 9}=S_{2, 9}=\tfrac{1}{2}\,.$$
 The confined lines are   $(3+4+5)$ and $(6+7+8)$, since neither of these two lines lift to the child theory, the line 9 can be braided around them by going to the child side.   Restricting this to the parent means 
 \begin{equation*}
     S_{k,9} = 0\,,\quad  k\in \{3,4,5,6,7,8\}.
 \end{equation*}
 To determine $S_{9,9}$ in the parent consider taking both of the 9's and bringing them to the wall, then we get $(9_1+9_2+9_3)$ next to each other. We can lift  them to the child side in three ways, and go around each other. The sum $\sum_{i,j}S_{9_i,9_j} = -\tfrac{1}{2}$ in the child, and therefore in the parent we have
 \begin{equation*}
     S_{9,9}= S_{(9_1+9_2+9_3),(9_1+9_2+9_3)}=-\frac{1}{2}\,.
 \end{equation*}

\begin{itemize}
    \item { \textbf{$S$-matrix of totally confined lines and the vacuum }}
\end{itemize}

We now determine the braiding of the totally confined lines with $\varphi$, and with themselves in the parent. This is the most complicated case. We first recognize that since $\varphi$  can go around either $(3+4+5)$ or $(6+7+8)$ by moving to the child side, then as per figure \ref{confinednotconfined}
we get the equations
  \begin{align}\label{centerrelationssu3}
      S_{3,0}+S_{3,1}+S_{3,2} &=0\,, \notag \\
      S_{4,0}+S_{4,1}+S_{4,2} &=0\,, \notag \\
      S_{5,0}+S_{5,1}+S_{5,2} &=0\,,
  \end{align}
 as well as 
 \begin{align}
          S_{6,0}+S_{6,1}+S_{6,2} &=0\,, \notag \\
      S_{7,0}+S_{7,1}+S_{7,2} &=0\,, \notag \\
      S_{8,0}+S_{8,1}+S_{8,2} &=0\,.
\end{align}
Our method of using the relative center properties is not quite enough to solve for the matrix elements. We now employ our knowledge of the fusion of the lines on the wall, which we assume were given to us in the beginning.
 For simplicity of writing, let $a = S_{3,0},\, b= S_{3,1},\, c=S_{3,2}$. Motivated by taking 3 and encircling it around $(3+4+5)$ we consider the following fusions:
\begin{subequations}\label{equationsforS3}
\begin{align}\label{s33}
    S_{3,3} &= S_{3,4\times 2}=\frac{S_{3,4}S_{3,2}}{S_{3,0}}\,,\\\label{s34}
    S_{3,4} &= S_{3,3\times 1}=\frac{S_{3,3}S_{3,1}}{S_{3,0}}\,,\\\label{s35}
    S_{3,5} &= S_{3,3\times 2}=\frac{S_{3,3}S_{3,2}}{S_{3,0}}\,.
\end{align} 
\end{subequations}
 Furthermore by inspecting other fusion relations we have
 \begin{subequations}\label{moreequationsforS3}
\begin{align}\label{s34second}
    S_{3,4} &= S_{3,5\times 2}=\frac{S_{3,5}S_{3,2}}{S_{3,0}}\,,\\\label{s35second}
    S_{3,5} &= S_{3,4\times 1}=\frac{S_{3,4}S_{3,1}}{S_{3,0}}\,.
\end{align}
\end{subequations} 
We can plug \eqref{s34} into \eqref{s33} to get $a^2 = bc$. Also, by setting \eqref{s34} equal to \eqref{s34second} and \eqref{s35} equal to \eqref{s35second} we get $c^2=ab$ and $b^2=ac$.  All together we have the system
\begin{equation}
    a+b+c=0,\,\quad  a^2=bc,\,\quad  b^2=ac,\,\quad  c^2=ab,
\end{equation}
which has the solution $\{a,b,c\}=\{a,a \omega, a \omega^{2} \}$ and $\{ a, a \omega^{2}, a \omega \}$, where $\omega$ is a cube root of unity.  We notice that if $a$ is real, which it is because $a=S_{3,0}$ is just the quantum dimension of 3, divided by ${D} = \sqrt{\sum_i \text{q-dim}^2_i}$, then the two solutions are complex conjugates.
 The next piece of information which we can draw from the fusion rules on the wall is from using the Verlinde formula.  Consider the fact that $3 \times 3 = 6+8$, then we have 
\begin{equation}
   1=  N^8_{3,3} = \sum_{a}\frac{S_{3,a}S_{3,a}S^*_{8,a} }{S_{0,a}}\,.
 \end{equation} 
 But $S^*_{8,a} = S_{3,a}$ because $3 \times 3 = 0+9$, so we can write the above formula as 
\begin{equation}\label{verlinde338}
    1 = \sum_{a} \frac{S^3_{3,a}}{S_{0,a}}\,.
\end{equation}
 We know that given $S_{3,0}$, then $S_{3,1} = S_{3,0}\,\omega$ and $S_{3,2} = S_{3,0}\,\omega^2$. Note that this also satisfies the first equation in \eqref{centerrelationssu3}. The same holds true for $S_{3,3}$ and $S_{3,6}$ and can be easily seen from the fusion rules, i.e.
 \begin{subequations}\label{CubeUnity}
  \begin{align}
     S_{3,4}&=S_{3,3}\,\omega\,, \quad S_{3,5} = S_{3,3}\,\omega^2\,, \\
     S_{3,7}&=S_{3,6}\,\omega\,, \quad S_{3,8} = S_{3,6}\,\omega^2\,.
 \end{align}
 \end{subequations}
To use \eqref{verlinde338}, we need to relate both $S_{3,3}$ and $S_{3,6}$ to $S_{3,0}$, so then the sum can be written with only a single unknown variable. In order to make the relations manifest we use the following fusion rules 
\begin{subequations}\label{fusionforS30}
\begin{align}
S_{3,3\times 3}&\rightarrow S_{3,0}(S_{3,6}+S_{3,8}) = S^2_{3,3}\,,\\
S_{3,3\times 4}&\rightarrow  S_{3,0}(S_{3,6}+S_{3,7}) = S_{3,3}S_{3,4}\,,\\
S_{3,3\times 5}&\rightarrow  S_{3,0}(S_{3,7}+S_{3,8}) = S_{3,3}S_{3,5}\,,\\
S_{3,3\times 6}&\rightarrow  S_{3,0}(S_{3,1}+S_{3,9}) = S_{3,3}S_{3,6}\,,\\
S_{3,3\times 7}&\rightarrow  S_{3,0}(S_{3,2}+S_{3,9}) = S_{3,3}S_{3,7}\,,\\
S_{3,3\times 8}&\rightarrow  S_{3,0}(S_{3,0}+S_{3,9}) = S_{3,3}S_{3,8}\,,\\
S_{3,3\times 9}&\rightarrow  S_{3,0}(S_{3,1}+S_{3,4}+S_{3,5}) = S_{3,3}S_{3,9}\,.
\end{align}
\end{subequations}
By using the relations in \eqref{CubeUnity} and the fact that $S_{3,9}=0$ we can simplify the equations in \eqref{fusionforS30} into
\begin{subequations}
\begin{align}
    S_{3,0}S_{3,6}(1+\omega^2)&=S^2_{3,3}\,,\label{S30simplifieda}\\
    S_{3,0}S_{3,6}(1+\omega)&=S^2_{3,3}\,\omega\,,\label{S30simplifiedb}\\
    S_{3,0}S_{3,6}(\omega+\omega^2)&=S^2_{3,3}\,\omega^2\,,\label{S30simplifiedc}\\
    S^2_{3,0}\,\omega&= S_{3,3}S_{3,6}\,, \label{S30simplifiedd}\\
    S^2_{3,0}\,\omega^2&= S_{3,3}S_{3,6}\,\omega\,,\label{S30simplifiede} \\
    S^2_{3,0}&= S_{3,3}S_{3,6}\,\omega^2\,. \label{S30simplifiedf}
\end{align}
\end{subequations}
The sum of equations \eqref{S30simplifieda} and \eqref{S30simplifiedc} along with \eqref{S30simplifiedf} gives 
\begin{equation}\label{S33intoS30}
    S^3_{3,0}\,(1+\omega^2)^{-1} = S^3_{3,3}\,;
\end{equation}
by cubing \eqref{S30simplifiedf} and using \eqref{S33intoS30} we find 
\begin{equation}\label{S36intoS0}
    S^3_{3,0}(1+\omega^2)=S^3_{3,6}\,.
\end{equation}
By the fact that $(3+4+5)$ are grouped together, then $S_{3,0}=S_{4,0}=S_{5,0}$ and $S_{0,6}=S_{0,7}=S_{0,8}$ by duality of $\{6,7,8\}$ with $\{5,4,3\}$. The fusion $S_{0,3\times 6}=S_{0,0}(S_{0,1}+S_{0,9})=S_{0,3}S_{0,6}$ gives 
\begin{equation}
    (S_{3,0}-S_{0,0})(S_{3,0}+S_{0,0})=\frac{1}{2}S_{0,0}\,,
\end{equation}
where all the quantities are positive.  Assuming that the two factors on the left of the equality correspond to either $\frac{1}{2}$ or $S_{0,0}$ on the right, it must therefore be that $S_{3,0}+S_{0,0}=\frac{1}{2}$ and $S_{3,0}-S_{0,0}=S_{0,0}$. 
We can therefore boil down \eqref{verlinde338} to
\begin{align}
    1 =& \frac{3S^3_{3,0}}{\frac{1}{2}S_{3,0}}+\frac{3S^3_{3,3}}{S_{3,0}}+\frac{3S^3_{3,6}}{S_{3,0}}\notag \\
    =& \frac{3S^3_{3,0}}{\frac{1}{2}S_{3,0}}+\frac{3S^3_{3,0}\,(1+\omega^2)^{-1}}{S_{3,0}}+\frac{3 S^3_{3,0}(1+\omega^2)}{S_{3,0}}
\end{align}
which gives $S_{3,0}=\frac{1}{3}$.
We summarize the relationships as follows,
\[\begin{tikzcd}
	{S_{3,3}} & {S_{3,4}} & {S_{3,5}} & {S_{3,6}} & {S_{3,7}} & {S_{3,8}} \\
	{S_{3,0}} \\
	{S_{0,0}=S_{1,0}=S_{2,0}}
	\arrow["\omega", from=1-1, to=1-2]
	\arrow["\omega", from=1-2, to=1-3]
	\arrow["{*}", from=1-3, to=1-4]
	\arrow["\omega", from=1-4, to=1-5]
	\arrow["\omega", from=1-5, to=1-6]
	\arrow[from=2-1, to=1-1]
	\arrow[from=2-1, to=1-4]
	\arrow["{\frac{1}{2}}", from=2-1, to=3-1]
\end{tikzcd}\]
 where the arrow from $S_{3,5}$ to $S_{3,6}$ reflects the fact that the $S$-matrix elements are conjugates of each other. The arrows from $S_{3,0}$ to $S_{3,3}$ and $S_{3,6}$ reflect equations \eqref{S33intoS30} and \eqref{S36intoS0}. 
 
 We can construct the analogues of \eqref{equationsforS3} and \eqref{moreequationsforS3}, by encircling 4 and 5 around $(3+4+5)$. We have 
 \begin{subequations}\label{equationsforS4S5}
\begin{align}\label{s43}
    S_{4,3} &= S_{4,4\times 2}=\frac{S_{4,4}S_{4,2}}{S_{4,0}}\,,& S_{5,3} &=S_{5,4\times 2} =  \frac{S_{5,4}S_{5,2}}{S_{5,0}}\,,   \\\label{s433}
    S_{4,4} &= S_{4,3\times 1}=\frac{S_{4,3}S_{4,1}}{S_{4,0}}\,, & S_{5,4} &=S_{5,3\times 1} =  \frac{S_{5,3}S_{5,1}}{S_{5,0}}\,, \\\label{s44}
    S_{4,5} &= S_{4,3\times 2}=\frac{S_{4,3}S_{4,2}}{S_{4,0}}\,. & S_{5,5} &=S_{5,3\times 2} =  \frac{S_{5,3}S_{5,2}}{S_{5,0}}\,,
\end{align} 
\end{subequations}
 as well as 
  \begin{subequations}\label{moreequationsforS4S5}
\begin{align}\label{s44second}
    S_{4,4} &= S_{4,5\times 2}=\frac{S_{4,5}S_{4,2}}{S_{4,0}}\,,  & S_{5,4} &=S_{5,5\times 2} =  \frac{S_{5,5}S_{5,2}}{S_{5,0}}\,,\\\label{S54}
    S_{4,5} &= S_{4,4\times 1}=\frac{S_{4,4}S_{4,1}}{S_{3,0}}\,.  & S_{5,5} &=S_{5,4\times 1} =  \frac{S_{5,4}S_{5,1}}{S_{5,0}}\,.
\end{align}
Just like the case with $S_{3,0}$ we find 
\begin{align}
    S_{4,1}&= S_{4,0}\,\omega\,, \quad S_{4,2}=S_{4,0}\,\omega^2\,, \\
     S_{5,1}&= S_{5,0}\,\omega\,, \quad  S_{5,2}=S_{5,0}\,\omega^2\,,
\end{align}
\end{subequations} 
where $S_{4,0}=S_{5,0}=S_{3,0}$ due to their quantum dimensions. The relations among $S_{4,-}$ and $S_{5,-}$ are summarized by:
\[\begin{tikzcd}
	{S_{4,3}} & {S_{4,4}} & {S_{4,5}} & {S_{4,6}} & {S_{4,7}} & {S_{4,8}\,,}
	\arrow["\omega", from=1-1, to=1-2]
	\arrow["\omega", from=1-2, to=1-3]
	\arrow["{*}", from=1-3, to=1-4]
	\arrow["\omega", from=1-4, to=1-5]
	\arrow["\omega", from=1-5, to=1-6]
\end{tikzcd}\]
 
 \[\begin{tikzcd}
	{S_{5,3}} & {S_{5,4}} & {S_{5,5}} & {S_{5,6}} & {S_{5,7}} & {S_{5,8}\,.}
	\arrow["\omega", from=1-1, to=1-2]
	\arrow["\omega", from=1-2, to=1-3]
	\arrow["{*}", from=1-3, to=1-4]
	\arrow["\omega", from=1-4, to=1-5]
	\arrow["\omega", from=1-5, to=1-6]
\end{tikzcd}\]
Lastly, recall that $S_{4,3}$ and $S_{4,5}$ can be related to $S_{3,0}$ by our previous analysis, so all the nontrivial $S$-matrix elements that we could not obtain from restricting the child theory, we can relate to $S_{3,0}$. 

We now make a concluding remark about reconstructing the parent $S$-matrix. When we were considering the totally confined lines, as well as the child theory, all of the lines were direct sums of simple lines in the parent theory.
In this sense, we already knew about the spectrum and fusion of the parent theory, though still, it can be nontrivial to construct the $S$-matrix elements as we have seen. But one tool we gain is the Verlinde formula, which is fundamentally important and also will be used in appendix \ref{reconstructingsu210}.
One can wonder if it is possible to completely construct the parent lines through only the fusion information of the wall category.


\section*{Acknowledgments}
It is a pleasure to thank Changha Choi, Diego Delmastro, Janet Hung, and Theo Johnson-Freyd, for many fruitful discussions. A special thanks to Jaume Gomis for initiating the author's interest in anyon condensation.
This research is supported in part by Perimeter Institute for Theoretical Physics. Research at Perimeter Institute is supported in part by the Government of Canada through the Department of Innovation,
Science and Economic Development Canada and by the Province of Ontario through the Ministry of Colleges and Universities.

\newpage 
\appendix

\section{Further Examples of Nonabelian Condensation}\label{morenonabelianexaples}

{$\boldsymbol{\SU(3)_2 \times (G_2)_1}$:}
We continue with an example of a product theory; we review this example because this type of theory arises frequently when one considers using the folding trick.
We first give the two constituent spectra 
\begin{align}
    \begin{array}{c|ccc}
    \SU(3)_2 & \lambda &h & \text{q-dim}\\\hline
         0& [0,0,2] & 0 & 1 \notag \\
         1& [0,2,0] & 2/3 & 1  \notag\\
         2& [2,0,0] & 2/3 & 1 \notag\\
         3& [1,1,0] & 3/5 & 1.618033988750 \notag\\
         4& [1,0,1] & 4/15 & 1.618033988750 \notag\\
         5& [0,1,1] & 4/15 & 1.618033988750 
    \end{array}
    \qquad 
     \begin{array}{c|ccc}
    (G_2)_1 & \lambda &h & \text{q-dim}\\\hline
         0& [0,0,1] & 0 & 1 \notag \\
         1& [0,1,0] & 2/5 & 1.618033988750\,. 
    \end{array}
\end{align}
The spectrum of the product theory consists of 12 lines given by
\begin{align}
    \begin{array}{c|ccc}
    \SU(3)_2 \times (G_2)_1 & \{\ell_1,\ell_2\} &h & \text{q-dim}\\\hline
         0& \{0,0\} & 0 & 1 \notag \\
         1& \{1,0\} & 2/3 & 1  \notag\\
         2& \{2,0\} & 2/3 & 1 \notag\\
         3& \{0,1\} & 2/5 & 1.618033988750 \notag\\
         4& \{1,1\} & 16/15 & 1.618033988750 \notag\\
         5& \{2,1\} & 16/15 & 1.618033988750 \notag\\
         6& \{3,0\} & 3/5 & 1.618033988750 \notag\\
         7& \{4,0\} & 4/15 & 1.618033988750 \notag\\
         8& \{5,0\} & 4/15 & 1.618033988750 \notag\\
         9& \{3,1\} & 1 & 2.618033988750 \notag\\
         10& \{4,1\} & 2/3 & 2.618033988750 \notag\\
         11& \{5,1\} & 2/3 & 2.618033988750\,, 
    \end{array}
\end{align}
from which we can form the algebra $\varphi = 0+9$. The modules constructed from this algebra are given by 
\begin{align}\label{modulesexample1}
     \varphi\times 0 &=   \varphi\,,     & \varphi\times 6 &= 6+(3+9_2)\,,   \notag \\  
     \varphi\times 1 &=  \varphi\,,  &  \varphi\times 7 &= 7+(4+10_2)\,, \notag \\  
     \varphi\times 2 &= 2+11_1\,,  &  \varphi\times 8 &= 8+(5+11_2)\,, \notag \\ 
     \varphi\times 3 &= 3+(6+9_2)\,,  & \varphi\times 9 &= 9_1+(0+9_2+3+6)\,, \notag \\ 
     \varphi\times 4 &= 4+(7+10_2)\,,  &  \varphi\times 10 &= 10_1+(1+4+7+10_2)\,, \notag \\ 
     \varphi\times 5 &= 5+(8+11_2)\,,  &  \varphi\times 11 &= 11_1+(2+5+8+11_2)\,. 
\end{align}
The quantum dimension of the last three lines on the table, are exactly off from the quantum dimensions of lines $3$ through $8$ by 1, hinting at the fact that those three lines will split. 
Indeed, if we greedily assign the quantum dimension of line 3 and 6 to $9_2$, then the vacuum $0+9_1$ has quantum dimension 1.  We similarly assign the quantum dimension for $10_2$ and $11_2$.
By grouping based on quantum dimensions we get the lines
\begin{align}
    \begin{array}{c|c}
        \ell& \text{q-dim}     \\ \hline
        \varphi = (0+9_1) & 1 \\
         (1+10_1)& 1 \\
         (2+11_1) & 1\\
         (3+6+9_2) & 1.618033988750\\
         (4+7+10_2) & 1.618033988750\\
         (5+8+11_2) & 1.618033988750\,,
    \end{array}
\end{align}
the last three are projected out because of the simple objects have different spins. The three remaining lines 
\begin{equation}
    \{\varphi = (0+9_1),\, (1+10_1),\,(2+11_1) \}
\end{equation}
are the ones in $(E_6)_1$.\\

\paragraph {$\boldsymbol{(F_4)_3}$:} The spectrum for this theory consists of 9 lines  given by 
\begin{align}
    \begin{array}{c|ccc}
    (F_4)_3 & \lambda &h & \text{q-dim}\\\hline
         0& [0,0,0,0,3] & 0 & 1 \notag \\
         1& [0,0,0,1,2] & 1/2 & 5.449489742783  \notag\\
         2& [0,0,0,2,1] & 13/12 & 8.898979485566 \notag\\
         3& [0,0,0,3,0] & 7/4 & 4.449489742783 \notag\\
         4& [0,0,1,0,1] & 1 & 9.898979485566 \notag\\
         5& [0,0,1,1,0] & 13/8 & 10.898979485566 \notag \\
         6& [0,1,0,0,0] & 3/2 & 5.449489742783 \notag\\
         7& [1,0,0,0,1] & 3/4 & 4.449489742783 \notag\\
         8& [1,0,0,1,0] & 4/3 & 8.898979485566 \notag
    \end{array}
    \end{align}
from which we form the algebra $\varphi = 0+4$. By inspecting the quantum dimension, we see that $4.449\ldots$ is the lowest that is not 1, and the other higher quantum dimensions can be partitioned into $4.449\ldots$ and 1. 
The modules constructed from this algebra are given by 
\begin{align}
    \varphi \times 0 &= \varphi\,, \notag\\
    \varphi \times 1 &= 1_1+ (1_2+2_2+4_3+5_4+6_2+7+8_2)\,,\notag\\
    \varphi \times 2 &= 2+ (1_2+2_2+3+4_3+4_3+5_3+5_4+6_2+7+8_1+8_2)\,,\notag\\
    \varphi \times 3 &=3+ (2_2+4_3+5_4+6_2+8_2)\,,\notag \\
    \varphi \times 4 &=4_1+(0+1+2_1+2_2+3+4_2+4_3+5_3+5_4+6+7+8_1+8_2)\,, \notag\\
    \varphi \times 5 &= 5_1+(1+2_1+2_2+3+4_2+4_3+5_2+5_3+5_4+6+7+8_1+8_2)\,, \notag\\
    \varphi \times 6 &=6_1+(1_2+2_2+3+4_3+5_4+6_2+8_2)\,,\notag \\
    \varphi \times 7 &=7+(1_2+2_2+4_3+5_4+8_2)\,,\notag \\
    \varphi \times 8 &= 8_1+(1_2+2_1+2_2+3+4_2+4_3+5_3+5_4+6_2+7+8_2)\,.
\end{align}
While the fusion structure is more complicated, one does notice the following grouping of lines to appear  
$$ (1_2+4_2+5_3+7+8_1)\,, \quad (2_2+3+4_3+5_4+6_2+8_2),$$
 both of which we greedy assign q-dim 4.49$\ldots$, which is that of line 3 and 7.  Together with the remaining lines we form the groupings given by
 \begin{align}
    \begin{array}{c|c}
        \ell& \text{q-dim}     \\ \hline
        \varphi = (0+4_1) & 1 \\
         (1_1+6_1)& 1 \\
         5_1 & 1\\
         5_2 &1 \\
         (1_2+4_2+5_3+7+8_1) & 4.449489742783\\
        (2_2+3+4_3+5_4+6_2+8_2) & 4.449489742783\,,
    \end{array}
\end{align}
 the first four 
 \begin{equation}
     \{ \varphi = (0+4_1),\, (1_1+6_1),\, 5_1,\,5_2 \}
 \end{equation}
 are the lines of $\Spin(26)_1$, while the last two are projected out.
 \paragraph \large{$\boldsymbol{(G_2)_4}$:} The spectrum for this theory consists of 9 lines  given by 
 \begin{align}
    \begin{array}{c|ccc}
    (G_2)_4 & \lambda &h & \text{q-dim}\\\hline
         0& [0,0,4] & 0 & 1 \notag \\
         1& [0,1,3] & 1/4 & 4.449489742783  \notag\\
         2& [0,2,2] & 7/12 & 8.898979485566 \notag\\
         3& [0,3,1] & 1 & 9.898979485566 \notag\\
         4& [0,4,0] & 3/2 & 5.449489742783 \notag\\
         5& [1,0,2] & 1/2 & 5.449489742783 \notag \\
         6& [1,1,1] & 7/8 & 10.898979485566 \notag\\
         7& [1,2,0] & 4/3 & 8.898979485566 \notag\\
         8& [2,0,0] & 5/4 & 4.449489742783 \notag
    \end{array}
    \end{align}
    from which we form the algebra $\varphi=0+3$. The modules constructed from this algebra are given by 
    \begin{align}
    \varphi \times 0 &= \varphi \notag\\
    \varphi \times 1 &= 1+ (2_2+3_3+4_2+6_4+7_2)\notag\\
    \varphi \times 2 &= 2_1+ (1+2_2+3_2+3_3+4_2+5+6_3+6_4+7_1+7_2+8)\notag\\
    \varphi \times 3 &=3_1+ (0+1+2_1+2_2+3_2+3_3+4_2+5+6_3+6_4+7_1+7_2+8)\notag \\
    \varphi \times 4 &=4_1+(1+2_2+3_3+4_2+5_2+6_4+7_2) \notag\\
    \varphi \times 5 &= 5_1+(2_2+3_3+4_2+5_2+6_4+7_2+8) \notag\\
    \varphi \times 6 &=6_1+(1+2_1+2_2+3_1+3_2+4+5+6_2+6_3+6_4)\notag \\
    \varphi \times 7 &=7+(1+2_1+2_2+3_1+3_2+4+5+6_3+6_4+7_2+8)\notag \\
    \varphi \times 8 &= 8_1+(2_2+3_3+5+6_4+7_2)\,.
\end{align}
By grouping based on quantum dimensions we greedily assign the dimension of line 8 and line 1, which is the lowest quantum dimension that is not 1, to the lines 
\begin{equation}
    (1+2_2+3_3+6_4+7_2)\,,\quad  (2_1+6_3+7_1+8)
\end{equation}
which appear repeatedly in the equations above.
In summary the groupings are
\begin{align}
    \begin{array}{c|c}
        \ell& \text{q-dim}     \\ \hline
        \varphi = (0+3_1) & 1 \\
         (4_1+5_1)& 1 \\
         6_1 & 1\\
         6_2 & 1\\
         (1+2_2+3_3+6_4+7_2) & 4.4494897427830\\
         (2_1+6_3+7_1+8) & 4.4494897427830\\
            (3_2+5_2)& 5.4494897427830\,.
    \end{array}
\end{align}
The last three lines are projected out due to the fact that the simple objects have different spins.  The first four lines give those of $\Spin(14)_1$.\\

\paragraph \large{$\boldsymbol{\SU(3)_5}$:} It will be clear after this example that as the number of lines becomes even larger, finding the modules for a condensation algebra becomes a tedious task.  The spectrum of this theory contains 21 lines given by
 \begin{align}
    \begin{array}{c|ccc}
    \SU(3)_5 & \lambda &h & \text{q-dim}\\\hline
         0& [0,0,5] & 0 & 1 \notag \\
         1& [0,5,0] & 5/3 & 1  \notag\\
         2& [5,0,0] & 5/3 & 1 \notag\\
         3& [1,4,0] & 3/2 & 2.414213562373 \notag\\
         4& [4,0,1] & 7/6 & 2.414213562373 \notag\\
         5& [0,1,4] & 1/6 & 2.414213562373 \notag \\
         6& [3,0,2] & 3/4 & 3.414213562373 \notag\\
         7& [0,2,3] & 5/12 & 3.414213562373 \notag\\
         8& [2,3,0] & 17/12 & 3.414213562373 \notag\\
         9& [0,3,2] & 3/4 & 3.414213562373 \notag \\
         10& [3,2,0] & 17/12 & 3.414213562373 \notag\\
    \end{array}
    \quad 
    \begin{array}{c|ccc}
    \SU(3)_5 & \lambda &h & \text{q-dim}\\\hline
    11& [2,0,3] & 5/12 & 3.414213562373 \notag\\
         12& [4,1,0] & 3/2 & 2.414213562373 \notag\\
         13& [1,0,4] & 1/6 & 2.414213562373 \notag \\
         14& [0,4,1] & 7/6 & 2.414213562373  \notag\\
         15& [1,1,3] & 3/8 & 4.828427124746 \notag\\
         16& [1,3,1] & 25/24 & 4.828427124746 \notag\\
         17& [3,1,1] & 25/24 & 4.828427124746 \notag\\
         18& [2,2,1] & 1 & 5.828427124746 \notag \\
         19& [2,1,2] & 2/3 & 5.828427124746 \notag\\
         20& [1,2,2] & 2/3 & 5.828427124746 \notag\\
    \end{array}
    \end{align}
    The modules for the algebra $\varphi = 0+18$, created by the nonabelian boson is
    \begin{align}\label{modulesofsu35}
        \varphi\times 0 &=   \varphi\,,     & \varphi\times 11 &= 11_1+(8_2+11_2+14+17_2+20_3)\,,   \notag \\  
     \varphi\times 1 &=  1+19_1\,,  &  \varphi\times 12 &= 12+(9_2+15_2+18_3)\,, \notag \\  
     \varphi\times 2 &= 2+20_1\,,  &  \varphi\times 13 &= 13+(10_2+16_2+19_3)\,, \notag \\ 
     \varphi\times 3 &= 3+(6_2+15_2+18_3)\,,  & \varphi\times 14 &= 14+(11_2+17_2+20_3)\,, \notag \\ 
     \varphi\times 4 &= 4+(7_2+16_2+19_3)\,,  &  \varphi\times 15 &= 15_1+(3+6_2+9_2+12 \notag \\
      \varphi\times 5 &= 5+(8_2+17_2+20_3)\,,&  &\quad +15_2+18_2+18_3)\,, \notag \\ 
     \varphi\times 6 &= 6_1+(3+6_2+9_2+15_2+18_3)\,,  &  \varphi\times 16 &= 16_1+(4+7_2+10_2+13 \notag \\
     \varphi\times 7 &= 7_1+(4+7_2+10_2+16_2+19_3)\,, & &\quad +16_2+19_2+19_3)\,, \notag \\
     \varphi\times 8 &= 8_1+(5+8_2+11_2+17_2+20_3)\,,  &  \varphi\times 17 &= 17_1+(5+8_2+11_2+14 \notag \\
       \varphi\times 9 &= 9_1+(6_2+9_2+12+15_2+18_3)\,, & & \quad +17_2+20_2+20_3)\,, \notag \\
    \varphi\times 10 &= 10_1+(7_2+10_2+13+16_2+19_3)\,.  &  \varphi\times 18 &= 18_1+(0+3+6_2+9_2+12 \notag \\
     & &  &\quad +15_1+15_2+18_2+18_3)\,, \notag \\
       & &  \varphi\times 19 &= 19_1+(1+4+7_2+10_2+13 \notag \\
       & & &\quad +16_1+16_2+19_2+19_3)\,,
    \end{align}
    \begin{align}
        \varphi\times 20 &= 20_1+(2+5+8_2+11_2+14+17_1+17_2+20_2+20_3)\,,
    \end{align}
By closely examining the repeating structures within the modules, we can see the following grouping of lines    
\begin{align}
    \begin{array}{c|c}
        \ell& \text{q-dim}     \\ \hline
        \varphi = (0+18_1) & 1 \\
        (1+19_1)& 1\\
        (2+20_1)& 1 \\
        (6_1+9_1)&1 \\
         (7_1+8_1)& 1 \\
         (10_1+11_1) & 1\\
    \end{array}
    \quad 
     \begin{array}{c|c}
        \ell& \text{q-dim}     \\ \hline
         (3+6_2+9_2+15_2+18_3) & 2.414213562373\\
         (15_1+18_2) & 2.414213562373 \\
         (4+7_2+10_2+13+16_2+19_3) & 2.414213562373\\
         (16_1+19_2) & 2.414213562373 \\
         (5+8_2+11_2+14+17_2+20_3)& 2.414213562373\\
         (17_1+20_2)& 2.414213562373\,.
    \end{array}
\end{align}
\\

\paragraph \large{$\boldsymbol{\Sp(16)_1}$:} We present this theory to give a nontrivial example of when nonabelian condensation for a line with non-integer spin can be performed after abelian condensation, in a consistent way. In the bulk of the paper, it was shown that for $(G_2)_2$ that there was no canonical way to group lines and assign quantum dimensions in any consistent way. But we will see in this simple example that the grouping of lines is canonical.
The spectrum consists of 9 lines given by
\begin{align}
    \begin{array}{c|ccc}
    \Sp(16)_1 & \lambda &h & \text{q-dim}\\\hline
         0& [0,0,0,0,0,0,0,0,1] & 0 & 1 \notag \\
         1& [0,0,0,0,0,0,0,1,0] & 2 & 1  \notag\\
         2& [0,0,0,0,0,0,1,0,0] & 77/40 & 1.902113032590 \notag\\
         3& [1,0,0,0,0,0,0,0,0] & 17/40 & 1.902113032590 \notag\\
         4& [0,0,0,0,0,1,0,0,0] & 9/5 & 2.618033988750 \notag\\
         5& [0,1,0,0,0,0,0,0,0] & 4/5 & 2.618033988750 \notag \\
         6& [0,0,0,0,1,0,0,0,0] & 13/8 & 3.077683537175 \notag\\
         7& [0,0,1,0,0,0,0,0,0] & 9/8 & 3.077683537175 \notag\\
         8& [0,0,0,1,0,0,0,0,0] & 7/5 & 3.236067977500\,. \notag
    \end{array}
    \end{align}
Upon condensing out the abelian boson  we are left with the lines 
\begin{align}
    \begin{array}{c|ccc}
    \Sp(16)_1/\bZ_2&\ell & h & \text{q-dim}\\\hline
         0& \varphi= (0+1) & 0 & 1 \notag \\
         1& (4+5)  & 4/5 & 2.618033988750  \notag\\
         2& 8_1 & 7/5 & 1.618033988750 \notag\\
         3& 8_2 & 7/5 & 1.618033988750\,,\notag\\
    \end{array}
    \end{align}
from which we sequentially condense $\tilde{\varphi}=0+1$, noticing that this is a nonabelian spin $\frac{4}{5}$ line that usually would have been abelian after the boson condensation.  Nevertheless, the modules are 
\begin{align}
    \tilde \varphi \times 0 &= \tilde \varphi\,, \notag \\
    \tilde \varphi \times 1 &= 1_1+(0+1_2+2+3)\,, \notag \\
    \tilde \varphi \times 2 &= 2+(1_2+3)\,, \notag \\
    \tilde \varphi \times 3 &= 3+(1_2+2)\,,
\end{align}
from which we can see that the remaining lines are $\tilde \varphi$ and $(1_2+2+3)$ with quantum dimension equal to the golden ratio.  As a remark, the modular invariants of $\Sp(16)_1$ only captures the abelian condensation, and not the second step. The spectrum of lines in $\Sp(16)_1/\bZ_2$ have spins that are all of a common denominator, so the set of $\cM$ contain more than just those which can be built from algebras. 
\\

\paragraph \large{$\boldsymbol{{\SU(4)_4}/{\bZ_4}}$:} We consider an example of a nonsimply connected group to prime ourselves for the next example in this appendix.  We will condense out an abelian line in $\SU(4)_4$, and follow up with a nonabelian condensation. After the abelian condensation the spectrum consists of 14 lines already given in \S\ref{S:condensingabelian}.
The algebra formed by the nonabelian boson, $\varphi = 0+6$ has as its modules  
\vspace{-2mm}
\begin{align}\label{modulesofsu4/4}
     \varphi\times 0 &=   \varphi\,,     & \varphi\times 8 &= 8+(6_3+7_2+11)\,, \notag \\   
     \varphi\times 1 &=  1+7_3\,,  &   \varphi\times 9 &= 9+(6_3+7_2+10)\,, \notag \\
     \varphi\times 2 &= 2+(3+4+5+12_2+13_2)\,,  & \varphi\times 10 &= 10+(6_3+7_2+9)\,, \notag \\ 
     \varphi\times 3 &= 3+(2+4+5+12_2+13_2)\,,  &  \varphi\times 11 &= 11+(6_3+7_2+8)\,, \notag \\ 
     \varphi\times 4 &= 4+(2+3+5+12_2+13_2)\,,  & \varphi \times 12 &= 12_1+(2+3+4+5+12_2+13_1+13_2)\,, \notag \\ 
     \varphi\times 5 &= 5+(2+3+4+12_2+13_2)\,,  & \varphi \times 13 &= 13_1+(2+3+4+5+12_1+12_2+13_2)\,.\notag 
\end{align}
\vspace{-10mm}
\begin{align}
    \varphi\times 6 &= 6_1+(0+6_2+6_3+7_1 +7_2+8+9+10+11)\,, & \phantom{aaaaaaaaaaaaaaaaaaaaaaaa}&\,\notag\\
    \varphi\times 7 &= 7_1+(1+6_2+6_3+7_2+7_3+8+9+10+11)\,,& &
\end{align}
%
%
%
%
%
%
The natural grouping of the lines from the modules is 
\begin{align}
    \begin{array}{c|c}
        \ell& \text{q-dim}     \\ \hline
        \varphi = (0+6_1) & 1 \\
        (1+7_3)& 1\\
        (12_1+13_1)& 1.414213562373 \\
         (6_2+7_1)& 2.414213562373 \\
         (6_3+7_2+8+9+10+11) & 2.414213562373\\
         (2+3+4+5+12_2+13_2) & 3.414213562373\,,
    \end{array}
\end{align}
The last two lines are confined due to the differing spins, so we find the remaining lines are 
\begin{equation*}
    \{\varphi = (0+6_1),\, (1+7_3),\, (12_1+13_1) \}\,.
\end{equation*}
\\

\paragraph \large{$\boldsymbol{\SU(2)_4^{o3}}$:} In this example we construct a theory where we show how anyon condensation can give insights into the symmetries of the theory that we may not have expected at first sight.   Consider $\SU(2)_4^3$, its abelian anyons form a $(\bZ_2)^3$ group. All of these are condensable, but we choose only to condense the $(\bZ_2)^2$ subgroup given by the lines $\{000, 110, 101, 011\}$.  Here, the numbers
denote the lines coming from each of the $\SU(2)_4$ factors, the spectrum was given in \S\ref{S:condensingabelian}
    The result we will call $\SU(2)_4^{o3}$ where the `o' stands for ``central product". The data of the spectrum consists of 17 lines and is given by
    \begin{align}
    \begin{array}{c|ccc}
    \SU(2)_4^{o3} & \{\ell_1,\ell_2,\ell_3\} &h & \text{q-dim}\\\hline
         0& \{0,0,0\} & 0 & 1 \notag \\
         1& \{0,0,1\} & 1 & 1  \notag\\
         2& \{0,0,4\} & 1/3 & 2 \notag\\
         3& \{0,4,0\} & 1/3 & 2 \notag\\
         4& \{0,4,4\} & 2/3 & 2 \notag\\
         5& \{0,4,4\} & 2/3 & 2 \notag\\
         6& \{2,2,2\} & 3/8 & 5.196152422706 \notag\\
         7& \{2,2,3\} & 7/8 & 5.196152422706 \notag\\
                  8& \{4,0,0\} & 1/3 & 2 \notag\\
    \end{array}
    \quad 
    \begin{array}{c|ccc}
     \SU(2)_4^{o3} & \{\ell_1,\ell_2,\ell_3\} &h & \text{q-dim}\\\hline
         9& \{4,0,4\} & 2/3 & 2 \notag\\
         10& \{4,0,4\} & 2/3 & 2 \notag\\
         11& \{4,4,0\} & 2/3 & 2  \notag\\
         12& \{4,4,0\} & 2/3 & 2 \notag\\
         13& \{4,4,4\} & 1 & 2 \notag\\
         14& \{4,4,4\} & 1 & 2 \notag\\
         15& \{4,4,4\} & 1 & 2 \notag  \\
         16& \{4,4,4\} & 1 & 2\,. 
    \end{array}
\end{align}
The 8-dimensional representation $\textbf{2}\otimes \textbf{2} \otimes \textbf{2}$ of $\SU(2)^3$ gives a map $\SU(2)_4^{o3} \to \Sp(8)_1$ which is conformal. The condensable anyons are any one of $\{13,14,15,16\}$, and one could wonder which algebra gives the conformal embedding. We will see that all four anyons can condense to give $\Sp(8)_1$.  The problem inherently has a triality due to the three $\SU(2)$ factors, but given the spectrum data and the fact actually four lines can condense prompts us to believe that as an MTC,  $\SU(2)_4^{o3}$ has an extra symmetry that is $S_4$. Since the theory has 17 lines, there are $17!$ permutations that are potentially a symmetry of the theory. A permutation will be a symmetry if it preserves the full modular data. One can see that there are $3!\cdot 4! \cdot6!$ permutations that preserve the spins and quantum dimensions.  Out of these, a brute force check shows that there are exactly 24 permutations that also preserve the fusion rules. Finally, by looking at how these permutations compose, it is straightforward to show that they correspond to the group $S_4$ \footnote{The $S_4$ preserves the $S$ and $T$ matrices of $\SU(2)_4^{o3}$, but in principle one should also check the F- and R-symbols.  We believe it should be possible to compute these symbols in terms of those of $\SU(2)_4$ }.

Instead of doing the complete analysis given above, we can see hints of an enlarged symmetry when we consider the theory after condensing each of the four nonabelian bosons. We present only the modules of $\varphi_1=0+13$, as the same procedure works for the other choices of condensate:
\begin{align}
    \varphi \times 0 &= \varphi\,, & \varphi \times 9 &= 9+(3+14)\,, \notag \\ 
    \varphi \times 1 &= 1+13_2\,, & \varphi \times 10 &=10+(5+12)\,, \notag \\
    \varphi \times 2 &= 2+(11+15)\,, & \varphi \times 11 &= 11+(2+15)\,, \notag \\
     \varphi \times 3 &= 3+(9+14)\,, & \varphi \times  \notag 12&= 12+(5+10)\,, \\
    \varphi \times 4 &= 4+(8+16)\,, & \varphi \times 13&=13_1+(0+1+13_2)\,, \notag \\
    \varphi \times 5 &= 5+(10+12)\,, & \varphi \times 14 &= 14+(3+9)\,, \notag \\
    \varphi \times 6 &= 6_1+(6_2+7_2)\,, & \varphi \times 15 &= 15+(2+11)\,,\notag \\
    \varphi \times 7 &= 7_1+(6_2+7_2)\,, & \varphi \times 16&= 16+(4+8)\,. \notag \\
    \varphi \times 8 &= 8+(4+16)\,,
\end{align}
In total, the modules for $\varphi_1 = 0+13$, $\varphi_2 = 0+14$, $\varphi_3 = 0+15$, and $\varphi_4 = 0+16$ give the organization of lines as follows  \footnote{A priori there is an ambiguity in splitting the quantum dimension of 6 and 7 into its constituents.  The way the dimensions were assigned is guided by the fact that there exists a conformal embedding.}:
\begin{align}
    \begin{array}{c|ccc}
    \varphi_1 & \ell & \text{q-dim}\\\hline
         0& (0+13_1)  & 1 \notag \\
         1& (1+13_2)  & 1 \notag \\
         2& (2+11+15)  & 1  \notag\\
         3& (3+9+14)  & 2 \notag\\
         4& (4+8+16) & 2 \notag\\
         5& (5+10+12) & 2 \notag\\
         6& 6_1& 1.732050807568 \notag \\
         7& 7_1& 1.732050807568 \notag \\
         8& 6_2+7_2& 3.464101615137
    \end{array}
    \quad \quad \begin{array}{c|ccc}
    \varphi_2 & \ell & \text{q-dim}\\\hline
         0& (0+14_1)  & 1 \notag \\
         1& (1+14_2)  & 1 \notag \\
         2& (2+12+16)  & 1  \notag\\
         3& (3+9+13)  & 2 \notag\\
         4& (4+10+11) & 2 \notag\\
         5& (5+8+15) & 2 \notag\\
         6& 6_1& 1.732050807568 \notag \\
         7& 7_1& 1.732050807568 \notag \\
         8& 6_2+7_2& 3.464101615137
    \end{array}
    \end{align}
\begin{align}
    \begin{array}{c|ccc}
    \varphi_3 & \ell & \text{q-dim}\\\hline
         0& (0+15_1)  & 1 \notag \\
         1& (1+15_2)  & 1 \notag \\
         2& (2+11+13)  & 1  \notag\\
         3& (3+10+16)  & 2 \notag\\
         4& (4+9+12) & 2 \notag\\
         5& (5+8+14) & 2 \notag\\
         6& 6_1& 1.732050807568 \notag \\
         7& 7_1& 1.732050807568 \notag \\
         8& 6_2+7_2& 3.464101615137
    \end{array}
    \quad \quad 
     \begin{array}{c|ccc}
    \varphi_4 & \ell & \text{q-dim}\\\hline
         0& (0+16_1)  & 1 \notag \\
         1& (1+16_2)  & 1 \notag \\
         2& (2+12+14)  & 1  \notag\\
         3& (3+10+15)  & 2 \notag\\
         4& (4+8+13) & 2 \notag\\
         5& (5+9+11) & 2 \notag\\
         6& 6_1& 1.732050807568 \notag \\
         7& 7_1& 1.732050807568 \notag \\
         8& 6_2+7_2& 3.464101615137\,.
    \end{array}
\end{align}
From the tables above the unconfined lines are
\begin{align}
     \begin{array}{c|c}
        \ell& \text{q-dim}     \\ \hline
        \varphi_1 = (0+13_1) & 1 \\
        (1+13_2)& 1\\
        6_1& 1.732050807568 \\
         7_1& 1.732050807568 \\
         (5+10+12) & 2\,
    \end{array}
    \quad 
     \begin{array}{c|c}
        \ell& \text{q-dim}     \\ \hline
        \varphi_1 = (0+14_1) & 1 \\
        (1+14_2)& 1\\
        6_1& 1.732050807568 \\
         7_1& 1.732050807568 \\
         (4+10+11) & 2\,
    \end{array}
    \end{align}
\begin{align}
     \begin{array}{c|c}
        \ell& \text{q-dim}     \\ \hline
        \varphi_1 = (0+15_1) & 1 \\
        (1+15_2)& 1\\
        6_1& 1.732050807568 \\
         7_1& 1.732050807568 \\
         (4+9+12) & 2\,
    \end{array}
    \quad 
        \begin{array}{c|c}
        \ell& \text{q-dim}     \\ \hline
        \varphi_1 = (0+16_1) & 1 \\
        (1+16_2)& 1\\
        6_1& 1.732050807568 \\
         7_1& 1.732050807568 \\
         (5+9+11) & 2\,
    \end{array}
\end{align}
where each choice of condensation gives a copy of $\Sp(4)_1$, hence the triality symmetry we were expecting should be enlarged to a group that can permute four objects.

\section{Reconstruction of $\SU(2)_{10}$}\label{reconstructingsu210}
One of the new features of this example is that when a line splits such that one part is confined and one part moves to the child, we have some different condition on the $S$-matrix element.  To see this explicitly, consider from the following table
\begin{align}
    \begin{array}{c|c}
        \ell& \text{confined/unconfined}     \\ \hline
        \varphi=0+6_1 & \text{unconfined} \\
         (4_1+10)& \text{unconfined} \\
         (3_1+7_1) & \text{unconfined}\\
         (1+5_2+7_2) & \text{confined} \\
         (3_2+5_1+9) & \text{confined} \\
         (2+4_2+6_2+8) &\text{confined}\,
    \end{array}
\end{align}
the element $S_{(1+5_2+7_2),\varphi}$. Since $\varphi$ can move past a totally confined line by going to the child theory, we would expect that 
\begin{equation}
    S_{1,\varphi} = S_{5,\varphi} = S_{7,\varphi}=0
\end{equation}
in the parent theory.  However, the last equality does not hold due to the fact that there is an unconfined line with $7_1$ as a constituent object. When it is not the case that $S_{a,b}$ is between lines where a single line splits on the wall and into the child, then the consistency relations in \S\ref{ungaguginganyons} still hold. We will now run through the cases for the $S$-matrix.
\begin{itemize}
    \item $S_{\text{confined,unconfined}}$ 
\end{itemize}
More explicitly, from $\varphi$, $(4_1+10)$, $(3_1+7_1)$ going around $(1+5_2+7_2)$ we see that 
\begin{subequations}
\begin{align}
~&S_{1,0}+S_{1,6}=S_{5,0}+S_{5,6}=0\,,  \label{Aroundc1main}
\\&S_{1,4}+S_{1,10}=S_{5,4}+S_{5,10}=0\,, \label{Aroundc1third}
\\&S_{1,3}+S_{1,7}=S_{5,3}+S_{5,7}=0\,.\label{Aroundc1}
\end{align}
\end{subequations}
From the unconfined lines brought around $(3_2+5_1+9)$ we have 
\begin{align}\label{Aroundc2}
~&S_{5,0}+S_{5,6}=S_{9,0}+S_{9,6}=0\,, 
\\&S_{5,4}+S_{5,10}=S_{9,4}+S_{9,10}=0\,,
\\&S_{5,3}+S_{5,7}=S_{9,3}+S_{9,7}=0\,.
\end{align}
Next consider the unconfined lines brought around $(2+ 4_2 +6_2+ 8)$
\begin{align}
~&S_{2,0}+S_{2,6}=S_{8,0}+S_{8,6}=0\,,\label{S20S26} 
\\&S_{2,4}+S_{2,10}=S_{8,4}+S_{8,10}=0\,,\label{S24S210} 
\\&S_{2,3}+S_{2,7}=S_{8,3}+S_{8,7}=0\,.\label{S23S27}
\end{align}

\begin{itemize}
    \item $S_{\text{confined,confined}}$ 
\end{itemize}
Here we apply the same logic as above for the $S$-matrix between two confined lines, using the intuition that one of confined line can be lifted to the parent theory making trivial braiding with other confined line in the wall. We list all of the relations for one confined line encircling another, in which the ``moving" line does not involve a simple object that splits into a component on the wall and a component in the child.
First consider $S_{(1+ 5_2+ 7_2),(1+ 5_2+ 7_2)}$, we expect three relations
\begin{align}\label{c1withc1}
~&S_{1,1}+S_{1,5}+S_{1,7}=0\,,
\\&S_{5,1}+S_{5,5}+S_{5,7}=0.
\end{align}
The next term $S_{(3_2+ 5_1+9),(3_2+ 5_1+9)}$ gives equations 
\begin{subequations}
\begin{align}
~&S_{5,3}+S_{5,5}+S_{5,9}=0\,,
\\&S_{9,3}+S_{9,5}+S_{9,9}=0.
\end{align} 
\end{subequations}
The last diagonal term is $S_{(2+ 4_2+ 6_2+8), (2+ 4_2+ 6_2+8)}$ and gives equations
\begin{subequations}
\begin{align}\label{c3withc3}
~&S_{2,2}+S_{2,4}+S_{2,6}+S_{2,8}=0\,,
\\&S_{8,2}+S_{8,4}+S_{8,6}+S_{8,8}=0.
\end{align}
\end{subequations}
We now look at the off diagonal terms of the $S$-matrix, starting off with $S_{(1+ 5_2+ 7_2), (3_2+ 5_1+9)}$, which gives the equations 
\begin{subequations}
\begin{align}
~& S_{1,3}+S_{1,5}+S_{1,9}=0\,,\label{c1withc2first}
\\& S_{5,3}+S_{5,5}+S_{5,9}=0\,,
\\& S_{1,5}+S_{5,5}+S_{7,5}=0\,,
\\& S_{1,9}+S_{5,9}+S_{7,9}=0\,,
\end{align} 
\end{subequations}
where the first two equations arise from $(1+ 5_2+ 7_2)$ encircling $(3_2+ 5_1+9)$ by moving into the parent, and the last two equations arise from $(3_2+ 5_1+9)$ encircling $(1+ 5_2+ 7_2)$ by moving into the parent. For the next off diagonal component we consider $S_{(3_2+ 5_1+9), (2+ 4_2+ 6_2+8) }$, which gives equations 
\begin{subequations}
\begin{align}
~&S_{5,2}+S_{5,4}+S_{5,6}+S_{5,8}=0\,,
\\&S_{9,2}+S_{9,4}+S_{9,6}+S_{9,8}=0\,,
\\&S_{3,2}+S_{5,2}+S_{9,2}=0\,,
\\&S_{3,8}+S_{5,8}+S_{9,8}=0\,.
\end{align} 
\end{subequations}
The final off-diagonal element $S_{(2+ 4_2+ 6_2+8),(1+ 5_2+ 7_2)}$ gives the equations
\begin{subequations}
\begin{align}\label{c3withc1}
~&S_{1,2}+S_{1,4}+S_{1,6}+S_{1,8}=0\,,
\\&S_{5,2}+S_{5,4}+S_{5,6}+S_{5,8}=0 \,,
\\\label{c3withc1third}& S_{2,1}+S_{2,5}+S_{2,7}=0\,,
\\& S_{8,1}+S_{8,5}+S_{8,7}=0\,.
\end{align} 
\end{subequations}
\begin{itemize}
    \item $S_{\text{unconfined,unconfined}}$ 
\end{itemize}

We first consider using $S_{\varphi,\varphi}=\frac{1}{2}$ from the child an obtaining relations for the parent theory. The equations we get are
\begin{subequations}
\begin{align}
~ S_{0,0}+S_{0,9}&={\frac{1}{2}}\,,
\\
S_{0,9}+S_{9,9}&={\frac{1}{2}}\,.
\end{align} 
\end{subequations}
We next consider 
\begin{align}
    \{S_{\varphi,(4_1+10)}=\frac{1}{2},\,\,  S_{\varphi,(3_1+7_1)}=\frac{1}{\sqrt{2}},\,\,S_{(4_1+10), (3_1+7_1)}=-\frac{1}{\sqrt{2}},\,\,\notag \\ S_{(4_1+10),(4_1+10)}=\frac{1}{2},\,\,S_{(3_1+7_1), (3_1+7_1)}=0\}\,,
\end{align}
which give the following relationships in the parent theory:
\begin{subequations}
\begin{align}
    ~&S_{0,4}+S_{0,10}={1\ov2},\quad S_{6,4}+S_{6,10}={1\ov 2}
,\quad S_{0,4}+S_{6,4}={1\ov2}
,\quad S_{0,10}+S_{6,10}={1\ov2}, \\
& S_{0,3}+S_{0,7}=\frac{1}{\sqrt{2}}\,,\quad S_{6,3}+S_{6,7}=\frac{1}{\sqrt{2}}\,, \quad S_{0,3}+S_{6,3}=\frac{1}{\sqrt{2}}\,,\quad S_{0,7}+S_{6,7}={1\ov\sqrt{2}}, \\ 
&S_{4,3}+S_{4,7}= -\frac{1}{\sqrt{2}}\,,\quad  S_{10,3}+S_{10,7}=-\frac{1}{\sqrt{2}}\,,\quad S_{4,3}+S_{10,3}=-\frac{1}{\sqrt{2}}\,,\notag \\ 
& S_{4,7}+S_{10,7}=-{1\ov \sqrt2}\,,\\
& S_{4,4}+S_{4,10}=\frac{1}{2}\,,\quad S_{10,4}+S_{10,10}={1\ov 2}\,, \\
& S_{3,3}+S_{3,7}=0\,,\quad S_{7,3}+S_{7,7}=0.
\end{align}
\end{subequations}
\subsection{Implementing the fusion rules}
\begin{itemize}
    \item Matrix elements of the form $S_{1,-}$
\end{itemize}
Having used found all the relations we can by utilizing the relative center, we now employ the fusion rules of the wall category. We first determining the $S$-matrix elements of form $S_{1,-}$.  Since $1 \times 1 = 0+2$ we can use the Verlinde formula for $N^{0}_{1,1} =1 = \sum_d \frac{S^2_{1,d} S^*_{0,d}}{S_{0,d}}$; we also have $N^2_{1,1}$ but for now we will set that aside. By using the fact that $S^*_{0,d}$ is real, then the Verlinde formula gives
\begin{equation}\label{fromverlinde}
     S^2_{1,{0}}+ S^2_{1,{1}} + \ldots + S^2_{1,10} =1\,.
\end{equation}
Another relation we will have to use frequently is \eqref{verlinde}, in particular we need 
\begin{align}
    S_{1,10\times 2} &= \frac{S_{1,10}S_{1,2}}{S_{1,0}} = S_{1,8}\,, \\ \notag 
    S_{1,10\times 8} &= \frac{S_{10,1}S_{1,8}}{S_{1,0}}= S_{1,2}\,,
\end{align}
these two equations imply that $S^2_{1,2} = S^2_{1,8}$ so $S_{1,2} = \pm S_{1,8}$.
We use this, along with the relations in  \eqref{Aroundc1main}, \eqref{Aroundc1third}, and \eqref{Aroundc1} to simplify \eqref{fromverlinde} to
\begin{equation}\label{N220rewrite}
    2\left(S^2_{1,0}+S^2_{1,2}+S^2_{1,3}+S^2_{1,4} \right)+S^2_{1,1}+S^2_{1,9}+ S^2_{1,5} = 1\,.
\end{equation}
To proceed we first solve for $S_{1,5}$,  From fusion we have the two equations 
\begin{align}
    {S_{1,0}}S_{1,1\times5} &= S_{1,1}S_{1,5} = (S_{1,4}+S_{1,6})S_{1,0}\,,\\
    {S_{1,0}}S_{1,9\times 5} &= S_{1,9}S_{1,5}=(S_{1,4}+S_{1,6})S_{1,0}\,,
\end{align}
which can be combined to give 
\begin{equation}\label{fusionconditionS15}
    S_{1,5}(S_{1,9}-S_{1,1}) = 0\,,
\end{equation}
so either $S_{1,5}=0$ or $S_{1,1}=S_{1,9}$. If $S_{1,9}=S_{1,1}$, and we assume that $S_{1,1}\neq0$, then  from \eqref{Aroundc1}, \eqref{c1withc1}, and \eqref{c1withc2first} we find $S_{1,1}=-S_{1,9}-2S_{1,5}$  so $S_{1,1} = -S_{1,5}$.
But then by \eqref{c1withc1} we get $S_{1,7}=0$, so $S_{1,3}=0$. Furthermore, from 
\begin{equation}
    S_{1,1\times 2} = \frac{S_{1,1}S_{1,2}}{S_{1,0}}\,
\end{equation}
then $S_{1,2}=0=S_{1,8}$, and it is then easy to derive that $S_{1,1}=S_{1,9}=0$, which contradicts our initial assumption.  Therefore we take $S_{1,5}=0$, so that $S_{1,1}=-S_{1,9}$. With this \eqref{N220rewrite} can be simplified to 
\begin{equation}\label{N220simplified}
    2\left(S^2_{1,0}+S^2_{1,1}+S^2_{1,2}+S^2_{1,3}+S^2_{1,4} \right) = 1\,.
\end{equation}
A natural next step to consider is replacing the different squares with as many of the same quantities as possible.  To do this consider the fusion having to do with $S_{1,-}$: 
\begin{align}\label{S1fusions}
    S_{1,1\times 1} &= S_{1,0}+ S_{1,2} = \frac{S^2_{1,1}}{S_{1,0}}\,, \\ \notag
    S_{1, {9}\times 9} &= S_{1,0}+ S_{1,2} = \frac{S^2_{1,9}}{S_{1,0}}\,,\\ \notag 
    S_{1,{2}\times 2} &= S_{1,0}+ S_{1,2}+ S_{1,4} = \frac{S^2_{1,2}}{S_{1,0}}\,, \\ \notag 
    S_{1,{10}\times 10} &= S_{1,0}+ S_{1,2}+S_{1,4} = \frac{S^2_{1,10}}{S_{1,0}}\,, \\ \notag 
    S_{1,\texb{3}\times 3} &=  S_{1,0}+ S_{1,2}+S_{1,4}+S_{1,6} = \frac{S^2_{1,3}}{S_{1,0}}\,, \\ \notag
     S_{1,{7}\times 7} &=  S_{1,0}+ S_{1,2}+S_{1,4}+S_{1,6} = \frac{S^2_{1,7}}{S_{1,0}}\,, \\ \notag
      S_{1,{4}\times 4} &=  S_{1,0}+ S_{1,2}+S_{1,8}+S_{1,4}+S_{1,6} = \frac{S^2_{1,4}}{S_{1,0}}\,, \\ \notag
      S_{1,{6}\times 6} &=  S_{1,0}+ S_{1,2}+S_{1,8}+S_{1,4}+S_{1,6} = \frac{S^2_{1,{6}}}{S_{1,0}}\,, \\ \notag
      { S_{1,5\times 5} }&={  S_{1,0} +S_{1,10}+S_{1,2}+S_{1,8}+S_{1,4}+S_{1,6} = \frac{S^2_{1,5}}{S_{1,0}}  }\,,\\ \notag
\end{align}
and recall that $S_{1,2}+S_{1,4}+S_{1,6}+S_{1,8} = 0 $ by \eqref{c3withc1}. Then we can write, $S^2_{1,3} = S^2_{1,0} - S_{1,0}S_{1,8}$.  We may simplify \eqref{N220rewrite} even further to be
\begin{align}
    2\Bigl[S^2_{1,0}+\left(S^2_{1,0}+S_{1,0}S_{1,2} \right)
  +\left(S^2_{1,0} +S_{1,0}S_{1,2}+ S_{1,0}S_{1,4}\right)   + \left(S^2_{1,0}-S_{1,0}S_{1,8} \right)+S^2_{1,0}\Bigr]&=1\,,   \label{simplifyVerlinde}\\  10S^2_{1,0}+S_{1,0}\left(4S_{1,2}-2S_{1,8}+2S_{1,4} \right) &= 1\,.
\end{align}
We desire some relations between $S_{1,0}, S_{1,4}, S_{1,8}$, we can consider 
\begin{subequations}
\begin{align}
    {S_{1,0}} S_{1,2\times 4} &= -S_{1,0} S_{1,8} = -{S_{1,10} S_{1,2}}\,,\\ 
     {S_{1,0}} S_{1, 8 \times 4} & = -S_{1,0} S_{1,2} = -S_{1,10} S_{1,8}\,,\\ 
    {S_{1,0}}  S_{1, 2 \times 8} & = S_{1,0} S_{1,10} +S_{1,10} S_{1,2} -S_{1,10} S_{1,4} \,.
\end{align}
\end{subequations}
By adding the first two equations we get 
\begin{equation}
    \left(S_{1,0}-S_{1,10} \right) \left(S_{1,2}+S_{1,8} \right)=0\,,
\end{equation}
from which we have either $S_{1,0}=S_{1,10}$ or $S_{1,2}=-S_{1,8}$.  but the last of \eqref{S1fusions} would cause the former choice to run into a contradiction.  We have thus determined $S_{1,2}=-S_{1,8}$ and so $S_{1,4}=-S_{1,6}=S_{1,0}$.  We now try to relate $S_{1,2}$ with $S_{1,0}$, to do this consider the fact that 
\begin{equation}
    S^2_{1,2}=S^2_{1,0}+S_{1,0}S_{1,2}+S_{1,0}S_{1,4}
\end{equation}
and can be simplified to 
\begin{equation}
    S_{1,2}(S_{1,2}-S_{1,0})=2S^2_{1,0}
\end{equation}
which is satisfied if $S_{1,2}=2S_{1,0}$. We summarize how all of $S_{1,-}$ is related to $S_{1,0}$ by the following equations
\begin{align}\label{allS1conditions}
    S^2_{1,1}&=3S^2_{1,0} \,,& S^2_{1,2} &= 4S^2_{1,0}\,,&  S^2_{1,3} &=3S^2_{1,0}\,,& S^2_{1,4}&=S^2_{1,0}\,, \notag \\
    S^2_{1,5}&=0\,, & S^2_{1,6}&=S^2_{1,0}\,, & S^2_{1,7}&=3S^2_{1,0}\,, &  S^2_{1,8}&=4S^2_{1,0}\,,\notag \\
    S^2_{1,9}&=3S^2_{1,0}\,, & S^2_{1,10}&=S^2_{1,0}\,,
\end{align}
and therefore \eqref{N220simplified} becomes $24S^2_{1,0}=1$, and thus $S_{1,0}=\frac{1}{\sqrt{24}}$.\\

We now repeat a similar process to find the elements of $S_{2,-}$.  We start off systematically by giving the fusion rules:
\begin{subequations}\label{4fusion}
\begin{align}\label{a}
    S_{2,0}\,S_{2,2\times 10} &= S_{2,0}(S_{2,8})\,,\\  \label{b}
    S_{2,0}\,S_{2,2\times 1} &= S_{2,0}(S_{2,1}+S_{2,3})\,,\\ \label{c}
   S_{2,0} \,S_{2,2\times 9} &= S_{2,0}(S_{2,9}+S_{2,7})\,,\\ \label{d}
    S_{2,0}\,S_{2,2\times 2} &= S_{2,0}(S_{2,0}+S_{2,2}+S_{2,4})\,,\\ \label{e}
    S_{2,0}\,S_{2,2\times 8} &= S_{2,0}(S_{2,10}+S_{2,8}+S_{2,6})\,,\\ \label{f} 
    S_{2,0}\,S_{2,2\times 3} &= S_{2,0}(S_{2,1}+S_{2,3}+S_{2,5})\,,\\ \label{g}
    S_{2,0}\,S_{2,2\times 7} &= S_{2,0}(S_{2,9}+S_{2,7}+S_{2,5})\,,\\ \label{h}
    S_{2,0}\,S_{2,2\times 4} &= S_{2,0}(S_{2,2}+S_{2,4}+S_{2,6})\,,\\ \label{i}
    S_{2,0}\,S_{2,2\times 6} &= S_{2,0}(S_{2,8}+S_{2,4}+S_{2,6})\,,\\ \label{j}
    S_{2,0}\,S_{2,2\times 5} &= S_{2,0}(S_{2,3}+S_{2,7}+S_{2,5}) \,.
\end{align}\label{k}
\end{subequations}
From \eqref{S23S27} applied to \eqref{j} then $S_{2,2}S_{2,5}=S_{2,0}S_{2,5}$  which gives us two conditions: either $S_{2,5}=0$ or $S_{2,2}-S_{2,0}=0$, or both. Let us consider first $S_{2,5}=0$ without putting conditions on $S_{2,2}-S_{2,0}$ just yet. A remarkable fact is that we can show that this leads to a contradiction down the line, and thus was the incorrect choice.   We go to \eqref{4fusion} and massage the equations based off the assumption $S_{2,5}=0$.
\begin{align}
    \eqref{a} &\rightarrow S_{2,2}S_{2,10} = S_{2,0}S_{2,8}\,, \\ 
    \eqref{b}+ \eqref{c} &\rightarrow (S_{2,2}-S_{2,0})(S_{2,1}+S_{2,9}) = 0\,, \label{second}\\
    \eqref{d}+ \eqref{e} &\rightarrow S_{2,2}(S_{2,2}+S_{2,8}) = S_{2,0}(S_{2,0}+S_{2,10})\,, \\\label{104}
    \eqref{f} &\rightarrow S_{2,2}S_{2,3}= (S_{2,1}+S_{2,3})S_{2,0}\,,\\
    \eqref{g} &\rightarrow S_{2,2}S_{2,7} = (S_{2,9}+S_{2,7})S_{2,0}\,,\\\label{condition}
    \eqref{h}- \eqref{i} &\rightarrow S_{2,2}(S_{2,4}-S_{2,6}) = (S_{2,2}-S_{2,8})S_{2,0}\,,\\
    \eqref{j} &\rightarrow 0\,.
\end{align}
Equations \eqref{h} and \eqref{i} can be added to get $S_{2,2}(S_{2,4}+S_{2,6}) = S_{2,0}(S_{2,4}+S_{2,6})$, and therefore 
\begin{equation}
    (S_{2,2}-S_{2,0}) (S_{2,4}+S_{2,6})=0\,.
\end{equation}
There are multiple possibilities to consider, either 
\begin{enumerate}
    \item $S_{2,2}- S_{2,0}=0\,,\quad S_{2,4}+S_{2,6}=0$\,,
    \item $S_{2,2}- S_{2,0}=0\,,\quad S_{2,4}+S_{2,6}\neq0\,,$ 
    \item  $S_{2,4}+S_{2,6} = 0\, ,\quad S_{2,2}-S_{2,0} \neq 0$\,.
\end{enumerate}
Suppose we consider the first of the above cases. But then \eqref{104} would imply that $S_{2,1}=0$, but it was solved already in \eqref{allS1conditions} that $S_{2,1}\neq 0$, so we have a contradiction. The second case also leads to a contradiction by the same reason as the first condition. One can also check that the third case is invalid as well. Thus our assumption that $S_{2,5}=0$ was incorrect. We amend this choice and instead let $S_{2,5} \neq 0$ but let ${S_{2,2}-S_{2,0}=0}$.  This does not run into the problem of earlier because if $S_{2,5} \neq 0$, then \eqref{f} is not simply $S_{2,3} = S_{2,1}+S_{2,3}$, but rather $S_{2,3} = S_{2,1}+S_{2,3}+S_{2,5}$. We use this to simplify the equations in \eqref{4fusion}
\begin{align}
    \eqref{a} &\rightarrow S_{2,2}S_{2,10} = S_{2,0} S_{2,8}\,, \\ \notag
    \eqref{b} &\rightarrow S_{2,3}=0\,, \\ \notag 
    \eqref{c} &\rightarrow S_{2,7}=0\,, \\ \notag 
    \eqref{d} & \rightarrow S_{2,0}+S_{2,4}=0\,, \\ \notag 
    \eqref{e} & \rightarrow S_{2,10}+S_{2,6}=0\,, \\ \notag 
    \eqref{f} & \rightarrow S_{2,1}+S_{2,5}=0\,, \\ \notag 
    \eqref{g} & \rightarrow S_{2,9}+S_{2,5}=0\,, \\ \notag 
    \eqref{h} & \rightarrow S_{2,2}+S_{2,6}=0\,, \\ \notag 
    \eqref{i} & \rightarrow S_{2,8}+S_{2,4}=0\,, \\ \notag 
    \eqref{j} & \rightarrow S_{2,3}+S_{2,7}=0\,. \\ \notag 
\end{align}
The important part now is to relate everything back to $S_{2,0}$ and $S_{2,1}$, the latter which we already obtained. In total we have
\begin{align}
    S_{2,0}&=S_{2,2}=-S_{2,4}=-S_{2,6}=S_{2,8}=S_{2,10}\notag \\
    S_{2,1}&=-S_{2,5}=S_{2,9}.
\end{align}
Now using the Verlinde formula in the form $N^{0}_{2,2} =1 = \sum_d \frac{S^2_{2,d} S^*_{0,d}}{S_{0,d}}$ we have 
\begin{align}
    1 = 6S^2_{2,0}+3S^2_{2,1}\notag \\
    1=\frac{1}{2}+6S^2_{2,0}\,,
\end{align}
thus $S_{2,0}=\frac{1}{\sqrt{12}}$.\\

We now skip to finding the matrix elements of $S_{5,-}$, this is because 5 behaves differently from the other lines. The fusion rules give 
\begin{subequations}\label{line5}
\begin{align}\label{line5a}
   S_{5,0}\, S_{5,5\times 10} &= S_{5,0}S_{5,5}\,,\\  \label{line5b}
   S_{5,0}\, S_{5,5\times 1} &= S_{5,0}(S_{5,4}+S_{5,6})\,,\\ \label{line5c}
   S_{5,0} \,S_{5,5\times 9} &= S_{5,0}(S_{5,4}+S_{5,6})\,,\\ \label{line5d}
    S_{5,0}\,S_{5,5\times 2} &= S_{5,0}(S_{5,3}+S_{5,7}+S_{5,5})\,,\\ \label{line5e}
   S_{5,0}\, S_{5,5\times 8} &= S_{5,0}(S_{5,3}+S_{5,7}+S_{5,5})\,,\\ \label{line5f} 
   S_{5,0}\, S_{5,5\times 3} &= S_{5,0}(S_{5,2}+S_{5,8}+S_{5,4}+S_{5,6})\,,\\ \label{line5g}
   S_{5,0}\, S_{5,5\times 7} &= S_{5,0}(S_{5,2}+S_{5,8}+S_{5,4}+S_{5,6})\,,\\ \label{line5h}
   S_{5,0}\, S_{5,5\times 4} &= S_{5,0}(S_{5,1}+S_{5,9}+S_{5,3}+S_{5,7}+S_{5,5})\,,\\ \label{line5i}
   S_{5,0}\, S_{5,5\times 6} &= S_{5,0}(S_{5,1}+S_{5,9}+S_{5,3}+S_{5,7}+S_{5,5})\,,\\ \label{line5j}
   S_{5,0} \,S_{5,5\times 5} &= S_{5,0}(S_{5,0}+S_{5,10}+S_{5,2}+S_{5,8}+S_{5,4}+S_{5,6}) \,.
\end{align}\label{line5k}
\end{subequations}
manipulating the equations gives 
\begin{align}
    \eqref{line5a} &\rightarrow S_{5,5}(S_{5,10}-S_{5,0})=0\,, \\ \notag
    \eqref{line5b}-\eqref{line5c} &\rightarrow S_{5,5}(S_{5,1}-S_{5,9})=0\,, \\ \notag 
    \eqref{line5d} &\rightarrow S_{5,5}(S_{5,0}-S_{5,2})=0 \,,\\ \notag 
    \eqref{line5e} & \rightarrow S_{5,5}(S_{5,0}-S_{5,8})=0\,, \\ \notag 
    \eqref{line5f} & \rightarrow S_{5,0}S_{5,3}=0\,, \\ \notag 
    \eqref{line5g} & \rightarrow S_{5,0}S_{5,7}=0\,, \\ \notag 
    \eqref{line5h} & \rightarrow S_{5,5}S_{5,4}=-S_{5,0}S_{5,5}\,, \\ \notag 
    \eqref{line5i} & \rightarrow S_{5,5}S_{5,6}=-S_{5,0}S_{5,5}\,, \\ \notag 
    \eqref{line5j} & \rightarrow S^2_{5,5}=S_{5,0}(S_{5,0}+S_{5,10})\,,\\ \notag 
\end{align}
We have some choices, from the first of the equations we could have $S_{5,5}=0$ and also $S_{10,1}-S_{10,0}=0$. But then that contradicts the last equation of the above. Now suppose that $S_{5,10}=S_{5,0}$ with $S_{5,5}\neq 0$. Then we get $S_{1,5}+S_{5,5}=0 $ from one of our previous equations.  However, we said before that $S_{1,5}$ around equation \eqref{N220simplified} this was already zero, so then $S_{5,5}$ would also have to be zero which is a contradiction. So we need to have $S_{5,10}\neq S_{5,0}$ and $S_{5,5}= 0$.
Because from earlier $S_{3,5}+S_{5,9}=0$, then $S_{5,9}=0$,
and furthermore from \eqref{c3withc3} and \eqref{Aroundc1third} we have $S_{5,4}=-S_{5,10}$ with $S_{5,10}=-S_{5,0}$ in \eqref{line5j}.
The relationships are summarized as 
\begin{align}
    S_{5,1}&=0\,,& S_{5,2}&=-S_{2,1}=-\frac{2}{\sqrt{24}}\,,& S_{5,3}&=0\,,&  S_{5,4}&=S_{5,0}\,,\\
    S_{5,5}&=0\,,& S_{5,6}&=-S_{5,0}\,, & S_{5,7}&=0\,, & S_{5,8}&=S_{2,1}\,,\\  
    S_{5,9}&= 0\,,& S_{5,10}&=S_{5,0}\,.
\end{align}
Then by the Verlinde formula we have 
\begin{equation}
    1= \sum_{a} S^2_{5,a}=1/3+4S^2_{5,0}
\end{equation}
so $S_{5,0} = \frac{1}{\sqrt{6}}$.

\bibliographystyle{JHEP}
\bibliography{AnyonCondensation.bib}

\end{document}